\documentclass[manuscript]{acmart}
\AtBeginDocument{%
  \providecommand\BibTeX{{%
    \normalfont B\kern-0.5em{\scshape i\kern-0.25em b}\kern-0.8em\TeX}}}

\copyrightyear{2022} 
\acmYear{2022} 
\setcopyright{acmlicensed}\acmConference[ASSETS '22]{The 24th International ACM SIGACCESS Conference on Computers and Accessibility}{October 23--26, 2022}{Athens, Greece}
\acmBooktitle{The 24th International ACM SIGACCESS Conference on Computers and Accessibility (ASSETS '22), October 23--26, 2022, Athens, Greece}
\acmPrice{15.00}
\acmDOI{10.1145/3517428.3544824}
\acmISBN{978-1-4503-9258-7/22/10}

\usepackage{array,multirow,graphicx}
\usepackage{tabularx}
\usepackage{makecell}
\usepackage{tabu}
\usepackage{caption}
\usepackage{subcaption}
\usepackage{csquotes}
\usepackage{wrapfig}
\usepackage{soul}

\usepackage{xspace}
\newcommand{\eg}{\textit{e.g.}}
\newcommand{\ie}{\textit{i.e.}}
\newcommand{\etal}{\textit{et al.}\@\xspace}
\begin{document}

\title{Blind Users Accessing Their Training Images in Teachable Object Recognizers}

% \affiliation{%

%% Authors: 
\author{Jonggi Hong}
\affiliation{%
  \institution{Smith-Kettlewell Eye Research Institute}
  \streetaddress{2318 Fillmore St.}
  \city{San Francisco}
  \country{United States}
  }
\email{jhong@ski.org}
\orcid{0000-0003-1060-6770}

\author{Jaina Gandhi}
\affiliation{%
  \institution{University of Maryland, College Park}
  \country{United States}
  }
\email{jaina.gandhi@gmail.com}

\author{Ernest Essuah Mensah}
\affiliation{%
  \institution{ University of Maryland, College Park}
  \country{United States}
  }
\email{eessuahm@umd.edu}

\author{Farnaz Zamiri Zeraati}
\affiliation{%
  \institution{University of Maryland, College Park}
  \country{United States}
  }
\email{farnaz@umd.edu}

\author{Ebrima Haddy Jarjue}
\affiliation{%
  \institution{University of Maryland, College Park}
  \country{United States}
  }
\email{ebjarjue@terpmail.umd.edu}

\author{Kyungjun Lee}
\affiliation{%
  \institution{University of Maryland, College Park}
  \country{United States}
  }
\email{kjlee@cs.umd.edu}
\orcid{0000-0001-8556-9113}

\author{Hernisa Kacorri}
\affiliation{%
  \institution{University of Maryland, College Park}
  \country{United States}
  }
\email{hernisa@umd.edu}
\orcid{0000-0002-7798-308X}

\renewcommand{\shortauthors}{Hong, et al.}
 
%% Paper length: between 8,000 - 10,000 words
%% Abstract length: 150 words

%% The abstract is now 144 words. Feel free to change it.
\begin{abstract}

Teachable object recognizers provide a solution for a very practical need for blind people – instance level object recognition.  They assume one can visually inspect the photos they provide for training, a critical and inaccessible step for those who are blind. In this work, we engineer data descriptors that address this challenge. They indicate in real time whether the object in the photo is cropped or too small, a hand is included, the photos is blurred, and how much photos vary from each other. Our descriptors are built into open source testbed iOS app, called MYCam. In a remote user study in ($N=12$) blind participants' homes, we show how descriptors, even when error-prone, support experimentation and have a positive impact in the quality of training set that can translate to model performance though this gain is not uniform. Participants found the app simple to use indicating that they could effectively train it and that the descriptors were useful. However, many found the training being tedious, opening discussions around the need for balance between information, time, and cognitive load.

% Iteration of training and evaluating a machine learning model is an important process to improve its performance. However, while teachable interfaces enable blind users to train and test an object recognizer with photos taken in their distinctive environment, accessibility of training iteration and evaluation steps has received little attention. Iteration assumes visual inspection of the training photos, which is inaccessible for blind users. We explore this challenge through MyCam, a mobile app that incorporates automatically estimated descriptors for non-visual access to the photos in the users' training sets. We explore how blind participants ($N=12$) interact with MyCam and the descriptors through an evaluation study in their homes. We demonstrate that the real-time photo-level descriptors enabled blind users to reduce photos with cropped objects, and that participants could add more variations by iterating through and accessing the quality of their training sets. Also, Participants found the app simple to use indicating that they could effectively train it and that the descriptors were useful. However, subjective responses were not reflected in the performance of their models, partially due to little variation in training and cluttered backgrounds.

\end{abstract}


\begin{teaserfigure}
  \includegraphics[width=0.95\columnwidth]{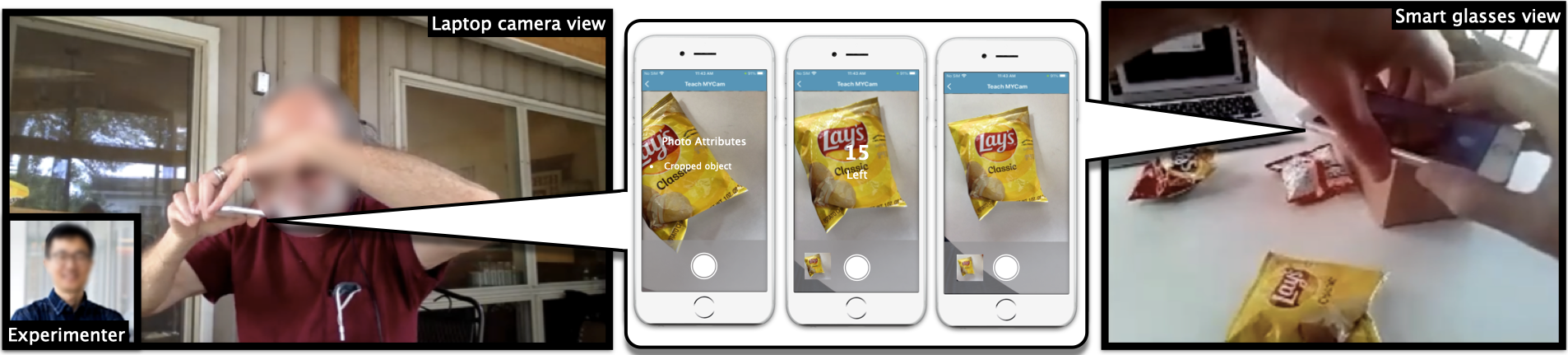}
  \caption{A blind participant in our study training the MYCam app in their homes to recognize Lays with real-time descriptors. A dual video conferencing captures participant’s activities via a laptop camera and smart glasses worn by the participant.
  }
  \label{fig:teaser}
\end{teaserfigure}
% Alt text:
% The study setup. On the left there is the laptop camera view showing a blind participant at their home with blurred face, while pointing the smartphone's camera downwards. At the bottom left of this picture there is the picture of the experimenter with blurred face. An arrow goes from the smartphone in the participant's hand to a box located at the middle part of this figure, including three screenshots of the MyCam app. The screenshots show the Lays object on a white background in the camera view, and the shutter button at the bottom of the screen. In the first screenshot from the left, the left upper corner of the Lays object is cropped and there is overlaid text showing photo attributes cropped object. On the right there is smart glasses view, showing the participant's hands pointing an iphone 8 to Lays on a white table. There is also a laptop and two other snacks in front of the participant. An arrow goes from the smartphone in this picture to the middle box including the screenshots.

\maketitle

\section{Introduction}
    Echoing the end-user programming paradigm, the idea of having end-users consciously provide training examples in AI-infused applications
    % \footnote{A term coined by Amershi \etal (2019)~\cite{amershi2019guidelines}.}
    has recently gained traction along with advances in neural networks. By leveraging prior work in transfer, meta, and few-shot learning (\eg,~\cite{vinyals2016matching, bronskill2020tasknorm, sun2019meta, bauer2017discriminative, zhang2021meta}), we are now able to build \textit{teachable} applications, where end-users can train models of their own. These applications facilitate personalization as they promise a better fit for real-world scenarios by significantly constraining the machine learning task to a specific user and their environment. Thus, it is no surprise to see early mentions of the term ``teachable'' in accessibility research (\eg~\cite{patel1998teachable}), where data is sparse and there is a high diversity even for a given disability~\cite{kacorri2020incluset, kacorri2020data}. A more recent example (and the focus of this work) is teachable object recognizers~\cite{kacorri2017people, sosagarcia2017hands, lee2019revisiting, lee2019hands, ahmetovic2020recog, massiceti2021orbit}, where blind users train their camera-equipped devices such as mobile phones to recognize everyday objects by providing a few photos as training examples.

   \textbf{Why is it important to access one's training examples?}
    In teachable applications such as teachable object recognizers, users are called to interact with the machine learning model and improve its performance by accessing and controlling their examples~\cite{zhu2018overview}. Personalization is often the ultimate goal. However, the interactive nature  of these applications can also help people uncover basic machine learning concepts and gain familiarity with AI (\eg,~\cite{hitron2019can, queiroz2020ai, carney2020teachable, hong2020crowdsourcing, dwivedi2021exploring}). Thus, they can also contribute to the larger goal of ``\textit{making the process of teaching machines easy, fast and above all, universally accessible}''~\cite{simard2017machine}. An underlying assumption for both improving a model and uncovering concepts via experimentation is that users can inspect their data and iterate the training and testing. By doing so, they could build an intuition about what works and what doesn't and perhaps why. However, this assumption does not often hold for assistive teachable applications.  Inspecting training examples typically requires similar skills to those the technology aims to fulfill~\cite{kacorri2017teachable, goodman2021toward}; thus, it is often inaccessible. For example,
    \textit{teachable object recognizers, where users teach the model to recognize objects on their behalf, assume that they can see the training images they are providing, which is almost never the case with blind users}.  Sure enough, blind participants in prior studies with this technology wanted to know more about their training examples~\cite{kacorri2017teachable} with one of them stating ``\textit{the most challenging and most fun is training the person}''.
    % [who will be teaching the machine]}''. 
    
    Existing approaches for real-time `\textit{alt text}' for individual images and `\textit{scene description}' for a series of images are not suitable for this task; they do not capture fine-grained differences across otherwise similar images (\eg, see Figure~\ref{fig:teaser}).
    % Hong \etal \cite{hong2020crowdsourcing} highlighted this challenge by showing that even sighted non-experts in machine learning could not improve their training sets through iterations as they did not know what to change.
    In this paper, we explore this \textbf{challenge of accessing one's training data}.
    % which can be uniquely faced by users of teachable assistive technologies.
    Within the context of teachable object recognizers for the blind, we study the potential and limitations of real-time `\textit{data descriptors}' that can capture users' training examples with photo- and set-level attributes. Specifically, we investigate whether these descriptors could be derived from visual attributes used to code training photos from sighted (\eg,~\cite{hong2020crowdsourcing, hong2019exploring}) and blind (\eg,~\cite{kacorri2017people, lee2019revisiting}) people. 
    To this end, we engineer photo-level descriptors that communicate to the user in real-time information about the photo they just took such as blurriness, presence of their hand, object visibility, and framing. We also engineer set-level descriptors that communicate information one would get from glancing over a group of training photos such as variation in object background, distance, and perspective; all factors that can affect model performance.
    
    Through a remote user study with 12 blind participants (with the setup shown in Figure~\ref{fig:teaser}), we demonstrate that our data descriptors support blind users in reducing photos with cropped objects in their training sets and increase variation.
    Many participants chose to iterate after inspecting their training sets and reflected by improving many photo attributes, which resulted in models that generalize better to photos from others, even though they reduced variation in background. 
    Aligned with prior studies, we also observe challenges among participants in crafting good testing examples that could further promote experimentation. Still, their models perform better when tested on their own photos compared to both aggregated test sets from all 12 blind  participants in our remote study and from 9 blind participants in an in-lab study~\cite{lee2019revisiting}. 
    Observations from our analysis of the photos and model performance are also confirmed by participants' subjective feedback. Responses support the potential of descriptors, with blind participants indicating that they were able to tell their meaning by looking at relative changes in values and finding them useful. However, errors in descriptors affected the reliability of the app for some. More so, some considered training being tedious referring both to time and cognitive load (\eg, optimize for multiple variables). Many made design recommendations that could further improve the effectiveness of the descriptors and the training process.

    % Contributions
   To the best of our knowledge, this is the first work to propose non-visual access to training data and to provide empirical results with blind participants on automatically estimating and incorporating descriptors for data inspection in teachable computer vision applications. Our analysis focuses on object recognizers, where `learning to train' is deemed as one of the main challenges among blind users~\cite{kacorri2017people, kacorri2017teachable}. However, we see how the underlying methods for extracting meaningful instance- and set-level descriptors can be adopted for other teachable applications both in assistive and informal AI learning contexts. Perhaps, they can also serve towards more accessible approaches for explainable AI interfaces, where there is an underlying assumption on people's ability to visually inspect explanations~\cite{stock2018convnets,samek2017explainable,simonyan2014deep}.

% Some texts about why we need descriptors instead of alt texts?
% \subsubsection{Rationale for Descriptors.}

% add text here 
% alt text -- image level
% scene descriptors 
    
\section{Related Work}
There is a rich literature exploring how computer vision can benefit people with disabilities (\eg,~\cite{reyes2012rehabilitation, khan2014computer, jiang2016enhanced, manresa2014design, campbell2019computer, tapu2019deep, bragg2019sign}). This is especially the case with assistive technologies for the blind, where computer vision is employed on smartphones (\eg,~\cite{ahmetovic2014zebrarecognizer, guo2016vizlens, zhao2018face, saha2019closing, kuribayashi2021linechaser, yamanaka2022one-shot}), smart glasses (\eg,~\cite{fiannaca2014headlock, zhao2019designing, lee2020pedestrian, son2020crosswalk}),
% systems for blind wheelchair users (\eg,~\cite{Ivanchenko2008Computer, Devigne2019shared}), 
and smart suitcases (\eg,~\cite{kayukawa2019bbeep, guerreiro2019cabot}). 
% Some have made it to real-world applications (\eg,~\cite{SeeingAI, TapTapSee, Envision, Aipoly, OrCam}).  
A common challenge we share with prior work is that aiming the camera and inspecting recognition errors typically requires similar skills to those the technology aims to fulfill (\ie, sight), even though the majority of prior work employs AI-infused systems pre-trained by engineers, not fine-tuned by the end-user. Thus, it is not a surprise to find that recognition errors affect blind users' experience~\cite{morris2020ai}; sometimes, to a degree where it can not be corrected even by human clarification~\cite{Salisbury2017TowardSS}. In fact, blind users may depend on the recognition especially when it is difficult to  verify its predictions. They may overtrust the predictions even when they know they can be error-prone~\cite{macleod2017understanding, lee2019revisiting} though, errors are especially non-acceptable when they can adversely affect interactions with others~\cite{abdolrahmani2017embracing, lee2020pedestrian}. Aligned with prior efforts aiming to support users' recovery from errors~\cite{amershi2019guidelines, hook2000steps, norman1994might}, we explore how to make training and resolving errors in teachable object recognizers more accessible to blind users.
Below, we focus on prior work that closely relates to ours and contrast it to our study. 

\subsection{Teachable Object Recognizers}

\begin{table}[t]
\small
\centering
\caption{Characteristics of related studies on teachable object recognizers juxtaposed with ours.}
       \begin{tabular}{l r c c c c c c c c c c}
            \toprule
              & & \cite{kacorri2017people} &\cite{sosagarcia2017hands} & \cite{lee2019revisiting} & \cite{hong2020crowdsourcing} & \cite{ahmetovic2020recog} & \cite{vartiainen2020learning} & \cite{dwivedi2021exploring} & \cite{theodorou2021disability} & This study \\
            \midrule
            \multirow{4}{*}{\textbf{Participants}}
            & Blind Adults          & 8         & 14        & 9         &           & 10        &           &           & 52        & 12         \\
            & Sighted Adults        & 2         & 10        & 2         & 100       &           &           &           &           &            \\
            & Sighted Children      &           &           &           &           &           & 6         & 14        &           &            \\
            \midrule
            \multirow{2}{*}{\textbf{Setting}}
            & Real-world            & $\bullet$ &           &           & $\bullet$ & $\bullet$ &           & $\bullet$ & $\bullet$ & $\bullet$  \\
            & Controlled            & $\bullet$ & $\bullet$ & $\bullet$ &           & $\bullet$ & $\bullet$ & $\bullet$ &           &            \\
            \midrule
            \multirow{2}{*}{\textbf{Input}}
            & Photo                 & $\bullet$ &           & $\bullet$ & $\bullet$ & $\bullet$ & $\bullet$ & $\bullet$ &           & $\bullet$  \\
            & Video                 &           & $\bullet$ &           &           &           &           &           & $\bullet$ &            \\
            \midrule
            \multirow{3}{*}{\textbf{Tasks}}
            & Train                 & $\bullet$ &           & $\bullet$ & $\bullet$ & $\bullet$ & $\bullet$ & $\bullet$ & $\bullet$ & $\bullet$  \\
            & Test                  & $\bullet$ & $\bullet$ & $\bullet$ & $\bullet$ & $\bullet$ & $\bullet$ & $\bullet$ & $\bullet$ & $\bullet$  \\
            & Iterate               &           &           &           & $\bullet$ &           & $\bullet$ & $\bullet$ &           & $\bullet$  \\
            \midrule
            \multirow{2}{*}{\textbf{Access}}
            & Framing               &           &           & $\bullet$ &           & $\bullet$ &           &           &           &            \\
            & Review                &           &           &           & $\bullet$ &           & $\bullet$ & $\bullet$ &           & $\bullet$  \\
        \bottomrule
    \end{tabular}%
\label{tab:related_work}
\end{table}

Looking at prior work on teachable object recognizers, we see diversity in research aims. Some, similar to this work, focus on the blind community. They explore the potential of teachable object recognizers as an assistive technology for blind users~\cite{kacorri2017people, sosagarcia2017hands}, build feedback mechanisms for better camera aiming~\cite{lee2019revisiting, ahmetovic2020recog}, and collect benchmarking datasets for evaluating approaches in transfer learning and meta learning~\cite{theodorou2021disability}. Our work is orthogonal and highly complementary to these efforts -- our shared goal is to improve blind users' experience with teachable object recognizers. 

We also see studies involving sighted people both adults (\eg,~\cite{hong2020crowdsourcing}) and children (\eg,~\cite{dwivedi2021exploring, vartiainen2020learning}). They aim to better understand the potential of teachable machines for enabling non-experts to uncover basic machine learning concepts as well as better understand common AI misconceptions they may have. Insights from these studies are very informative for our efforts in making the \textit{`learning to train'} challenge more accessible to blind adults and perhaps in the future to blind children that may want to participate in similar informal learning activities as in Dwivedi \etal~\cite{dwivedi2021exploring}.

Table~\ref{tab:related_work} provides a more detailed overview from a sample of these prior studies over the past five years (2017-2021) with the number of participants being typically smaller for in-person studies with blind people and sighted children.  As the performance of teachable object recognizers and users' behavior in taking photos can be affected by environmental factors such as background, light condition, and selection of objects, many studies collected inputs from participants' environments to incorporate these factors~\cite{kacorri2017people, hong2020crowdsourcing, ahmetovic2020recog, theodorou2021disability}.  We also opted for this approach in our study; the study was conducted in the homes of blind participants while we control for factors such as study procedure and object stimuli. 

The majority of prior work on teachable object recognizers facilitates training through photos~\cite{kacorri2017people, lee2019revisiting, hong2020crowdsourcing, carney2020teachable, ahmetovic2020recog}, except for one~\cite{theodorou2021disability}, where blind users are called to use short videos. In our study, we also used photos so that the outcomes of our study could be applicable (and comparable) to the majority of existing approaches. More so, collecting videos may increase the burden on the user, especially when they are given several instructions and tasks to do~\cite{theodorou2021disability} as in the case of our study. In addition, video-based assistive technology can pose a greater privacy risk for blind users~\cite{akter2020privacy} as it is more likely to capture unwanted objects and unnecessary information in a video. Perhaps, live photos~\cite{olson2021livephoto}, could be the middle ground between the two. We further reflect on the potential of this approach in the discussion section.

In their early explorations, Kacorri~\etal~\cite{kacorri2017people} highlighted some of the main challenges that blind users may face when training a teachable object recognizer and testing its performance.  They revolve around camera framing (\ie, adjusting the distance between the camera and an object and centering the object), capturing the side of the object with the most distinctive visual features (\ie, product logos), and reviewing the training photos after taking them (\ie, quality and characteristics). Lee \etal~\cite{lee2019revisiting} and Ahmetovic \etal~\cite{ahmetovic2020recog}  aimed to resolve the camera framing challenge by developing real-time audio/haptic feedback that helps blind users estimate the proper distance and position of the object in the image frame~\cite{lee2019revisiting, ahmetovic2020recog}.  However, the challenge of reviewing photos for iteration has not been addressed yet.  

Typically, studies included a training and testing step for exploring participants' interactions with the teachable interfaces. Very few of them~\cite{hong2020crowdsourcing, dwivedi2021exploring}, though, allowed people to reflect and iterate giving them access to their training data for review. We believe iteration is a critical step for understanding the potential of \textit{descriptors} for making the review process more accessible. Thus, in our study, we also provide blind participants with an opportunity to reflect and the option to iterate after reviewing their images with the descriptors. 
After all, our goal is to examine how data descriptors that provide non-visual access to training photos, either individually or as a set, can be helpful  during the iterative process of training and testing, as well as how blind users may interact with them.

\begin{figure}[b]
    \centering
    \includegraphics[width=\textwidth]{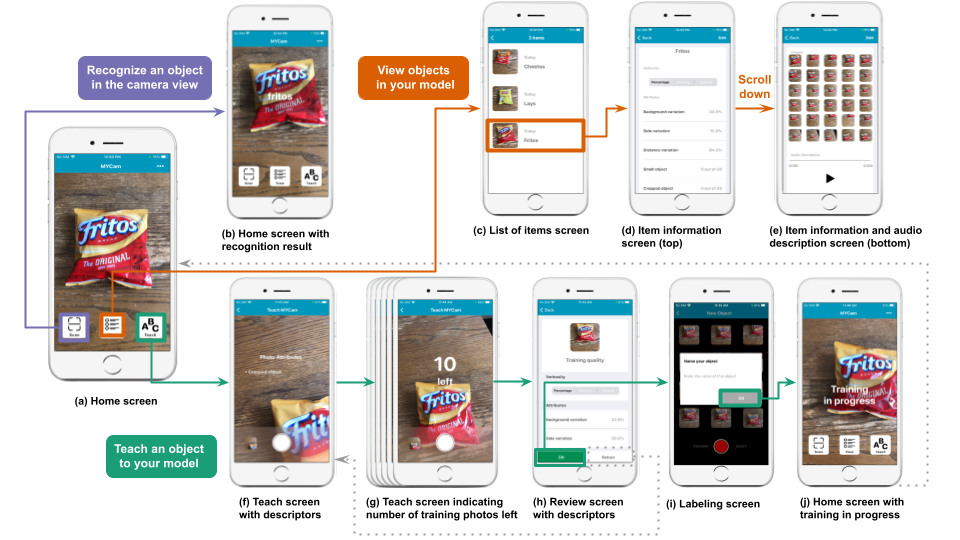}
    \caption{The user flow of MyCam. MyCam has three main parts: Recognizing an object in the camera view (purple thread), reviewing and editing the information of the objects (red thread), and teaching an object to the model (green thread).}
    \label{fig:tor_ui_1}
\end{figure}

\section{MyCam: A Testbed Teachable Object Recognizer with Descriptors}
To explore the potential and limitations of real-time \textit{descriptors} derived from visual attributes for accessing one's training data, we build MYCam. MYCam serves as a testbed for deploying descriptors in a teachable object recognizer. In the background, it sends users' photos to a server, where an image recognition model is being fine-tuned by the user. While privacy is one of the promises of teachable object recognizers~\cite{kacorri2017teachable} (\ie, by processing photos entirely on the user's mobile device), we find that the state-of-the-art on-device training is not there yet. As a screen-reader accessible iOS mobile app, MYCam enables remote studies with blind participants. This was critical for us; due to the pandemic, we had to move our study from the lab to blind participants’ homes. By open sourcing both the MYCam app (available at \url{https://iamlabumd.github.io/MYCam-Mobile/}) and our proof-of-concept implementation of the descriptors (available at \url{https://iamlabumd.github.io/MYCam-Server/}), we are hoping that others can contribute to further advance this work.

\subsection{Design Rationales}
\textit{\textbf{DR1: Prioritize Blind Users.}} Both the form factor and interaction modalities of MYCam are informed by prior work with blind users and teachable object recognizers as well as broader real-world object recognition applications. We opted for an iOS app since prior work in the United States, the location of our study participants, suggests that blind smartphone users overwhelmingly favor the iPhone~\cite{morris2014blind} though the actual numbers may be changing these past years~\cite{morales2019what}. 
When users open the app, they enter the main screen (Figure~\ref{fig:tor_ui_1}a), which shows a camera preview. We opted for the default camera app in iOS maximizing both compatibility with VoiceOver and user familiarity with it. The recognition mode for MYCam was modeled after existing real-world applications, such as Seeing AI~\cite{SeeingAI}, where users can immediately ask the app to recognize what is captured by the camera with a double-tap; the \textit{Scan} button is activated by default.
In this case, the app takes a photo, sends it to the personalized object recognition model in the server, and indicates the predicted label both via speech and visually (Figure~\ref{fig:tor_ui_1}b).  
To mitigate potential errors that can't be verified non-visually, the app says \textit{``Don't know''} when uncertain (approach for uncertainty is discussed in Implementation).

\textit{\textbf{DR2: Simplify the Machine Teaching Flow.}}
% \subsubsection{Training}
Users can add a new object to the recognition model via the \textit{Teach} button on the Home screen.  
The app displays the (rear-facing) camera preview with the shutter button at the bottom center and a thumbnail image of the last photo in the lower-left corner (Figure~\ref{fig:tor_ui_1}f). Users are asked to take 30 photos with the count indicated in real-time via speech and visually (Figure~\ref{fig:tor_ui_1}g); in Kacorri \etal~\cite{kacorri2017people}, blind participants indicated that they would like to obtain feedback from the camera on the number of photos taken. The number of training examples (\ie, 30) is also informed by the same study~\cite{kacorri2017people} with blind participants spending on average 65 seconds (\textit{SD}=35.2) to take 30 photos and often providing variation in their training examples. More so, $k$-shot learning results in the literature are often reported for $k=1$, $5$, or $20$. Thus, 30 examples could allow for bootstrap estimates for future comparisons in this field. As discussed in Related Work, the majority of prior work in teachable object recognizers opt for photos rather than videos -- we followed this approach in hope that photos provide blind users with more control over their training examples in terms of both conscious variation incorporated and privacy concerns mitigated (\eg, presence of their hand, bystanders, or surroundings in the camera frame). In the one study where videos were used, blind users had to be explicitly trained to record videos and follow specific filming techniques~\cite{theodorou2021disability}. Given the emphasis of this study on the descriptors, we decided to simplify the machine teaching and not require any explicit training steps for the users.
After the training examples, a dialogue box with a text field shows up prompting users to enter the name of the object (Figure~\ref{fig:tor_ui_1}i). More so, in this screen users can opt to add an audio description. (Both object name and description can be edited at a later time, as shown in Figures~\ref{fig:tor_ui_1}d and~\ref{fig:tor_ui_1}e.)
Once this step is completed, the app notifies the user with a ``Training in progress'' message (Figure~\ref{fig:tor_ui_1}j). At this point \textit{Scan} and \textit{Teach} buttons are made inactive. They are activated once training on the server is complete and the user is notified. 

\textit{\textbf{DR3: Enable Access to Training Data with Descriptors.}} Some of the main concerns of blind participants about teachable object recognizers in Kacorri \etal~\cite{kacorri2017people} were: ``\textit{knowing whether the photos were good, knowing the area of a package where the label or distinguishing information resides,.., and deciding on the distance between the object and camera lens.}'' We observe that this information wanted by the participants can be provided both at a photo level and at a higher level across a set of photos. Thus, we devise two types of descriptors, shown in Table~\ref{tab:tor_study_attributes}. These are all derived from visual attributes used to code training photos from sighted (\eg,~\cite{hong2020crowdsourcing, hong2019exploring}) and blind (\eg,~\cite{kacorri2017people, lee2019revisiting}) people. Photo-level descriptors are binary, they indicate whether the object is too small or partially included in the frame (cropped), whether the photo is blurred, and if user's hand is included in the frame. Set-level descriptors are indicated as a percentage. They draw from parallels to how humans recognize objects independent of size, viewpoint, and location~\cite{palmeri2004visual}. 

\begin{table}[b!]
\small
\centering
\caption{Photo-level and set-level descriptors. The descriptors are informed by prior studies with sighted and blind people who have no machine learning expertise looking at the way they synthesize their data for training~\cite{kacorri2017people, lee2019revisiting, hong2019exploring,hong2020crowdsourcing, dwivedi2021exploring}.}
\begin{tabu}{| m{2.7cm} | m{8cm} |} 
 \hline
 \multicolumn{2}{|c|}{\textbf{Photo-level descriptors}} \\ 
 \hline
 Small object & The bounding box of the object is smaller than 1/8 (12.5\%) of the image. \\
 \hline
 Cropped object & The object is partially included in the image. \\
 \hline
 Blurry photo & The photo is too blurry to recognize textures or texts.\\
 \hline
 Hand in photo & A user's hand is visible in the image.\\
 \tabucline[2pt]{-}
 \multicolumn{2}{|c|}{\textbf{Set-level descriptors}} \\ 
 \hline
 Variation in size & A set of images shows objects with different sizes.\\
 \hline
 Variation in perspective & A set of images shows different sides of objects. \\
 \hline
 Variation in background & A set of images show backgrounds with different textures or items. \\
 \hline
 \end{tabu}

\label{tab:tor_study_attributes}
\end{table}
% Alt text:
% The description for each of the photo-level and set-level attributes
% Photo-level descriptors
% Small object, The bounding box of the object is smaller than 1/8 (12.5\%) of the image
% Cropped object, The object is partially included in the image
% Blurry photo, The photo is too blurry to recognize textures or texts
% Hand in photo, A user's hand is visible in the image
% Set-level descriptors
% Variation in size, A set of images shows objects with different sizes
% Variation in perspective, A set of images shows different sides of objects
% Variation in background, A set of images show backgrounds with different textures or items

As shown in Figure~\ref{fig:tor_ui_1}f, users access the photo-level descriptors after every photo that they take so that they can identify problems in the photo (\eg, object being cropped) right away; since this gets repetitive, a photo-level descriptor is communicated only when true. Users can access this detailed information also later, when reviewing their trained objects (Figure~\ref{fig:tor_ui_1}e). Photo-level descriptors are also provided in aggregate together with the set-level descriptors (\eg, photo blurred in 50\% of the training examples for an object). Users can access these aggregates along with the set-level descriptors at the end of a training session (Figure~\ref{fig:tor_ui_1}h), where they are called to select either \textit{OK} to proceed or \textit{Retrain} to retake the photos from scratch. Both photo-level and set-level descriptors can be accessed at a later time when reviewing and editing trained objects, as shown in Figure~\ref{fig:tor_ui_1}d.

\subsection{Implementation}
\begin{figure}[t]
    \centering
    \includegraphics[width=\columnwidth]{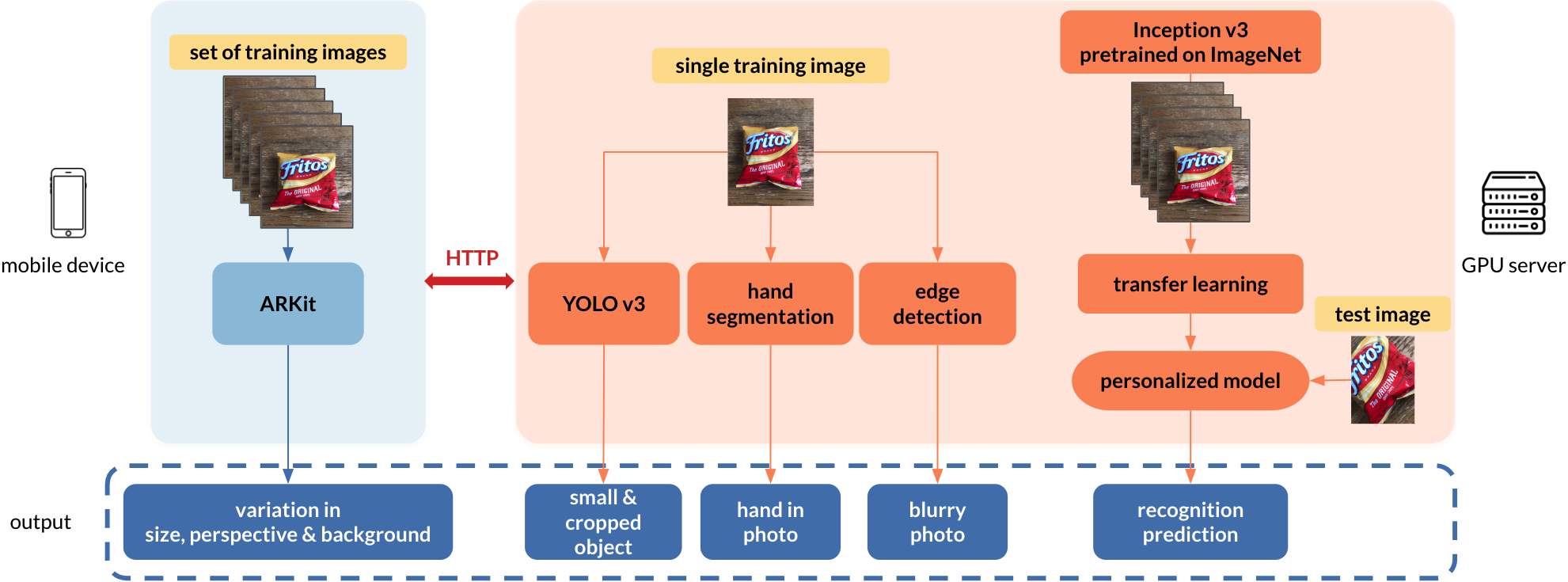}
    \caption{The architecture of the MYCam system indicating approaches for estimating the descriptors and recognizing the object.}
    \label{fig:mycam_system}
\end{figure}

We built the MYCam testbed on Apple iPhone 8 with the object recognition models and descriptor estimators running on our server on an NVIDIA GeForce GTX 1080 Ti GPU; the two communicate through HTTP. The architecture of the system, indicating how both descriptors and recognition predictions are obtained, is illustrated in Figure~\ref{fig:mycam_system}. The estimation of the descriptors in the current implementation of MYCam is error-prone; our approaches merely serve as proof of concept. Prior to making these approaches more robust, we wanted to examine whether blind users can leverage such descriptors in the first place for accessing their training data and experimenting with the model. 

% Alt text:
% Architecture of the MyCam system. The mobile device section which is on the left side of the image includes a set of training images connected to the ARKit box, an arrow goes from this to the output section at the bottom of the image pointing out the output which is "variation in size, perspective, and background. The server section is on the right side of the image and the mobile device and the server communicate by HTTP. On the server side, arrows go from a single training image to YOLO v3, hand segmentation, and edge detection and the outputs are "small/cropped object", "hand in photo", and "blurry photo", respectively. On the server side, there is also a sequence starting from Inception v3, then a set of training images, then transfer learning and at the end of the sequence we have the personalized model that inputs the test image and outputs the recognition result.

\subsubsection{Descriptors.}
\label{descriptors}  
In all previous studies that informed our descriptors, researchers coded the attributes of photos manually through visual inspection of the photos from participants. Given that this is a time-consuming process, methods like Wizard of Oz do not deem appropriate in this early exploration of descriptors for facilitating accessible non-visual experimentation. Thus, we opt for methods that attempt to automatically estimate them, even though, developing techniques for more accurate estimations is beyond the focus of this paper and is briefly discussed in Section~\ref{Discussion}. 
Specifically, we employ state-of-the-art computer vision techniques such as world tracking in ARKit, a YOLOv3 object detection model~\cite{redmon2018yolov3}, and hand segmentation models~\cite{lee2019hands} to estimate the descriptors.  To speed up the calculations for real-time interactions, the object detection, hand segmentation, and edge detection run on our server.

\begin{itemize}
    \item \textbf{Small object}:  Given a bounding box of an object in an image from YOLOv3~\cite{redmon2018yolov3}, the object is considered too small if the size of its bounding box is smaller than 1/8 (12.5\%) of the image.
    
    \item \textbf{Cropped object}: If the YOLOv3~\cite{redmon2018yolov3} bounding box  is at the edge of the image, the object is considered cropped.
    
    \item \textbf{Blurry photo}: The original RGB image is converted to grayscale (pixels values range: 0-255). We use Laplacian edge detection~\cite{kimmel2003regularized} to produce an image with the edges in the grayscale image. In this last image, we then calculate variance in pixel values to quantify blurriness. If the variance is lower than a threshold, the photo is considered blurry. In this study, we set the threshold at 3.0; we found it classifies the blurriness most accurately when tested on photos collected in a prior study with blind participants~\cite{lee2019revisiting}.
    
    \item \textbf{Hand in photo}: The server detects the pixels from a hand via a hand segmentation model that has been previously tested with blind participants~\cite{lee2019hands}. If the proportion of pixels of a hand(s) in an image is greater than a threshold, it considers the photo to show a hand. The threshold is 0.3\%, which detected photos with hands most accurately when tested with the photos collected in a prior study by Lee \etal~\cite{lee2019revisiting}
    
    \item \textbf{Variation in size}: When users take a photo, we detect the position of the smartphone with ARKit.  As the size of the object depends on the distance between the phone and the object, we used the standard deviation of the differences between the phone positions ($SD_{pos}$) to measure the variation in size indirectly. We set the maximum value of the variation as 0.15 ($SD_{max}$) that we could observe with the photos collected by a sighted person in our research team through an internal test. The app presents the variation in size as percentage ($SD_{pos}/SD_{max}*100$).
    
    \item \textbf{Variation in perspective}: We detect the sides of an object using ARKit. For this, we pre-trained the 3D object detection model in ARKit with the three object stimuli in our study. The model provides an enclosing bounding box of an object with six sides in 3D space when it detects the object regardless of the object shape. The model finds the main side of the bounding box based on the object orientation. We calculate the variation in perspective based on the number of object sides shown in a training set. For consistency with other descriptors, we present the variation in perspective as a percentage with scaling ($n*15 \%$ where $n$ is the number of sides in photos).
    
    \item \textbf{Variation in background}: Assuming that the backgrounds captured in photos can vary as a user moves the camera to different places or changes its orientation, we used the location and orientation of the camera to measure the variation in the background.  We calculate the standard deviations of differences in both orientation (using 1-cosine similarity) and the location of the camera in the 3D coordinate system in ARKit. The greater value of the two standard deviations is selected as a variation in background. Like variation in size, we set the maximum value as 0.15 through an internal test. We present the variation in background as a percentage. 
\end{itemize}

\subsubsection{Object Recognition Model.}  The base model for object recognition is Inception V3 pre-trained on ImageNet~\cite{deng2009imagenet}.  When users train the app, it fine-tunes the last layer of the base model using transfer learning with photos taken by the users.  The transfer learning works with a gradient descent algorithm with 500 iterations and a 0.01 learning rate. The training takes around 80 seconds with 90 photos of three objects. When users recognize an object with a personalized model, the time from taking a photo to notifying the recognition result is around 100 milliseconds.  To make the model distinguish the objects in a user's training set and tell the difference from other objects that it has not been trained on, we employed an approach of quantifying the confidence level of the discriminability based on the entropy of confidence scores~\cite{zhang2012online}.  Specifically, when the entropy value is greater than 2.0 or the confidence score is lower than 0.4, the app says \textit{"Don't know"} in synthesized speech instead of the label predicted by the model.  We decided the thresholds of the entropy and confidence score through internal tests such that the app could differentiate the three objects for the user study in the Section~\ref{user_study} from other items (\eg, pen, keyboard, mouse, keys) with the thresholds most accurately.

\section{User Study}
\label{user_study}
To explore the potential and limitations of descriptors in the context of a teachable object recognizer, we conducted a remote user study with blind participants. The study took place in participants' homes to minimize safety concerns during the COVID-19 pandemic. The study was approved by the Institutional Review Board at the University of Maryland, College Park (IRB \#1255427-1).
In designing this remote study, we came across many challenges, including how to provide remote guidance and observe participants' interactions with MYCam and their objects. We quickly found that having just the third-person camera view from the laptop was not enough. Thus, as shown in Figure~\ref{fig:teaser}, we added a first-person view with smart glasses.  We iterated via several pilot tests that involved blind and sighted researchers in our team to anticipate the logistics (\ie, study equipment delivery) and communication methods (\ie, laptop and smart glasses) required for this remote study. Lessons learned from accessing blind participants’ interactions via smart glasses (with this study serving as part of a larger case study) are discussed in depth in Lee \etal~\cite{lee2022lab}.

% \hl{The pilot tests revealed that the laptop camera often did not capture the participants nor their interactions with the objects and the app. The first-person perspective in the smart-glasses could help the experimenter monitor the participants interaction with a smartphone}~\cite{lee2022lab}.
% % To make sure that the participants can position the laptop computer, wear the smart glasses, and find the objects for the user study, we went through pilot tests with a blind researcher in our team.
 
\subsection{Participants}
We recruited 12 blind participants (6 women, 6 men, 0 nonbinary) from campus email lists and local organizations. As shown in Table~\ref{tab:participants_demographics}, their ages ranged from 32 to 70 ($M = 54.3, SD = 15.2$).  Three participants reported being totally blind, five having some light perception, and four being legally blind. P1 and P2 reported having an ``\textit{auditory processing disorder}'' and difficulty in hearing ``\textit{very high sound}'', respectively.  All participants reported using smartphones several times a day and taking a photo or recording a video at least once a month. As for their familiarity with machine learning, two participants reported being somewhat familiar, eight being slightly familiar, and two being not familiar at all---we used a 4-point scale for this question: (1) not familiar at all (have never heard of machine learning), (2) slightly familiar (have heard of it but don’t know what it does), (3) somewhat familiar (have a broad understanding of what it is and what it does), (4) extremely familiar (have extensive knowledge on machine learning). While all participants had experience taking photos before, many indicated that they had challenges related to image framing (9), focusing (2), holding a camera steadily (2), and controlling the lighting (2). Many participants indicated prior experience with other camera-based assistive mobile applications such as Aira~\cite{Aira}, Be My Eyes~\cite{BeMyEyes}, Google Lookout~\cite{GoogleLookout}, Microsoft Seeing AI~\cite{SeeingAI}, Mediate Labs Supersense~\cite{Supersense}, Super Lidar~\cite{SuperLidar}, and Voice Dream Scanner~\cite{VoiceDreamScanner}.

\begin{table}[h!]
    \small
    \centering
    \caption{Participants' demographics and experience with machine learning, photo taking, and camera-based assistive apps.}
      \vspace{-1ex}
    \begin{tabu}{| l l l l l l l l |} 

        \hline
        ID  & Age & Gender & Level of vision & Onset & Machine learning & Photo taking & Experience with assistive apps\\
        \hline
        P1  & 39  & Woman & Light perception & Birth        & Not familiar at all & Once a day   & Aira, Be My Eyes, Seeing AI\\
        P2  & 67  & Man   & Legally blind    & 55           & Slightly familiar & Once a month & Seeing AI\\
        P3  & 62  & Woman & Totally blind    & Birth        & Somewhat familiar  & Several times a month & Seeing AI, Be My Eyes\\
        P4  & 32  & Man   & Legally blind    & 20           & Slightly familiar  
        & Several times a day   & None\\
        P5  & 66  & Man   & Light perception & 46           & Slightly familiar  & Once a week & Seeing AI, Supersense, Super Lidar\\
        P6  & 61  & Man   & Light perception & 41           & Somewhat familiar  
        & Several times a week & Seeing AI \\
        P7  & 70  & Man   & Legally blind    & Birth        & Slightly familiar  
        & Several times a week & None\\
        P8  & 50  & Woman & Legally blind    & 45           & Slightly familiar  
        & Several times a week & Seeing AI\\
        P9  & 69  & Woman & Totally blind    & 55           & Not familiar at all  & Several times a day & VD Scanner, Be My Eyes, Seeing AI\\
        P10 & 66  & Woman & Light perception & Birth        & Slightly familiar  & Several times a week & None\\
        P11 & 33  & Woman & Light perception & Birth        & Slightly familiar  
        & Once a month & Seeing AI, VD Scanner\\
        P12 & 36  & Man   & Totally blind    & Birth        & Slightly familiar  & Several times a day & Seeing AI, VD Scanner, Lookout\\
        \hline
    \end{tabu} 
    % \\
    % \vspace{1ex}
    % {\small *ML: machine learning}
\label{tab:participants_demographics}
\end{table}

\subsection{Procedure}
Participants communicated with the experimenter remotely via dual Zoom video conferencing~\cite{zoom} connected both via a laptop and a pair of Vuzix Blade smart glasses~\cite{VuzixBlade} that we delivered prior to their study sessions (see Lee \etal~\cite{lee2022lab}).
At the beginning of the study, we briefly explained the concept of a teachable object recognizer. Here, we provided a minimal description of how to take photos to train or test the app to mitigate priming in photo-taking strategies for training and testing an object recognizer. The description given at the beginning of the study reads as follows:

% The study consists of three tasks: 1) training a teachable object recognizer app with photos of objects, 2) testing the performance of the app, 3) reviewing and editing the information of the objects.  At the beginning of the study, we explained the concept of teachable object recognizer briefly with minimal description of how to take photos to train or test the app so that we do not make bias in participants' strategies for taking photos for training and testing an object recognizer. The description of the app given at the beginning of the study reads as follows:

\begin{displayquote}
``The idea behind the app is that you can teach it to recognize objects by giving it a few photos of them, their names, and if you wish, audio descriptions. Once you've trained the app and it has them in its memory, you can point it to an object, take a photo, and it will tell you what it is. You can always go back and manage its memory.''
\end{displayquote}

Then, participants were asked to perform three tasks: (1) train the app with their own photos and labels of three snacks that served as object stimuli shown in Figure~\ref{fig:tor_study_objects}, (2) use the app again to recognize those objects later \ie~to test the performance of the app, and (3) review and edit the information of the already trained objects.
% During the user study, participants trained and tested the app with photos of three objects in Figure~\ref{fig:tor_study_objects}.
For the first task, the order of objects for training was fully counterbalanced between participants. When participants trained the app with the first object, the experimenter provided step-by-step instructions on the MYCam user interface (\eg, the position and functionality of buttons as well as the audio feedback that indicates the steps of training). Then, participants trained the app with the second and third objects and asked the experimenter for help when necessary.
When participants were testing the app for the first time, the experimenter also gave detailed instructions on the MYCam interface for testing. After that, participants were free to test their models for as long as they wished (taking as many photos).
When reviewing their trained objects in the third task, participants could access both information related to the descriptors as well as their own object labels and any recorded audio descriptions.
% After each task, we asked questions about their experiences with the app and the review process.

After reviewing a training set with the descriptors, participants decided whether they would collect the photos again or not for that object. We made retraining optional for two reasons: (1) to avoid collecting data from participants who are not motivated to experiment by retraining a model (as this could add a confounding factor in our analysis) and (2) to be able to contrast the attributes of training sets for who decided to retrain their models and those who did not.

Throughout the study, we encouraged participants to think out loud and to ask questions at any time.  After each task, participants were asked to answer questions related to their experience with the descriptors and MYCam and questions captioning usability satisfaction~\cite{lewis1995ibm}.  All questions in this study were either open-ended or on a 5-point Likert scale (\ie, strongly disagree, disagree, neutral, agree, strongly agree).  

% ------- stopped here ----------- I will resume after addressing comments (Jonggi)

\subsection{Object Stimuli}
\begin{figure}[t]
    \centering
    \includegraphics[width=0.4\columnwidth]{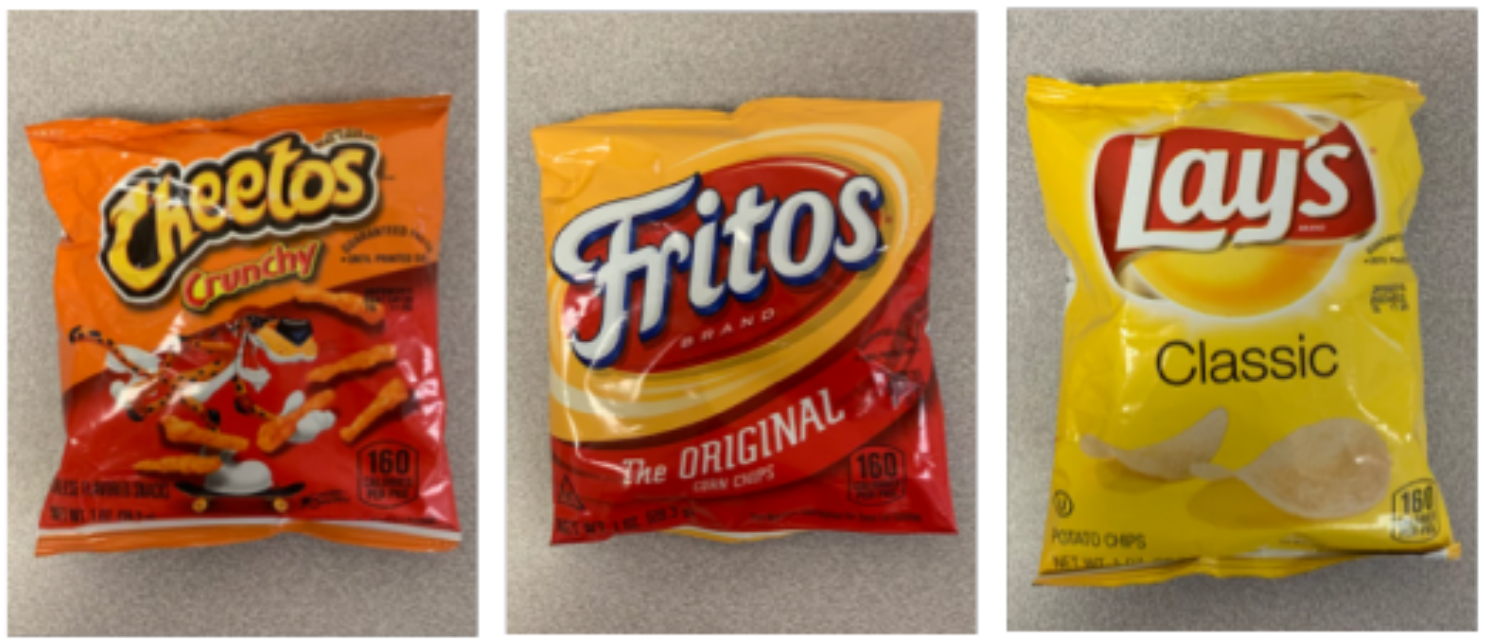}
    \caption{Object stimuli in the study chosen for a challenging fine-grained classification task: Fritos, Cheetos, and Lays.}
    \label{fig:tor_study_objects}
\end{figure}

Accounting for blind people's need for recognizing objects with similar sizes, weights, and textures with fine-grained labels~\cite{kacorri2017people, sosagarcia2017hands}, we selected three snacks, shown in Figure~\ref{fig:tor_study_objects}, with the same size, texture, and nearly identical weights for our user study. 
% This object choice was also observed in prior work that investigated non-ML experts' perception of machine teaching in the context of object recognition~\cite{hong2020crowdsourcing}.
As prior work shows that end-users' strategies of collecting training photos are often inconsistent between objects~\cite{hong2020crowdsourcing}, we expect that the choice of three similar objects allows us to observe blind people's teaching strategies in the context of fine-grained object recognition.
With these snacks, we simulated a scenario in which a blind user interacts with the app to recognize different objects that the blind user may feel difficult to distinguish using only the tactile sensation. It was engineered to be a challenging scenario for machine learning models since these objects were similarly shaped and colored, had reflective surfaces, and were deformable.  Unique and personal objects without logos or texts on them (\eg, key, mug cup) can be potentially used with a teachable object recognizer and perhaps could be fit for a more realistic scenario. However, for this study, we included only commercial products to allow for comparison and replicability similar to prior studies regarding teachable object recognizers~\cite{kacorri2017people, sosagarcia2017hands, lee2019revisiting}.
% Also, the order of the snacks was fully counterbalanced among participants to minimize the order effects in the study.

% Alt text:
% This figure shows three snacks, Fritos, Cheetos, and Lays with a plain white background.

\section{Results}
\begin{figure}[b]
    \centering
    \includegraphics[width=\textwidth]{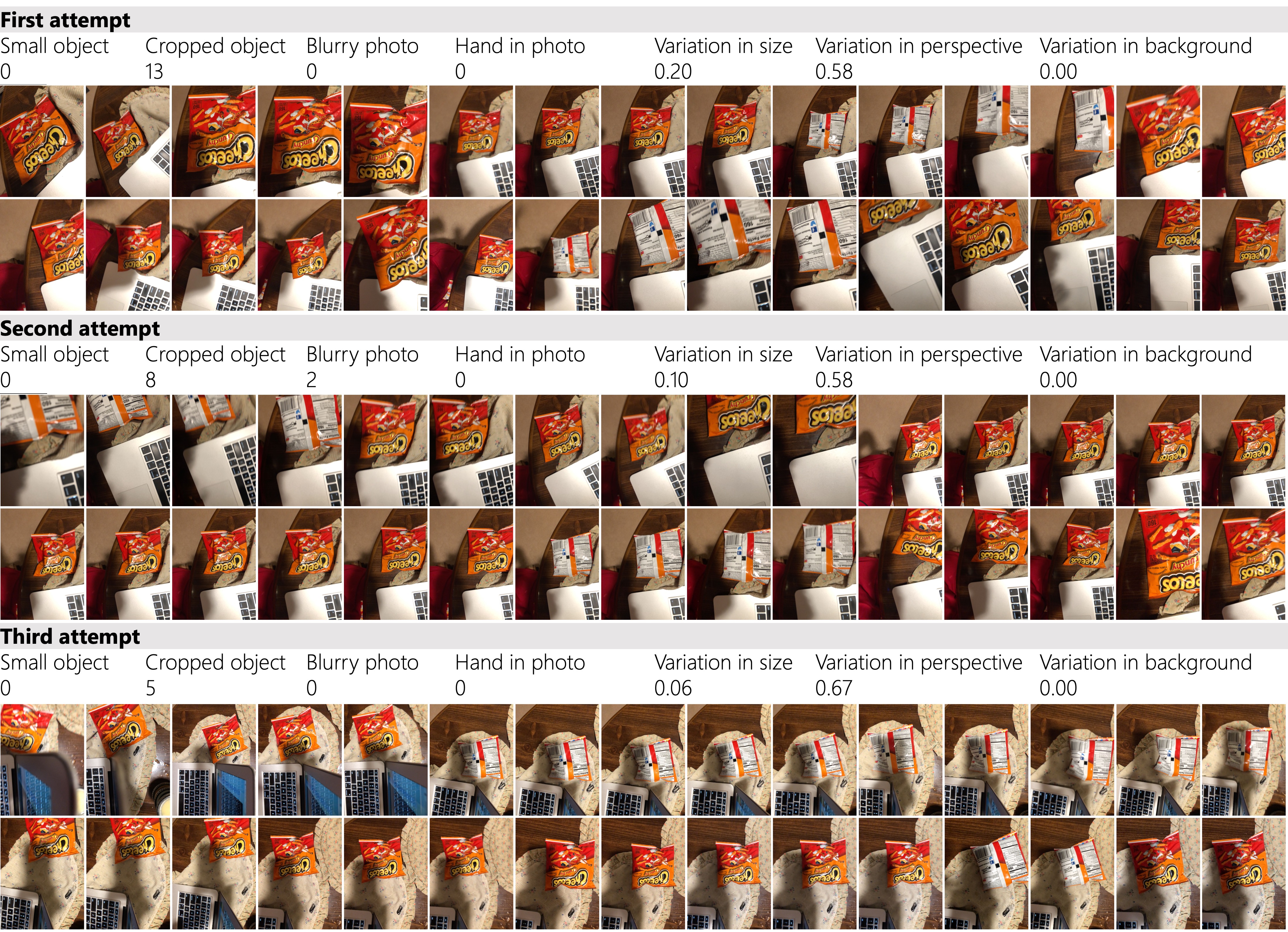}
    \caption{Photos of Cheetos from P10 and manually annotated attributes to be compared with automatically estimated descriptors.}
    \label{fig:descriptors-samples}
\end{figure}
% Alt text:
% [The 30 photos in the first attempt]
% Small object: 0
% Cropped object: 13
% Blurry photo: 0
% Hand in photo: 0
% Variation in size: 0.20
% Variation in perspective: 0.58
% Variation in background: 0.00

% [The 30 photos in the second attempt]
% Small object: 0
% Cropped object: 8
% Blurry photo: 2
% Hand in photo: 0
% Variation in size: 0.10
% Variation in perspective: 0.58
% Variation in background: 0.00

% [The 30 photos in the third attempt]
% Small object: 0
% Cropped object: 5
% Blurry photo: 0
% Hand in photo: 0
% Variation in size: 0.06
% Variation in perspective: 0.67
% Variation in background: 0.00

Participants spend on average 143.8 seconds ($SD=72.4$) taking 30 photos of an object. Five out of 12 participants re-train the object recognizer after inspecting their training sets with descriptors. Examples of the photo-collection attempts and their annotated attributes (\ie, ground-truth attributes annotated by a researcher through visual inspection) are shown in Figure~\ref{fig:descriptors-samples}. Through the analysis of the participants' photos and the performance of the personalized object recognition models, we show how descriptors may relate to the participants' strategies for collecting training photos when they decided to retrain their models. We also show the impact of these changes in training photos on the performance of the models. We observe promising trends in the characteristics of photos (\ie, adding more variations and reducing problematic photos) over time and iterations. Participants' subjective feedback also indicate that our descriptors can be a promising approach for providing access to one's training data in this context. 

\subsection{Correlation Between Estimated Descriptors and Annotated Attributes}

\begin{figure}[b]
    \centering
    \includegraphics[width=0.8\textwidth]{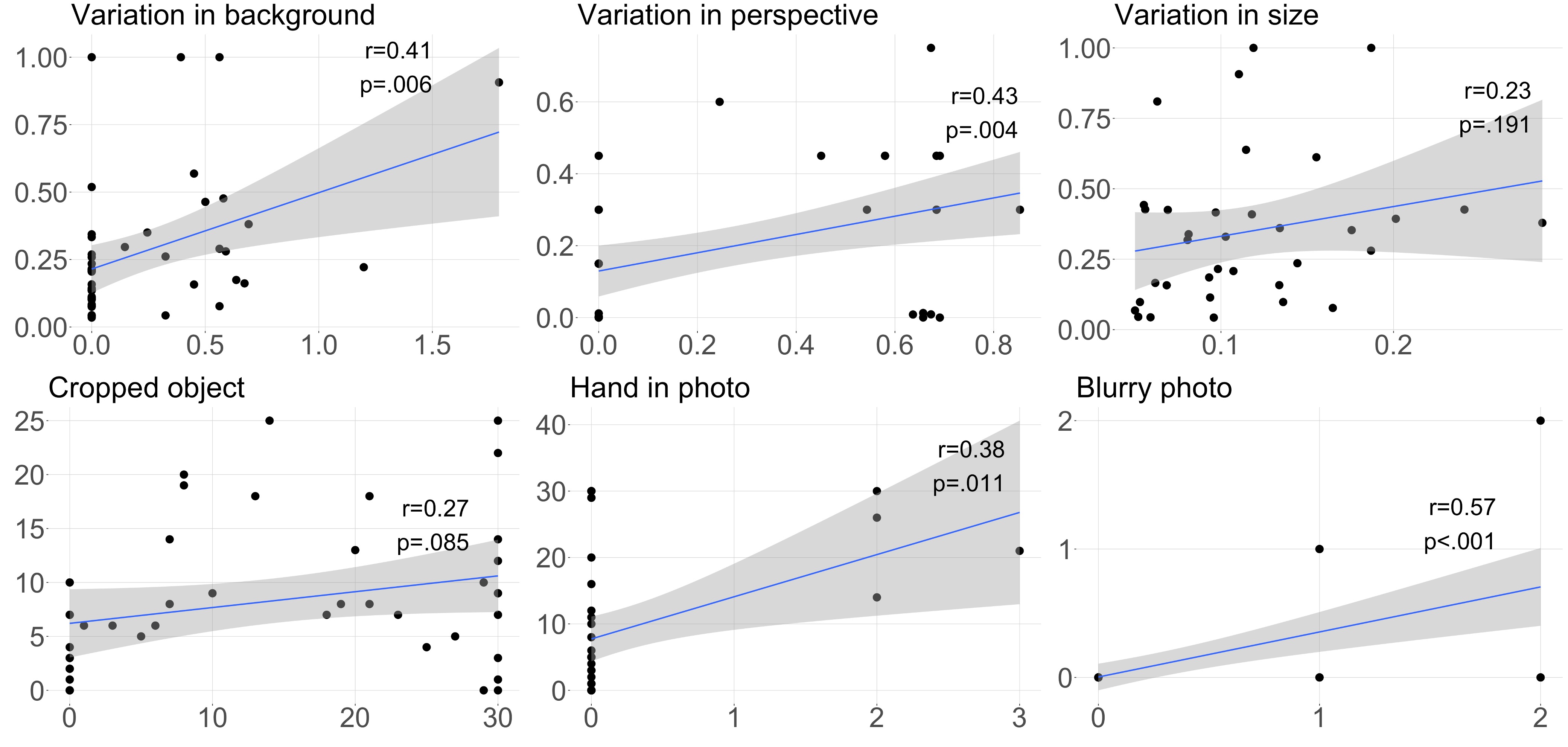}
    \caption{Scatter plots indicating correlations between manual annotations (x-axis) and estimations (y-axis) for each descriptor.}
    \label{fig:train-descriptor-correlation}
\end{figure}
% Alt text: 
% This figure includes six scatter plots with correlation coefficients and p values.
% Variation in background: r=0.41, p=0.06
% Variation in perspective: r=0.43, p=0.004
% Variation in size: r=0.23, p=0.191
% Cropped object : r=0.27, p=0.085
% Hand in photo : r=0.38, p=0.011
% Blurry photo : r=0.57, p<0.001
We report the performance of our approach in estimating descriptors
% (photo and set-level descriptors for the training photos) 
as it is a critical context for interpreting the remainder of the results. More so, it can provide a glimpse at future efforts for estimating such descriptors in a real-world context. Here we measure performance by computing the correlation between the estimated descriptors and annotated attributes. Given that prior work indicates high inter-rater agreement for the annotation of these attributes~\cite{hong2020crowdsourcing}, we had a single researcher in our team performing this task. To quantify the variation of background and perspective, the researcher grouped the photos within a set based on their similarity in terms of background and object side. We used the groups to calculate the Shannon-Wiener Diversity Index~\cite{shannon2001mathematical}, a measure of variation in background and perspective. The researcher also coded the photos with a cropped object, participants' hands, and blurriness. For the attributes related to the size of the object (\ie, variation in size, small object), the researcher annotated the bounding boxes of the objects. The variation in size was considered as the standard deviation of the proportions that the bounding boxes occupy in photos. The proportions range from 0.0 (\ie, the object is not captured) to 1.0 (\ie, a bounding box covers the entire photo). A photo with a small object is defined as one having a bounding box smaller than 12.5\% of the photo.

As shown in Figure~\ref{fig:train-descriptor-correlation}, the correlation coefficients between estimated descriptors and annotated attributes ranged from 0.23 to 0.57, highlighting that this is a challenging task. The correlation for "small object" is not shown since only three of all photos had small objects that are not detected by our descriptor estimator.
Even though we employed naive approaches for estimating the descriptors as a proof of concept, all pairs had positive correlations. This indicates that even with partial access there can be an opportunity for reflection and experimentation \ie, if participants considered relative changes rather than absolute values.
% to compare their training approaches through iterations. 
Below, we see some empirical evidence in support of this premise.

\subsection{Changes in Annotated Descriptors For Participants who Choose to Retrain}
Five participants (P1, P3, P5, P8, P10) decided to retrain with a new set of photos for an object after reviewing their initial training sets; one of them (P3) trained the same object three times, each time with a new set of photos. A participant (P10) retrained with new sets of photos for two of the three objects. No participant retrained all three objects. 

\begin{figure}[t]
  \begin{subfigure}[b]{0.49\textwidth}
    \includegraphics[width=0.9\columnwidth]{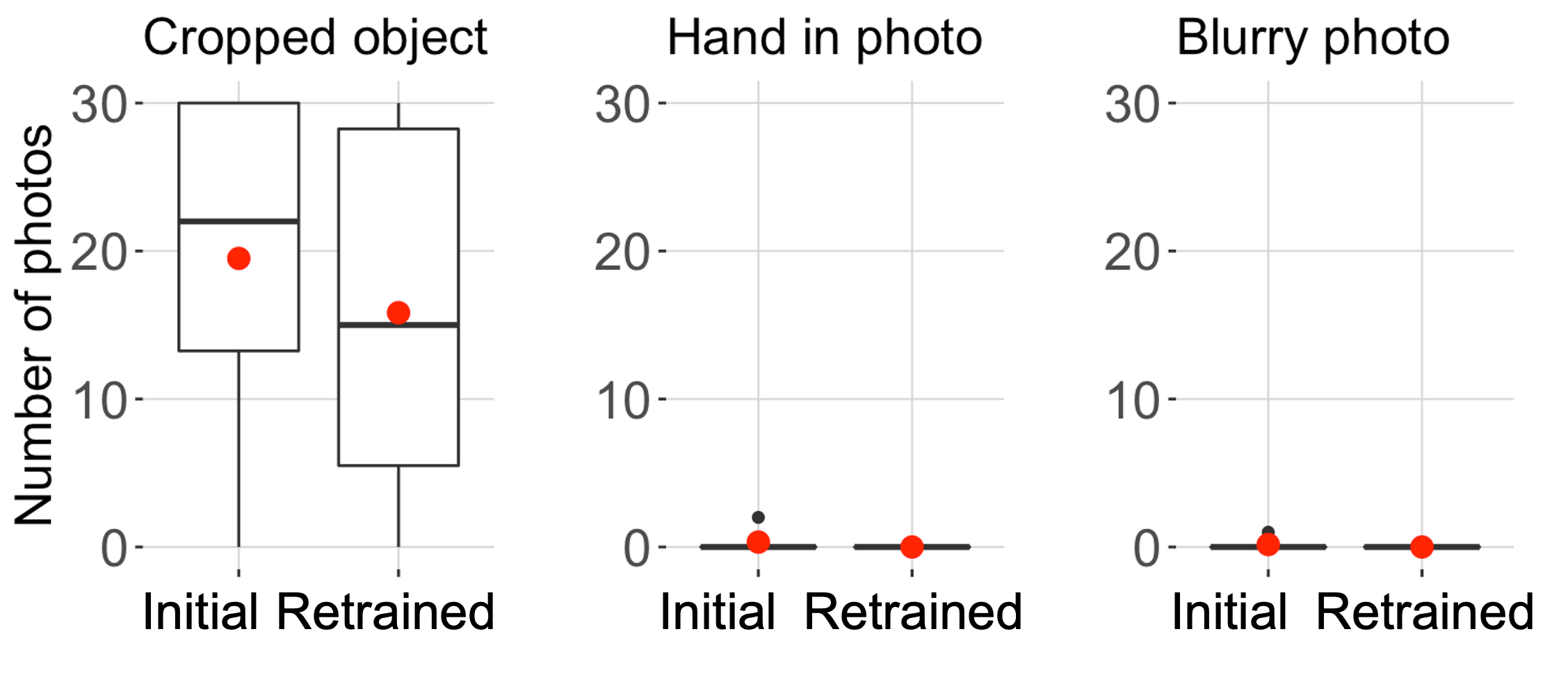}
    \caption{The aggregate of photo-level descriptors.}
    \label{fig:descriptors_photolevel_retrainnoretrain}
  \end{subfigure}
  \hfill
  \begin{subfigure}[b]{0.49\textwidth}
    \includegraphics[width=0.9\columnwidth]{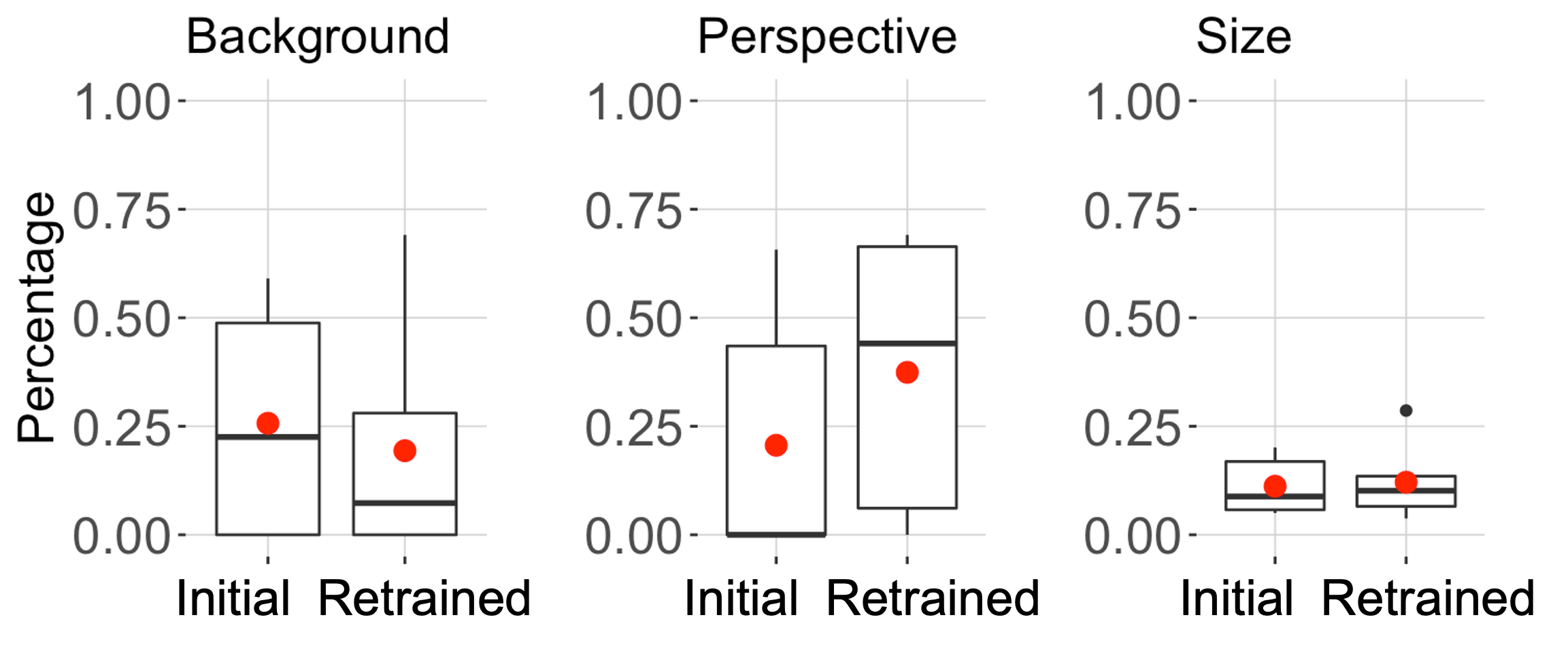}
    \caption{Set-level descriptors.}
    \label{fig:descriptors_setlevel_retrainnoretrain}
  \end{subfigure}
  \caption{Contrasting descriptor values in initial attempts to retraining attempts for P1, P3, P5, P8, and P10. Red dots indicate means.}
    \label{fig:descriptors_retrainnoretrain}
\end{figure}

As shown in Figure~\ref{fig:descriptors_retrainnoretrain}, we contrast the estimated descriptors for initial attempts to those during retraining attempts. When the participants decided to retrain, their new training sets had fewer photos with cropped objects, no hands included, almost no blurred photos, and higher variation in perspective and size on average compared to their initial photos.   This is a promising trend providing some evidence on participants' attempt to respond and adhere to the descriptors though it may have come at the cost of lower variation for background. 

% cropped initial 117 photos
% cropped retrained 95 photos

% hand initial 2 photos
% hand retrained 0 photos

% alt text:
% Box plots showing the number of photos with photo-level descriptors and percentages of set-level descriptors in initial and retrained sets.
% Cropped object (number of photos)
% Set, Mean, Upper quartile, lower quartile
% Initial, 7.17, 3, 10
% Retrained, 12.83, 0, 25
% Hand in photo (number of photos)
% Set, Mean, Upper quartile, lower quartile
% Initial, 5.47, 0, 6
% Retrained, 21.17, 12, 30
% Variation in background (percentage)
% Set, Mean, Upper quartile, lower quartile
% Initial, 0.29, 0.10, 0.38
% Retrained, 0.32, 0.08, 0.28
% Variation in perspective (percentage)
% Set, Mean, Upper quartile, lower quartile
% Initial, 0.18, 0.01, 0.30
% Retrained, 0.13, 0.00, 0.15
% Variation in size (percentage)
% Set, Mean, Upper quartile, lower quartile
% Initial, 0.39, 0.17, 0.48
% Retrained, 0.30, 0.08, 0.39

Specifically, the average numbers of photos with cropped objects and users' hands were fewer at 15.83 ($SD=13.41$) and 0.00 ($SD=0.00$) in their new training photos versus the initial at 19.50 ($SD=12.52$) and 0.33 ($SD=0.81$), respectively. The number of blurry photos was 0.00 ($SD=0.40$) and 0.17 ($SD=0.41$) in retrained and initial, respectively. The system did not detect any photos with a too-small object in either set.
As for variation, mean variation in perspective and size in retrained were 0.37 ($SD=0.33$) and 0.12 ($SD=0.09$), respectively, which is higher compared to those in the initial sets at 0.20 ($SD=0.32$) and 0.11 ($SD=0.07$). However, this trend was reversed for variation in background. This descriptor was on average lower in retrained at 0.19 ($SD=0.28$) compared to the initial at 0.26 ($SD=0.28$). 

\subsection{Changes in Annotated Attributes for All Participants Over Time}
Many participants chose not to retrain. Perhaps the interactive nature of the descriptors created opportunities for early reflection and experimentation; not just at the end of training. To explore this, we measured trends over time at different levels of granularity; for this analysis, we use the manually annotated attributes, which serve as the ground truth, rather than the estimated descriptors.
% We expected descriptors to affect the attributes of both a photo and a set of photos, since participants received photo-level descriptors for every individual photo as well as both photo- and set-level descriptors after collecting 30 photos of each object. We analyzed photos individually to identify changes in photos. We also compared the sets of photos within objects (\ie, sets collected for a certain object by selecting "Retrain" button in the app) and between objects (\ie, sets of photos with different objects).

\subsubsection{Fine-grained Changes Across 90 Training Photos}
\begin{figure}[t]
  \begin{subfigure}[b]{0.40\textwidth}
    \includegraphics[width=\columnwidth]{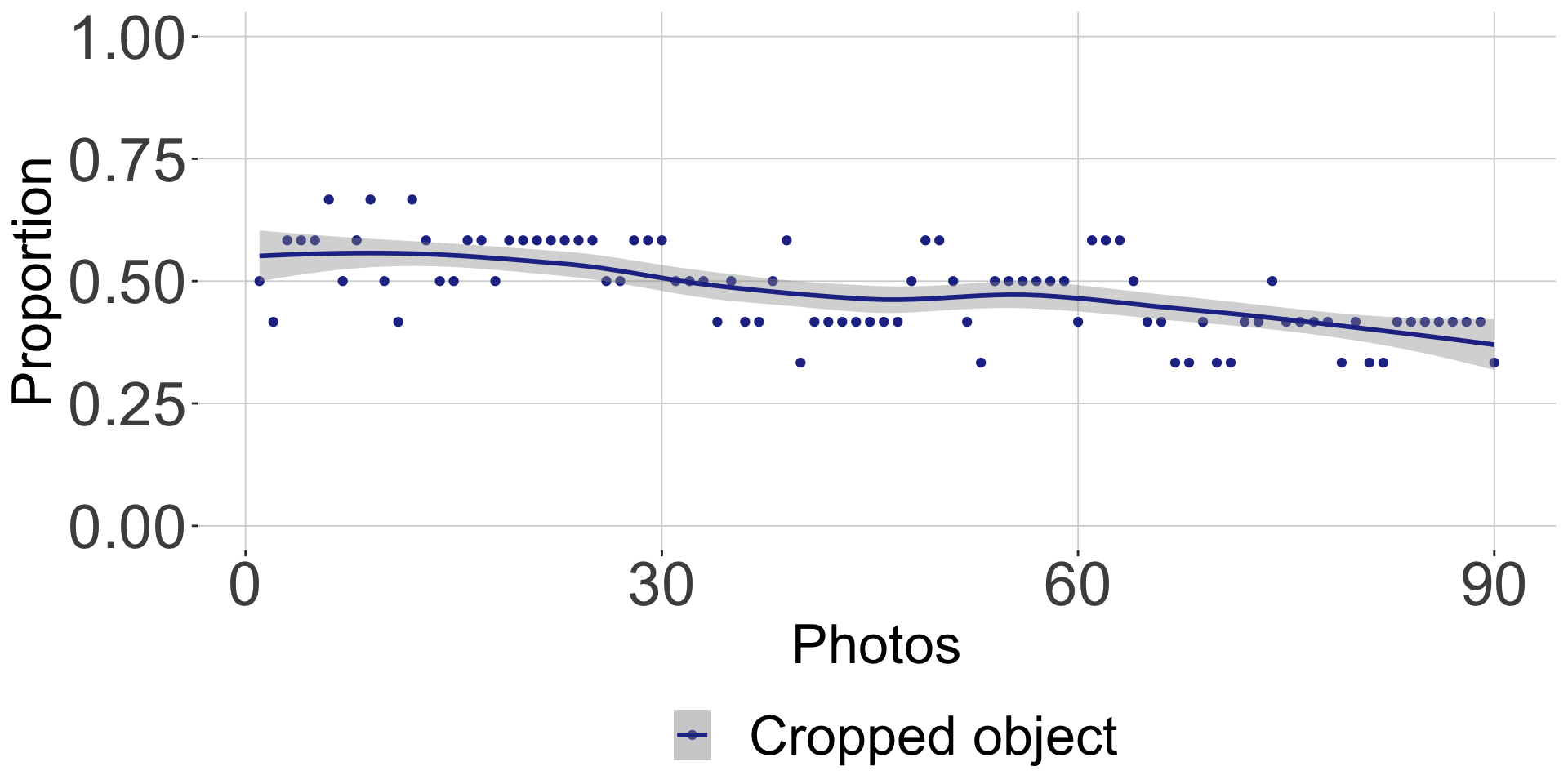}
    \caption{Cropped object.}
    \label{fig:descriptors_change_over_photos_crop}
  \end{subfigure}
  \begin{subfigure}[b]{0.40\textwidth}
    \includegraphics[width=\columnwidth]{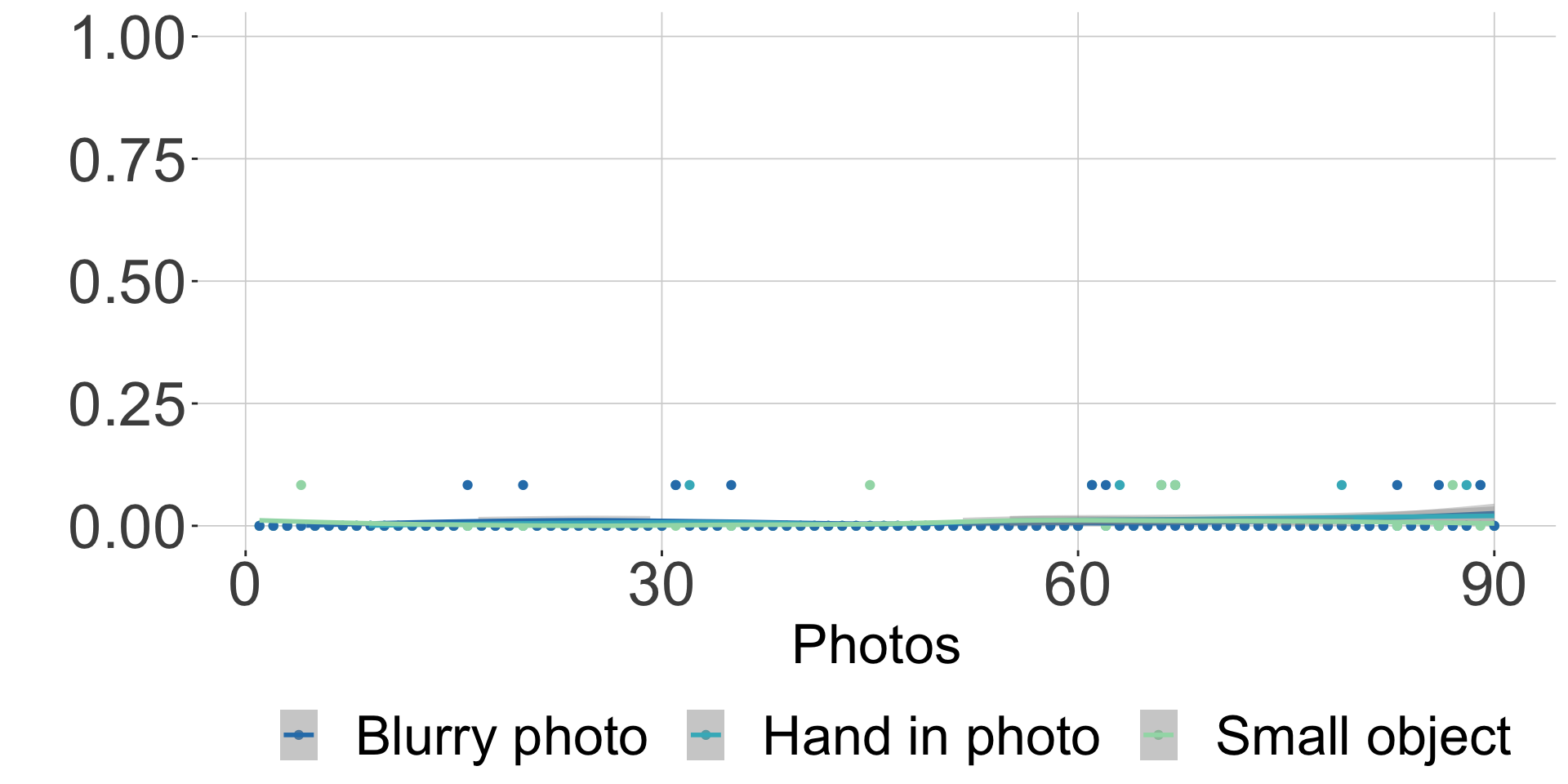}
    \caption{Blurry photo, hand in photo, too small object.}
    \label{fig:descriptors_change_over_photos_bht}
  \end{subfigure}
  \caption{The average values of annotated photo-level attributes for individual photos among 12 participants. The charts include photos of the first three training sets (1-30: first set, 31-60: second set, 61-90: third set). The lines are fitted to dots using LOWESS smoothing.}
  \label{fig:descriptors_change_over_photos}
\end{figure}
% Alt text:
% Two scatter plots showing the points for proportion of photos with photo-level attributes and lines fitted to the points. The proportion of photos with cropped object monotonically decrease. Other attributes are nearly zero throughout photos.

With photo-level descriptors participants' could gauge potential image quality issues right away; MYCam indicates them immediately after a photo is taken. As shown in Figure~\ref{fig:descriptors_change_over_photos}a, we observe a dropping trend in the number of images where the object was cropped as participants progressed in the study. This is promising for a descriptor that merely provides binary feedback (\ie, whether the object is cropped or not) instead of directional guidance on how to move a camera to fully capture an object (\eg, Lee~\etal~\cite{lee2019revisiting}). The proportion of photos with cropped objects was around 0.56 at the beginning (1st photo in 1st training), decreasing to 0.37 by the last photo (30th photo in 3rd training).  Whereas the proportion of training examples with participants' hands included, objects too small, or blurry photos were nearly zero throughout the study (Figure~\ref{fig:descriptors_change_over_photos_bht}).

\subsubsection{Coarse-grained Changes Across 3 Training Sets}
\begin{figure}[t]
  \begin{subfigure}[b]{0.33\textwidth}
    \includegraphics[width=\columnwidth]{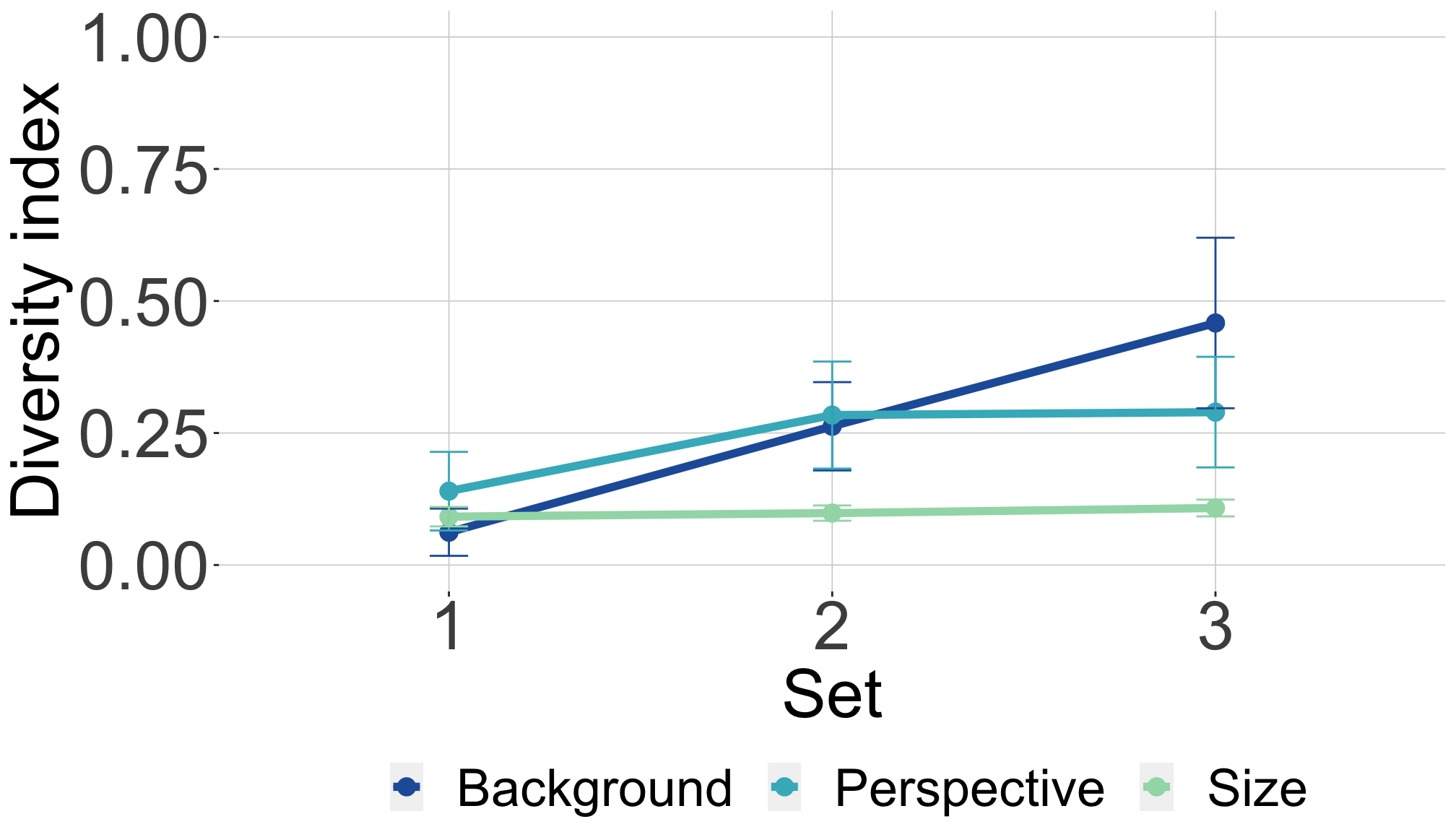}
    \caption{Set-level descriptors.}
    \label{fig:descriptors_change_over_sets_between_SL}
  \end{subfigure}
  \hfill
  \begin{subfigure}[b]{0.33\textwidth}
    \includegraphics[width=\columnwidth]{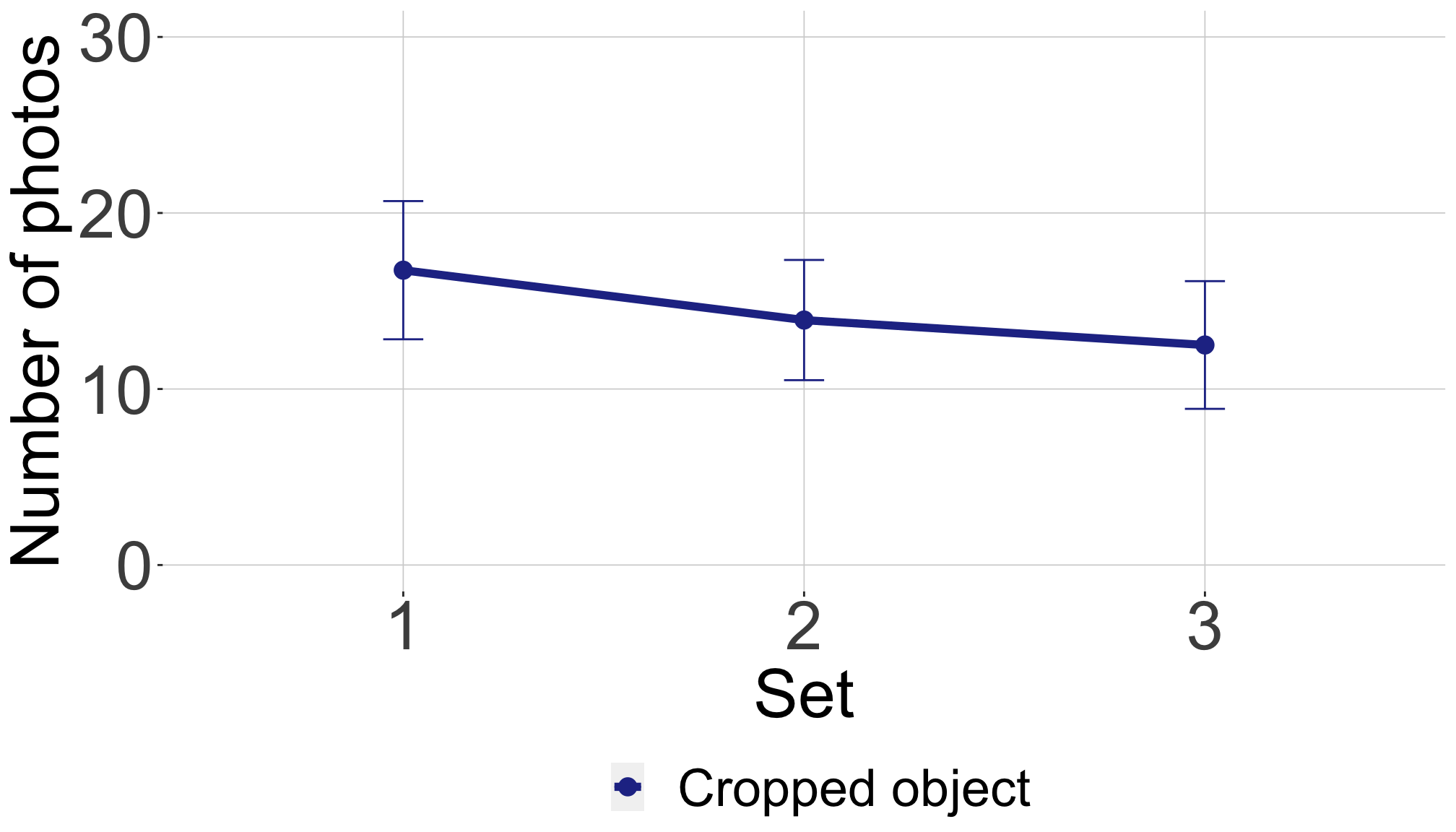}
    \caption{Cropped object.}
    \label{fig:descriptors_change_over_sets_between_CO}
  \end{subfigure}
  \hfill
  \begin{subfigure}[b]{0.33\textwidth}
    \includegraphics[width=\columnwidth]{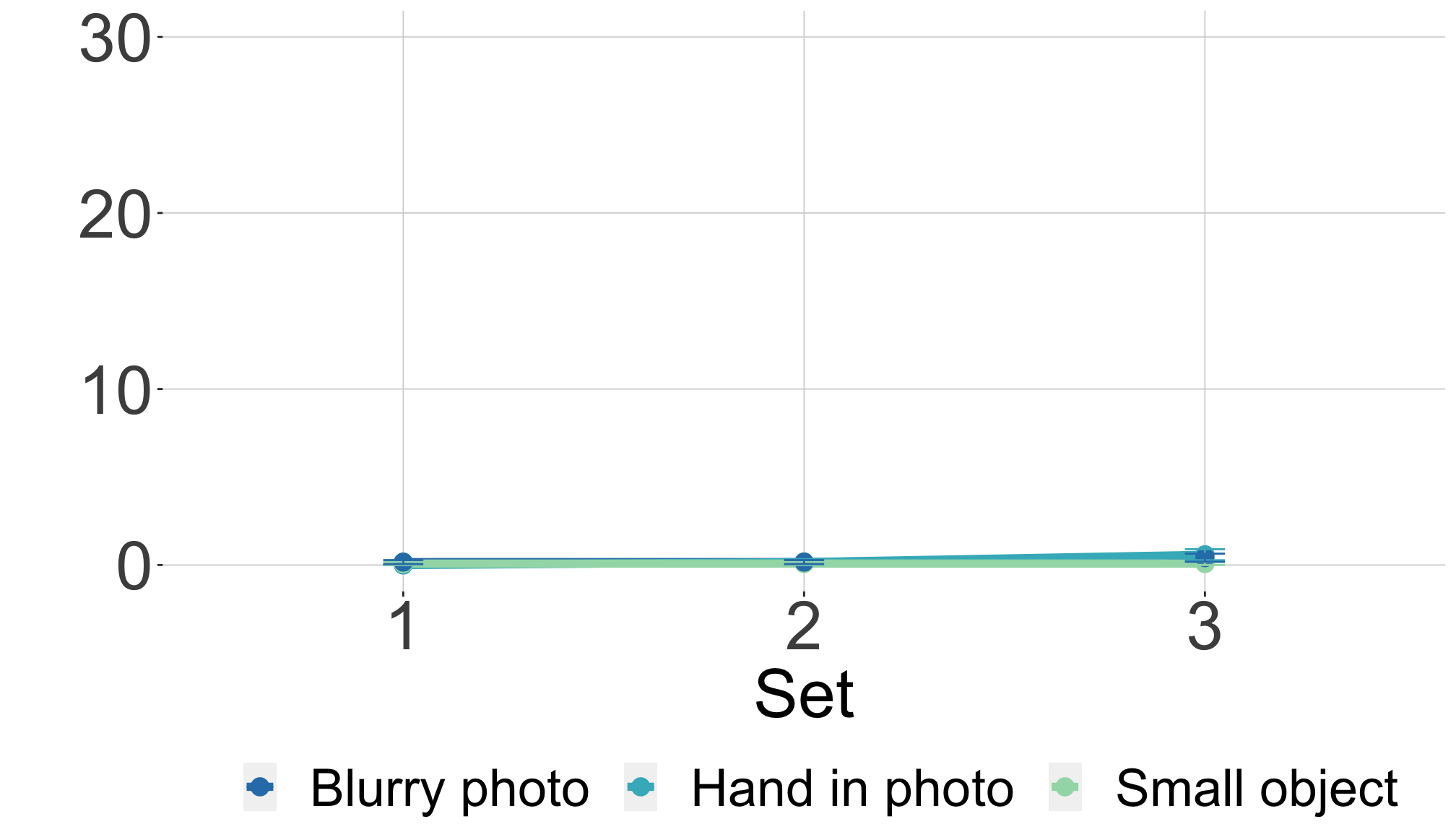}
    \caption{Blurry photo, hand in photo, small object.}
    \label{fig:descriptors_change_over_sets_between_PL}
  \end{subfigure}
  \caption{The average annotated values of set-level attributes and the annotated number of photos with photo-level attributes for all 12 participants across three training sets (a training set per object).}
  \label{fig:descriptors_change_over_sets_between}
\end{figure}

With photo-level descriptors and set-level aggregates participants’ could gauge potential issues related to their
teaching strategies or image quality at the end of each training attempt; MYCam shows them immediately after 30 photos are taken. Participants may or may not choose to go back and retrain. But they may also choose to reflect when training the next object, especially since our object stimuli were engineered to be very similar. As shown in Figure~\ref{fig:descriptors_change_over_sets_between},
participants increased the variation among their training examples and reduced the number of photos with cropped objects. A one-way repeated-measure ANOVA indicate a significant effect of order of sets on variation in background ($F(2,22)=4.59, p=0.022,$ partial $\eta^2=0.18$) and in perspective ($F(2,22)=3.61, p=0.044,$ partial $\eta^2=0.05$). We did not observe a statistically significant effect of the other attributes. However, we do observe a tendency for an increase in the number of photos that were blurry or where the participant's hand was included. Perhaps these descriptors were not deemed as that problematic or they were ranked lower in priority as teaching strategies evolved. Participants' feedback below can shed a bit more light on these observations. 

% Alt text:
% Line graphs showing the diversity index of set-level attributes and number of photos with photo-level attributes.
% (a)
% Set Descriptor Value SD
% 1	Background	0.06211104	0.04472982	
% 2	Background	0.26272858	0.08361987	
% 3	Background	0.45835233	0.16145514	
% 1	Perspective	0.13983881	0.07445077	
% 2	Perspective	0.28395566	0.10133139	
% 3	Perspective	0.28955221	0.10471891	
% 1	Size	0.09158828	0.01845046	
% 2	Size	0.09833468	0.01447150	
% 3	Size	0.10798020	0.01588949	
% (b)
% Set Descriptor Value SD
% 1	Cropped object	16.75000000	3.92568272	
% 2	Cropped object	13.91666667	3.41444865	
% 3	Cropped object	12.50000000	3.62754613	
% (c)
% Set Descriptor Value SD
% 1	Hand in photo	0.00000000	0.00000000	
% 2	Hand in photo	0.16666667	0.16666667	
% 3	Hand in photo	0.58333333	0.31281550	
% 1	Small object	0.08333333	0.08333333	
% 2	Small object	0.08333333	0.08333333	
% 3	Small object	0.08333333	0.08333333	
% 1	Blurry photo	0.16666667	0.11236664
% 2	Blurry photo	0.16666667	0.11236664	
% 3	Blurry photo	0.41666667	0.22890826	

\subsection{Performance of Participants' Object Recognition Models}
%\subsection{Evaluating the Performance of the Trained Models With Different Test Sets}
After finalizing their training for all objects, participants were called to test the performance of their models; we explicitly did not allow for intermediate train-test iterations in an attempt to limit interference from that type of experimentation in the observed behaviors. For the purpose of our analysis, we report model performance not only on participants' final training sets but also dive deeper and look at their photos chosen to test their models and how well their model generalizes \eg, if tested with photos taken by others.

% During the study, participants tested the models trained with the final training sets (\ie, the last photo-collection attempt for each object) and no testing was done with the models trained with initial training sets (\ie, first or second attempts). This was done to investigate the effect of the descriptors on participants' data-collection strategies in iterations without the interference of the test results with their initial training sets. After the study, we evaluated the models trained with both final and initial training sets using three different test sets. The test sets include individual participants' personal test samples, aggregated test samples from all participants, and test photos from nine blind people in a prior study~\cite{lee2019revisiting}. While personal test samples reflect each participant's idiosyncratic environment, we found that the samples had questionable validity due to several limitations, such as a small number of samples, cluttered background, limited variation in background, and limited variation in light condition as pointed by Zimmermann \etal~\cite{zimmermann2019youth} Kacorri \etal~\cite{kacorri2017people}. The other two test sets resolved some of these limitations with more variations and larger number of samples, though they would include samples that deviate from each participant's environment. This section reports on the analysis of participants' test sets and performance of the models trained with the first and last training sets.

\subsubsection{Model Performance with Testing Images from Self.}
% \subsubsection{Testing Strategies}
% \label{testing_strategies}
We found that participants used a very small number of photos ($M=3.7$, $SD=3.2$) to check if their models were working properly. Some (4 out of 12) included photos where the object was more than half cropped. Others (4 out of 12) captured multiple objects in the frame. Some of these observations could be perhaps explained by our study setup (\eg, participants were done with taking photos for the day or objects were in close proximity due to study setup). However, prior work in teachable object recognizers employing different study designs also indicates that model testing and evaluation can be challenging for end users~\cite{hong2020crowdsourcing, dwivedi2021exploring}. 
These challenges are critical as perceived and actual performance may be different when the models are actually used after testing.    
% Upon examining the test photos, we found that they had quality problems. Although the test photos can reflect the participants' idiosyncratic environments and behavior, the quality problems affect the validity of the participants' evaluations. 
% During the testing task, participants took 3.7 photos per object on average ($SD=3.2$) as shown in Figure~\ref{fig:results_test_num}.
% Considering that the test data need samples with different visual contexts (\eg, sides, sizes of objects, background, light condition) to test the object recognizer thoroughly, participants took fewer photos for testing than necessary for a thorough evaluation. We also observed problems related to image framing. The test sets from four out of 12 participants included photos with less than half of the objects. We also observed that test samples from four participants captured two or three snacks at the same time, making it hard for the models to distinguish which one they wanted to recognize (Figure ~\ref{fig:photos-test-cluttered}). 

\begin{figure}[t]
    \centering
    \begin{subfigure}[b]{0.49\textwidth}
        \centering
        \includegraphics[width=\textwidth]{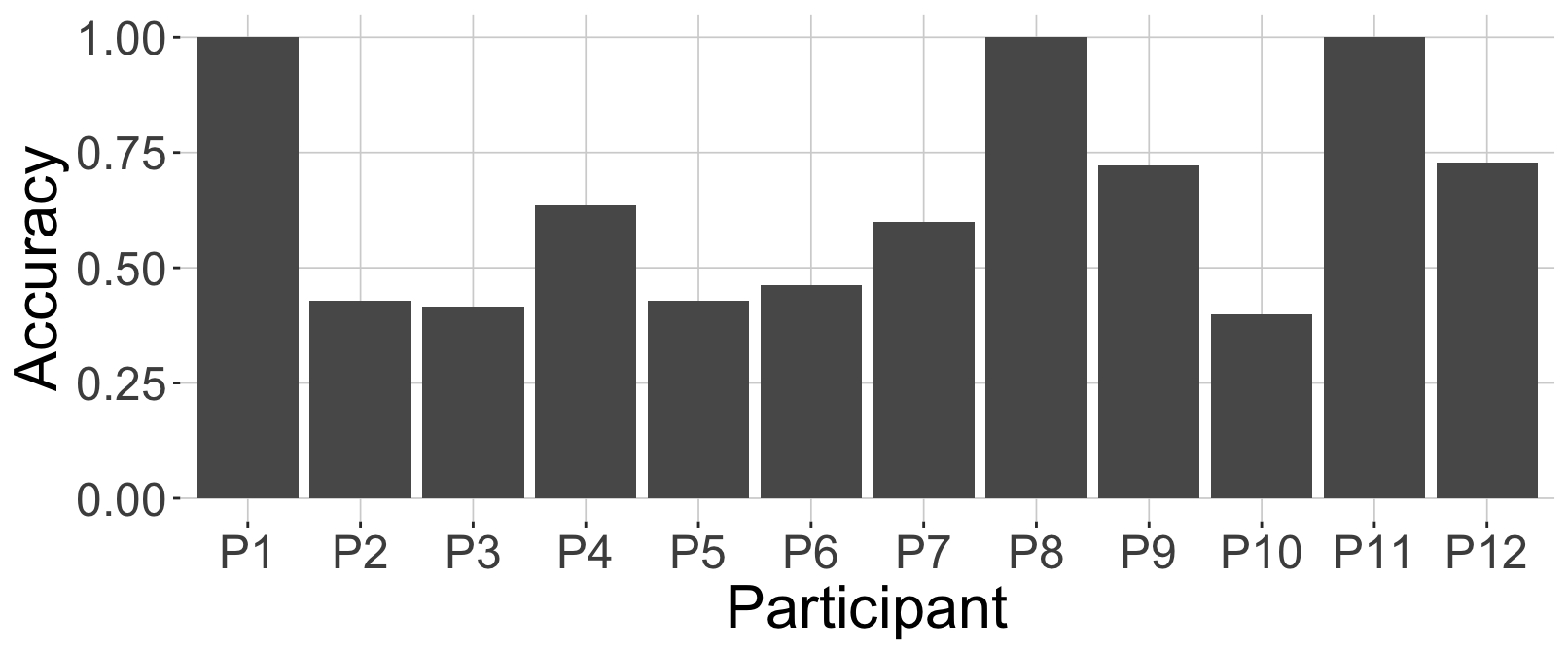}
        \caption{Object recognition accuracy on one's own testing images.}
        \label{fig:test_accuracy}
    \end{subfigure}
    \hfill
    \begin{subfigure}[b]{0.49\textwidth}
        \centering
        \includegraphics[width=\textwidth]{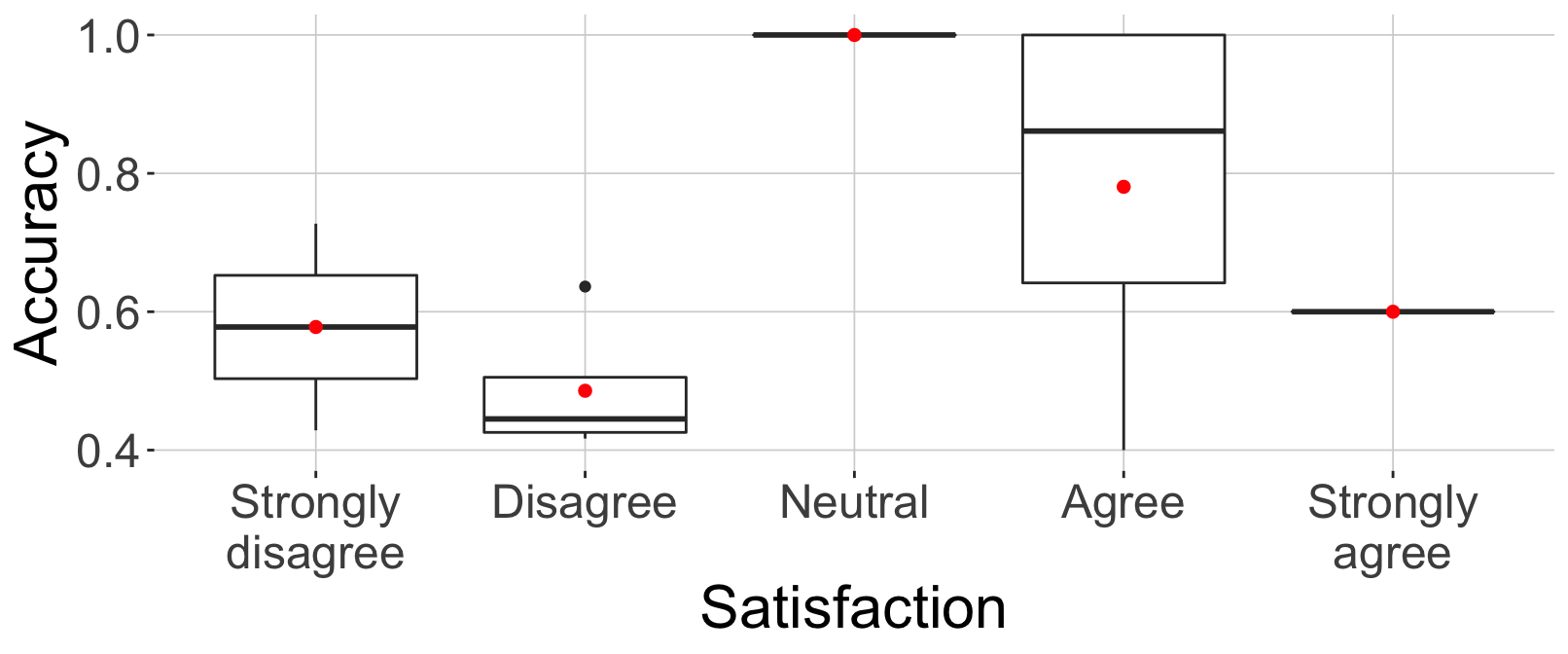}
        \caption{Accuracy against participant satisfaction with performance. 
        % Red dots indicate means.
        }
        \label{fig:test_accuracy_satisfaction}
    \end{subfigure}
    
    \caption{When testing their models, participants' experiences varied (a), which seems to be reflected in their satisfaction scores (b).
    % The number of tests per object and proportion of errors.
    }
    \label{fig:results_test_evaluation}
\end{figure}

% \subsubsection{Evaluation of the Models Trained with First and Last Sets with Individual Participant's Photos.}
%\subsubsection{Evaluating the Models Trained with First and Last Sets of the Individual Participant's Photos.}
Overall, we find that when testing on one's own data the average accuracy (\ie, the number of correct predictions divided by total test images) of the models was 0.65 ($SD=0.24$) with a breakdown across participants shown in Figure~\ref{fig:test_accuracy}. These results may seem surprisingly low for a 3-way classification task. However, beyond being a fine-grained classification, the task can be particularly challenging with objects of deformable shapes, same-size, reflective-surface, and similar colors that can be hard to distinguish. Among the high-performing models are those of P1 and P8 who choose to iterate on their training (they tested the models with 3 and 7 photos, respectively); though the same is not to be said for the models of P3, P5, and P10 who also iterated on their training (they tested the models with 12, 28, and 10 photos, respectively). When juxtaposing model performance with participants' subjective responses on the satisfaction of their models (Figure~\ref{fig:test_accuracy_satisfaction}), we find that those whose models did not perform well disagree with this statement and those whose models perform better agree. This alignment, however, did not hold for those on the edges (\textit{strongly disagree} and \textit{strongly agree}). Participants' feedback in the next section, provides a potential explanation.

% Figure~\ref{fig:accuracy_comparison} shows the accuracy of the models trained with the first and last sets. Two participants (P1, P10) got 100\% accuracy with both training sets. For one participant (P8), the model exhibited improved performance when training photos were recollected. Surprisingly, for another two participants (P3, P10), the model accuracy was lower with the last training sets. This is different from our expectation that the models trained with last sets would perform better than models trained with the first sets because we observed better quality (\eg, more variations and fewer photos with cropped objects) in the last training sets as shown in the Section~\ref{change_over_sets_within_objects}. One possibility that could account for this result is that the test samples were collected within a limited amount of time and space. The benefits of descriptors such as variations in the photos for model improvement may instead occur over long-term use of teachable object recognizer. Another possible reason could be the limitations of the test sets, such as small number of samples and multiple objects of interest in photos as mentioned in Section~\ref{testing_strategies}.

\subsubsection{Model Performance with Testing Images from Others.}
\begin{figure}[t]
    \centering
    \begin{subfigure}[b]{0.49\textwidth}
        \centering
        \includegraphics[width=\textwidth]{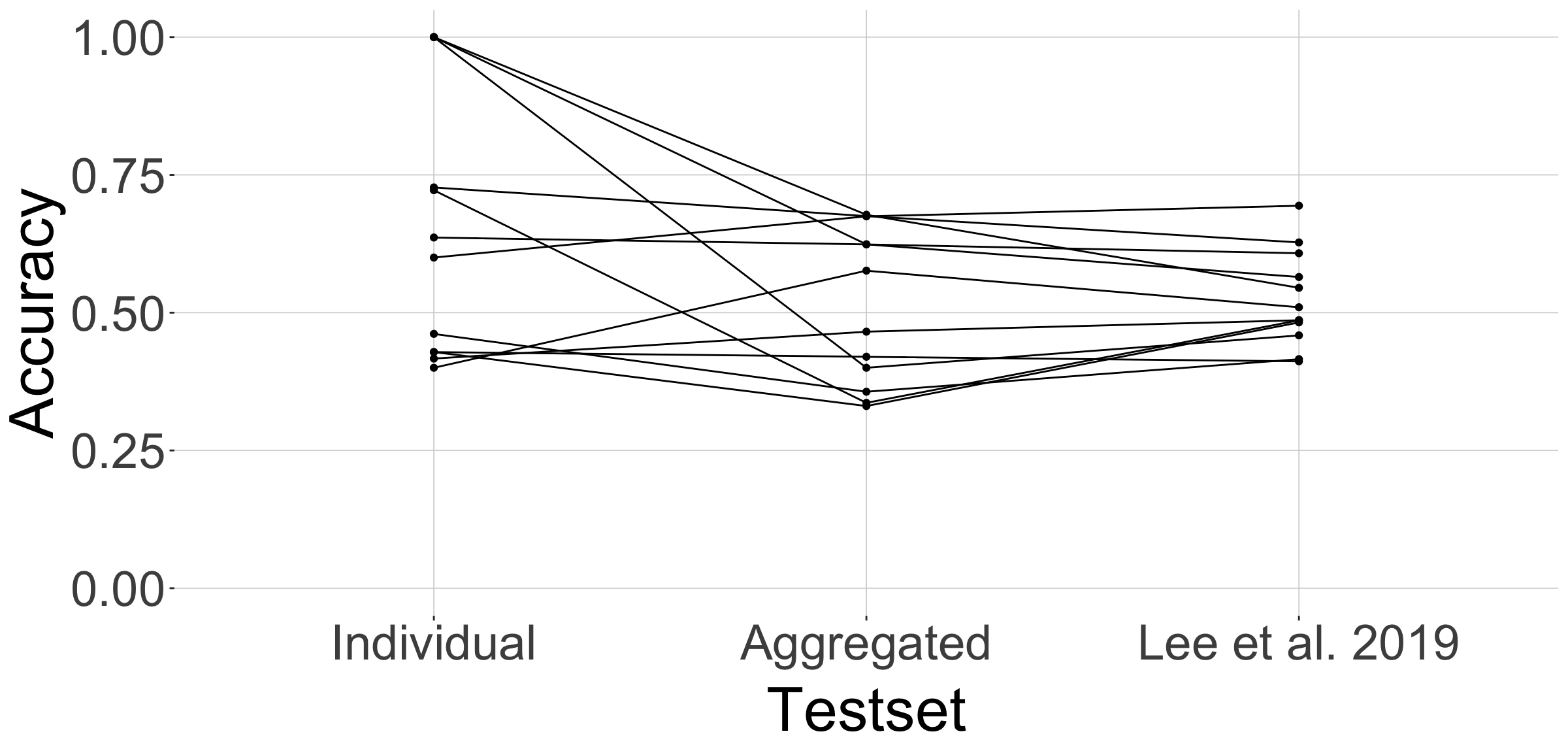}
        \caption{Accuracy per participant.}
        \label{fig:accuracy_datasets_individual}
    \end{subfigure}
    \hfill
    \begin{subfigure}[b]{0.49\textwidth}
        \centering
        \includegraphics[width=\textwidth]{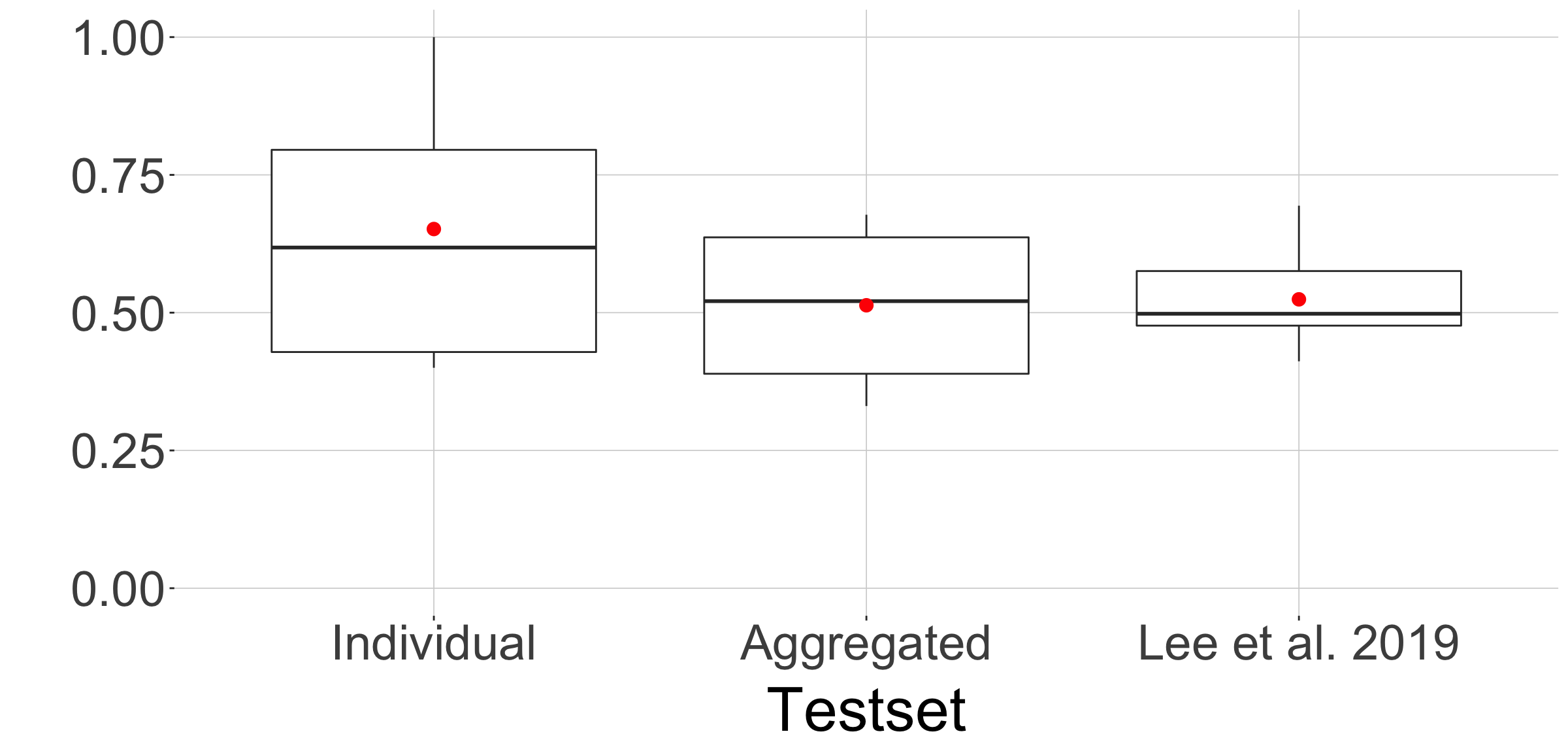}
        \caption{Summary statistics of accuracy.}
        \label{fig:accuracy_datasets_box}
    \end{subfigure}
    \caption{Model accuracy when tested on individual test images, aggregated test images from all 12 blind participants in this remote study, and aggregated test images from all 9 blind participants in a prior in-lab study~\cite{lee2019revisiting}.}
    \label{fig:accuracy_datasets}
\end{figure}
% Alt text
% (a) a line graph where a line is mapped to each participant, showing the accuracy of models measured with three test sets. (b) a box plot with the aggregated accuracy measured with three test sets.
% Participant, Individual, Aggregated, Lee et al. 2019
% P1, 1, 0.624, 0.5647059
% P2, 0.4285714, 0.3305785, 0.4823529
% P3, 0.4166667, 0.4655172, 0.4862745
% P4, 0.6363636, 0.6239316, 0.6078431
% P5, 0.4285714, 0.42, 0.4117647
% P6, 0.4615385, 0.3565217, 0.4156863
% P7, 0.6, 0.6747967, 0.6941176
% P8, 1, 0.677686, 0.545098
% P9, 0.7222222, 0.3363636, 0.4862745
% P10, 0.4, 0.5762712, 0.5098039
% P11, 1, 0.4, 0.4588235
% P12, 0.7272727, 0.6752137, 0.627451

%\subsubsection{Comparing the Accuracy of the First and Last Trained Models, Tested with Individual Test Sets and General Datasets.}
One of the promises of a teachable object recognizer is that it works well for each individual since the training and test sets are collected by the same person and it is highly likely that they are going to exhibit similar patterns~\cite{kacorri2017people, sosagarcia2017hands}. This was also the case in our study. As shown in Figure~\ref{fig:accuracy_datasets}, for 9 out of 12 participants, the accuracy of the model was higher when tested with an individual participant's test set than an aggregated test set from all participants in our study and photos from another study with blind participants~\cite{lee2019revisiting} on the same objects. The accuracy of the model with individual test sets was 0.65 ($SD=0.24$). The accuracy was lower at 0.51 ($SD=0.14$) and 0.52 ($SD=0.09$) when pooling test sets across all participants in the current study and testing photos from a prior study~\cite{lee2019revisiting}, where nine blind participants trained and tested a teachable object recognizer, respectively. However, we observed that the iteration can make the models generalize better. Among the five participants who did retraining, four and three participants had higher accuracy after retraining when their models were tested with the aggregated test set and the set from the prior study~\cite{lee2019revisiting}, respectively.

% The models trained with last training sets performed better with general datasets. When tested with the aggregated test set and the test set from the study by Lee\etal, models had higher accuracy with the last sets for four and three participants, respectively, as shown in the Figure~\ref{fig:accuracy_comparison}. This result can be applicable to long-term usage of teachable object recognizer where blind users take photos with various background, different sides of object, different distances, or different light conditions.

\subsection{Subjective Feedback from Participants}
% We report on participants' subjective feedback on their experience in training with descriptors and testing teachable object recognizer using MYCam.

\label{subjective_feedback}
\subsubsection{Overall Experience}  
\begin{figure}[b]
    \centering
    \includegraphics[width=0.85\columnwidth]{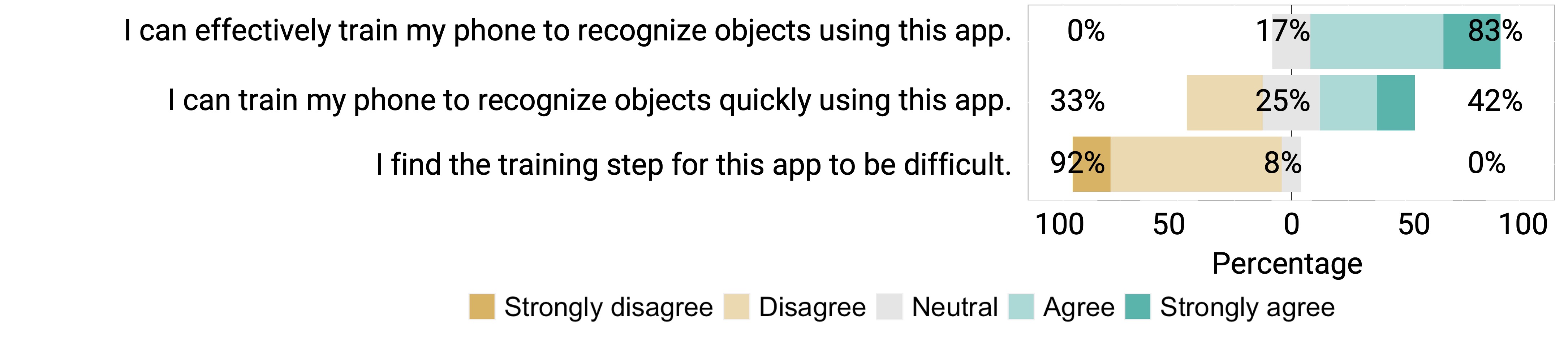}
    
    \caption{Participants' feedback on training with the MYCam testbed.}
    \label{fig:train_subjective}
\end{figure}

To provide more context on participants' feedback for the descriptors, we illustrate in Figure~\ref{fig:train_subjective} their responses related to the MYCam testbed. Overall, participants agreed that they could train the object recognition model effectively with MYCam and disagreed on training being difficult, though they were divided on whether it could be done quickly. This is promising. 
Specifically, ten participants agreed or strongly agreed that they could train their models effectively with some pointing to the need for onboarding. P1 and P10, for example, who are not familiar at all and slightly familiar with machine learning said \textit{``after a while, I learned that I could train it''} and \textit{``It’s pretty easy. You have to teach me though. But if you teach me then it’s pretty easy to follow instructions and finish the process.''} respectively. On the other hand, P11 and P12 were neutral.  P11 mentioned that taking 30 photos is time-consuming, saying \textit{``I don’t really feel like I was all that effective because it takes a while to train for each one.''}  The errors in descriptors affected the reliability of the app, making a participant think the training process was less effective even though the two models work independently of each other. P12 said \textit{"I don’t think that the app is correct, especially when I know, for example, that my hand was not in the photo...I don’t have a lot of confidence in the app's accuracy.''} 

When asked whether they could train the app quickly, five participants agreed, four disagreed, and three were neutral.  Seven participants indicated that taking 30 photos is tedious. For example, P10 said, \textit{"The process is pretty straightforward. But I have to spend, like, quite long time to train the three objects."} When asked about the difficulty of the training task, all but one who remained neutral, disagreed or strongly disagreed that the task was difficult.  P11 who was neutral, found it not difficult but tedious. Surprisingly, this sentiment of the task being tedious was not presented in the initial study with teachable object recognizers~\cite{kacorri2017people} even though the number of training photos was identical.  We suspect this difference reflects more on our implementation of the descriptors in the MYCam testbed rather than the process of training itself. In MYCam users could not opt in or out of the photo-level descriptors during training leading to higher training times; specifically, the time for taking photos for training an object was doubled from 65 seconds ($SD=35.2$) reported in that first study~\cite{kacorri2017people} to  143.8 seconds on average ($SD=72.4$) in our study.

% Alt text
% Likert chart showing the participants' responses to three questions in percentage.
% Question	Strongly disagree	Disagree	Neutral	Agree	Strongly agree
% I can effectively train my phone to recognize objects using this app.	0	0	16.666667	58.33333	25
% I am able to train my phone to recognize objects quickly using this app.	0	33.33333	25	25	16.66667
% I find the training step for this app to be difficult.	16.66667	75	8.333333	0	0

\subsubsection{Descriptors}  
\begin{figure}[b]
    \centering
    \includegraphics[width=0.8\columnwidth]{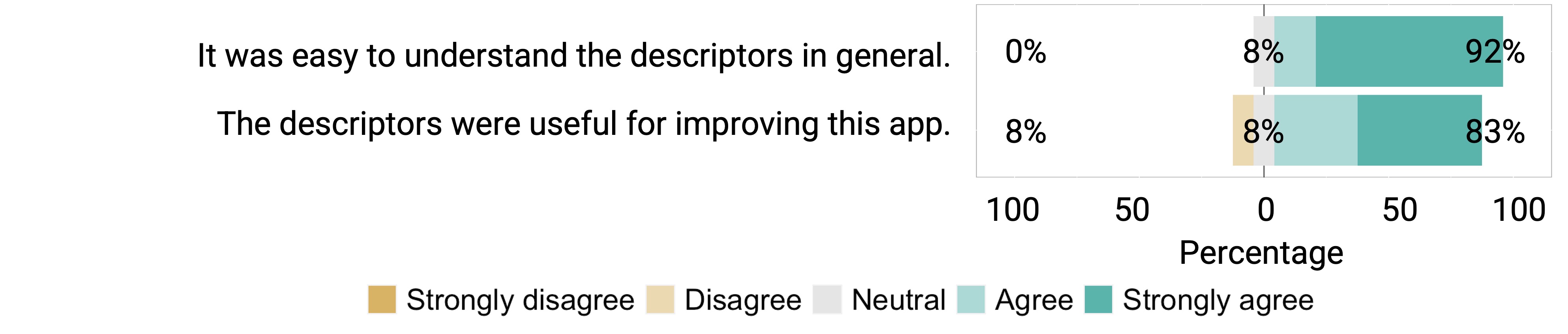}
    
    \caption{Participants' feedback on the descriptors.}
    \label{fig:descriptor_subjective}
\end{figure}
% Alt text
% Likert chart showing the participants' responses to three questions in percentage.
% Question	Strongly disagree	Disagree	Neutral	Agree	Strongly agree
% It was easy to understand the descriptors in general.	0	0	8.333333	16.66667	75
% The descriptors were useful for improving this app.	0	8.333333	8.333333	33.33333	50
As shown in Figure~\ref{fig:descriptor_subjective}, all but one participant (P1, who was neutral) agreed or strongly agreed that the \textbf{descriptors were easy to understand}. P6 said \textit{``I understood what it was telling me. I didn’t have questions about what I was supposed to do.''} Participants' responses indicate user reflection based on descriptor changes across multiple attempts, strengthening some of our observations in the previous sections. P2 elaborated \textit{``It gives you directions. The explanation (descriptors) afterwards, in the analysis, told me that my photographs were not always good. So I have to learn to take better photographs.''} Some participants found it difficult to understand the absolute values of provided in the descriptors and were wondering whether they should have a specific value as a goal. For example, the values of descriptors were somewhat ambiguous to P1 who said, \textit{``I guess just knowing exactly what they’re referring to what numbers are really preferable.''} P4 also mentioned the challenge in understanding the values of descriptors, but then mentioned that over repeated data collection during the study, he figured out their purpose.  P4 said \textit{``I wasn't aware of any of those fields when we did the first object [...] For the second and third objects, I could take a little bit more variation in the photos or to better train the application.''} This is interesting feedback as the descriptors are there merely to provide access to what one could infer via a visual inspection not per se dictate optimal characteristics for the training set. The difference of course is that when a sighted person glances over their training photos they may or may not make an inference on potentially problematic photos or lack of variation (see Hong \etal~\cite{hong2020crowdsourcing}), but a blind person always hears the descriptors. This explicit presence of the descriptors calls for the need for more context. While ``ideal values'' are use-case depended, during onboarding users could perhaps be provided with some rationale or examples.    
% highlights the need for supporting blind users for understanding the meaning of values in set- and photo-level descriptors. As set-level descriptors are generated with complicated calculations, blind users would need better instructions regarding the set-level descriptors than the ones used in this study.

Ten participants agreed or strongly agreed that the \textbf{descriptors were useful}.  P10 (who was neutral on this) and P11 thought descriptors helped them understand how to collect training examples for the object recognizer.  P10 said, \textit{``(I agree) because I know the quality of the photos, the different aspects of the photos that I take.''}  P11 said, \textit{``It helped me understand what the camera needed in order to recognize the objects.''}  Participants also mentioned that descriptors were useful to identify problems in their training sets.  P10 elaborated \textit{``you have to get feedback or you’re not going to improve [...] it helps you to understand what you’re doing wrong.''}  P2 had a similar idea: \textit{``the explanation (descriptors) afterward, in the analysis, told me that my photographs were not always good, so I have to learn to take better photographs.''}  P12 thought they were not useful because they were error-prone.  P12 said, \textit{``I don’t think that the app is correct, especially when I know, for example, that my hand was not in the photo, or that the object is not cropped because the previous objects were cropped.''} This feedback highlights the need for improving the estimation of descriptors in future work.

Participant feedback suggests that it would have been helpful to include more explicit guidance on how to improve the training photos. For example, P7 suggested similar feedback to Lee~\etal~\cite{lee2019revisiting} and Ahmetovic~\etal~\cite{ahmetovic2020recog} along the descriptors, elaborating \textit{``Cropped, it did not help me know what to do differently. If it said, maybe move up, move down and move camera left, move the camera, right. That would have been more useful.''} Our current implementation of this photo-level descriptor actually can be re-purposed to provide  such feedback. More so, P6 mentioned that the interface for replacing problematic photos in a training set would improve the app. He said \textit{``I would assume the training process can self-evaluate itself and it should sum that up for me and tell me what photos I should replace. [...] you need to replace those bad pictures unless you don’t need them for the training.''} This is an intriguing approach, one that we aim to explore.

\subsubsection{Model Performance}
\begin{figure}[b]
    \centering
    \includegraphics[width=0.75\columnwidth]{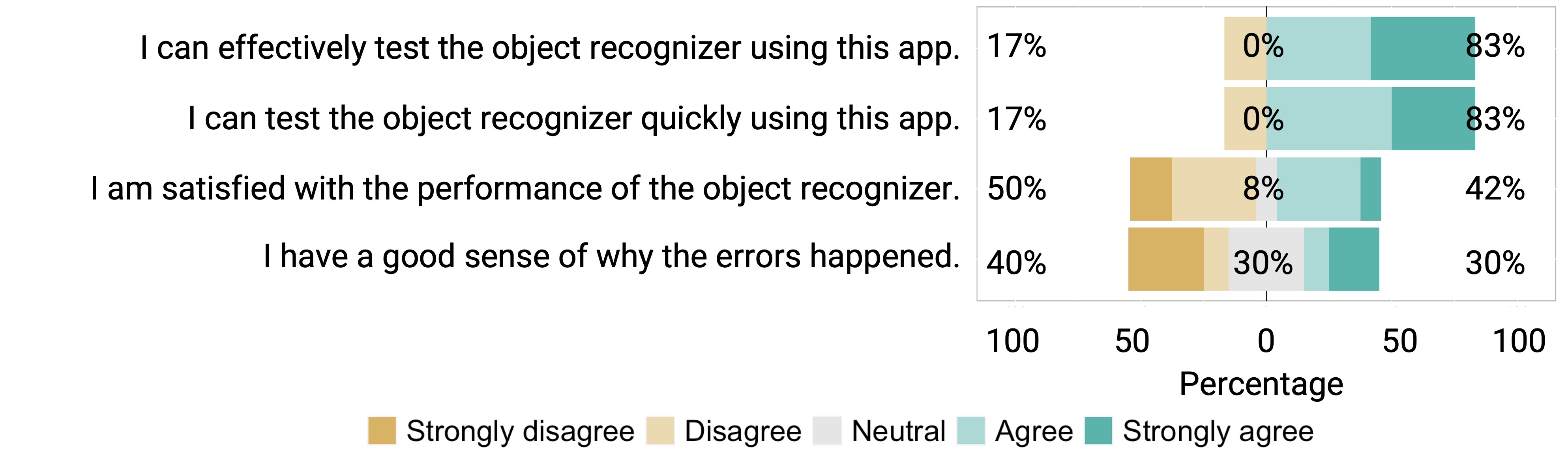}
    \caption{Participants' feedback on the performance of their object recognition models.}
    \label{fig:test_subjective}
\end{figure}

When we asked participants if they were satisfied with the performance of the object recognizer, opinions were divided; five participants agreed or strongly agreed, six participants disagreed or strongly disagreed, and one participant was neutral, as shown in Figure~\ref{fig:test_subjective}.  When accounting for the performance of their model in their subjective responses  (Figure~\ref{fig:test_accuracy_satisfaction}), we observe that participants were not satisfied if the accuracy was lower than 0.6.  However, it did not all come down to  model performance. Open-ended feedback indicates that satisfaction is also related to effort. 
% One of the factors was the amount of effort deeded to train the object recognizer. 
P11 remained neutral even though she did not observe any recognition errors (her model had the highest accuracy) but attributed this to the training task being tedious (see Section 5.5.1). P11 said, \textit{``Because it took so much work to get that small amount of performance.''} 

P7 and P10 agreed that they were satisfied with the performance though their model accuracy was only 0.6 and 0.4, respectively. As legally blind P7, expressed that the performance is good enough to supplement his vision. P10 believed that she just did not train the app properly. She said, \textit{``I think it recognized objects, but if you don't train it properly, then it's not going to recognize anything [...] the Fritos bag was the one that didn't work out, but that was probably my fault.''}

% Alt text
% Likert chart showing the participants' responses to three questions in percentage.
% I can effectively test the object recognizer using this app.	0	16.66667	0	41.66667	41.666667
% I am able to test the object recognizer quickly using this app.	0	16.66667	0	50	33.333333
% I am satisfied with the performance of the object recognizer.	16.66667	33.33333	8.333333	33.33333	8.333333
% I have a good sense of why the errors happened.	30	10	30	10	20

While the majority (9 out of 12) of participants observed recognition errors during testing, many could not explain why. Six participants were neutral or disagreed with the statement that they have a good sense of why the recognition errors occurred. Their responses were simply \textit{``I have no idea.''} or \textit{``I don't know.''}  Though P7 and P10 strongly agreed and agreed, respectively, their rationale was vague.  P10 said \textit{``I think it was my fault. I think it was my training. Other than that, I don’t know.''}  P9 strongly agreed and contributed the recognition errors to imperfect descriptors in training, elaborating \textit{``The reason is because I was teaching it, and I wasn't 100\% sure that it was 100\% accurate. It makes sense that while I was teaching it, I was a little bit off, so its recognition was a little bit off. It kept telling me that the hand was in the photos.''} We believe that these observations motivate the need for accessible computer vision explanations. 

\section{Discussions}
\label{Discussion}
Our user study, exploratory in nature, shows both promising results and future research directions for supporting blind users' interactions with teachable machines. In this section, we first reflect on lessons learned, while discussing implications for designing descriptors to access one's training data in teachable object recognizers and broader teachable applications either assistive or educational. We then discuss limitations in our study that may affect the generalizability of our findings as well as future work for better estimating such descriptors and exploring their potential for explainability.

\subsection{Implications}
Our study provides evidence that descriptors derived from visual attributes used to code training photos in teachable object recognizers, can provide blind users with a means to inspect their data, iterate, and improve their training examples. Challenges often involve onboarding, time needed for training, as well as descriptor accuracy and interpretation. 

Insights from this work are complementary to prior studies exploring the feasibility of training~\cite{kacorri2017people, sosagarcia2017hands, massiceti2021orbit} and camera aiming~\cite{lee2019hands, ahmetovic2020recog, theodorou2021disability} in teachable object recognizers for the blind. More so, the underlying methods for extracting meaningful descriptors, \ie, instance- and set-level characteristics that can be coded by quickly inspecting the training data and that point to noise and variation, respectively, can be adopted for other teachable applications. This is especially critical for those assistive applications where training typically requires similar skills to those the technology aims to fulfill. For example, teachable sound detectors for Deaf/deaf and hard of hearing people~\cite{bragg2016personalizable, goodman2021toward} could benefit from visualizations of the sound examples in a way that allows users to quickly inspect potential noise in a training example and variation across the training set (\eg, better start and end of a recording, multiple sound sources, variation, and other characteristics that hearing users could leverage for experimentation just by listening to the audio).  Indeed, Goodman \etal~\cite{goodman2021toward} observed that Deaf/deaf users collected similar-sounding examples during training and thus, could benefit from an interface that visualizes features of their training sets; data descriptors could fill that need.

Accessing one's training data is also critical for making informal learning activities that typically employ teachable machines with children more inclusive. Learning objectives for AI education in K12 (\eg~\cite{touretzky2019ai}) highlight the
use of interactive systems for exposing children to AI prior to using those that leverage block-based programming~\cite{touretzky2020ai4k12talk}. Dwivedi \etal~\cite{dwivedi2021exploring}
suggest that future teachable interfaces for such activities benefit from classification tasks that allow children to quickly inspect the data and uncover patterns. Thus, it is not a surprise to see many learning activities for exposing children to AI often leverage teachable image classification applications~\cite{google2017teachable, dwivedi2021exploring, vartiainen2020learning, carney2020teachable, google2022teachablev2}. However, in these initial explorations, none of these applications are inclusive of blind children. Our data descriptors could help increase their accessibility \eg, by leveraging our shared code for MYCam and the descriptors.  
Further, we see how researchers working in teachable object recognizers and broader contexts, could benefit from the following insights:  

\begin{description}
    \item[Balancing with demand on time and cognitive load.]{MYCam's set-level descriptors are given at the end of training but image-level descriptors are played every time the user takes a training photo (they can also be accessed when reviewing at a later time). Participants' feedback indicates that, although the image-level descriptors are informative, they add to the training time and cognitive load. Indeed, if we were to compare our study with the times reported in Kacorri \etal~\cite{kacorri2017people} training with descriptors (4.8 seconds per photo on average) took more than double the time without them (2.2 seconds on average), respectively. Still, this was much less when compared to another study, where blind users took photos with real-time camera aiming guidance (\ie, audio and haptic feedback for camera aiming); there, they spent on average 10.3 seconds per photo~\cite{lee2019revisiting} but did not reflect much on the time needed to train. The difference between the two is: MYCam feedback when taking photos is passive and requires listening to a list of descriptors and optimizing simultaneously multiple variables whereas the feedback in Lee \etal~\cite{lee2019revisiting} is interactive and requires listening to an audio cue or sensing vibration and optimizing a single variable (including the object in the frame). This challenge of maximizing information while minimizing cognitive load is not new and calls for better interactivity with the descriptors via opt in/out mechanisms (\eg, play descriptors via press and hold), verbosity controls, or audio haptic feedback. For example, P10 expressed that while descriptors provide hints to problems, they do not directly instruct users on how to solve them. For example, when an object is cropped in a photo, participants did not get feedback on in which direction the camera should move even though this information can be made available from the current implementation. We expect that combining the descriptors with camera guidance (\eg,~\cite{lee2019revisiting, ahmetovic2020recog}) could be helpful.}
    
    \item[Balancing descriptors with instructions.] {There is a rich literature on the value of tutorials, instructions, and in-context interactive assistance for supporting users with technology; a comprehensive review for blind users and smartphone devices can be found in Rodrigues \etal~\cite{rodrigues2019understanding}. Some prior studies with blind users have shown that real-time descriptions can lead to better accuracy and confidence compared to instructions at the start of a task (\eg,~\cite{giudice2020use}). While we did not compare the two, participants' feedback indicate that data descriptors would be complementary and not a substitute for tutorials and instructions. In addition to the real-time feedback, participants call for support in navigating the app and interpreting descriptor values. In our study, the experimenter provided some of this information. For example, for the set-level descriptors the experimenter said: ``you can check how much variation your photos have. For example, a 10\% variation in the background means that most of your photos have similar backgrounds.'' However, participants mentioned that they could better understand the absolute values of descriptors after experimenting. We suspect that the level of understanding for these values would affect both the quality of the training sets as well as how reliable the system is perceived by the users.}
    
    \item[Editing a training set based on descriptors.]{
    The current design of MYCam focuses on informing the users of the attributes of their training sets rather than instructing how to spot potential issues in their training sets or making the data collection process efficient. Participants had to diagnose problems for themselves based on descriptors and replace the entire training set with new photos if they wanted to fix something.
    Participants suggested adding functions to edit (\eg, delete) at a photo level \eg, right after taking a photo that is deemed noisy or while reviewing the training set at the end to make the iterations more efficient. For example, P8 suggested having an interface that filters out bad images based on descriptors or enables users to replace them instead of starting from scratch. This opens up interesting venues for approaches such as active learning and data valuation.}
\end{description}

\subsection{Limitations and Future Work}
There are many limitations that could impact the generalizability of our findings. Our observations come from a small sample even though $N=12$ is most common in human-computer interaction studies~\cite{caine2016local}. Our study is remote, yet participants are recruited from a relatively small area in the US in proximity to the authors' institution. The study is conducted in participants' homes, yet it shares more characteristics with a controlled in-lab study rather than a real-world deployment: object stimuli were predefined and small in number, the duration of the study was relatively short, MYCam was deployed on one of our devices, participants were being observed and had real-time support from the experimenter, and they were somewhat confined in terms of space. 

Specifically, participants were asked to wear smart glasses and communicate with the experimenter through a laptop computer in front of them.  Though these devices were necessary for communication and data analysis, they limit participants' behavior \eg, in walking around with the phone and taking photos in different locations and illuminations.  For example, when participants wanted to vary the backgrounds in photos, they took pictures with different parts of a table.  However, if they could move around outside the user study setup, perhaps they would choose completely different locations for background variation.  More so, using MYCam on one of our iPhone 8 devices instead of their own mobile devices could have affected our observations. All but one participant owned an iPhone; most participants were familiar with iOS apps. However, the difference between their personal phones and our device (\eg, in terms of size and camera location) could have affected the quality of photos and overall perception of the descriptors. We expect that the use of MYCam in a real-world scenario would have resulted in a richer set of contexts in users' photos (typically a table in our study).  Though we limited the objects to three snacks with similar textures and weights, blind people may choose to train on personal objects that may not be products in the market with a larger number of object instances. As the performance of an object recognizer depends on the number of classes and visual difference between the objects, these differences could have affected the performance of a personalized model and blind users' experiences with it.

Our experimental setup was in part restricted by our implementation of the descriptors, which is meant to serve as a proof of concept and is somewhat tied to a predefined set of object stimuli (\eg, for the ARKit to work and for establishing different thresholds).  Although the estimated descriptors had a positive correlation with the manually annotated attributes and enabled participants to inspect their training sets, they were error-prone. When some of the participants noticed the errors in descriptors, they deemed them as well as the object recognition model unreliable. This suggests that it is imperative to further advance approaches related to descriptor estimation for a better user experience.

Due to the lack of datasets for benchmarking our descriptor-based approach, we had to manually create our own dataset for comparison. As in other AI-based systems evaluation, having benchmark datasets is useful to assess systems for generating descriptors in a more widely accepted way. One potential step in this direction would be to invite blind data contributors, who can inspect their personal training data and agree to data sharing, to contribute to such benchmark datasets employing approaches similar to Theodorou \etal~\cite{theodorou2021disability}. 

Last, in this study, images were used for the purpose of training. This approach can provide more control for the blind users over their training sets regarding both incorporating variations and mitigating privacy risk concerns~\cite{akter2020privacy} as it would be less likely for a blind user to capture unnecessary information in an image. For example, blind users usually use their hands as a reference to center the object in the camera frame, but they are often not willing to include their hands in the final photo to preserve user anonymity~\cite{lee2019revisiting}. Also, in the one study where videos were used, blind users had to be trained to follow some instructions and filming techniques~\cite{theodorou2021disability}. On the other hand, video increases the number of collected images since it is a collection of frames. Also, the use of video increases the chance of the object being in the frame at some point~\cite{theodorou2021disability}. Perhaps a way to get the best of the two worlds could be live photos as they are easy to capture (like photos), and they include multiple frames over one to three seconds~\cite{olson2021livephoto}.

\section{Conclusion}
In this work, we examined the challenge of accessing one's training examples in teachable object recognizers, where visual inspection of training photos is not accessible to blind users with the ultimate goal of making machine teaching more inclusive. To this end, we engineered real-time descriptors that indicate to the blind user whether the photo they just took is blurry, if their hand is in it, if the object is cropped, and whether their photos overall vary in object background, distance, and perspective; all factors that can affect model performance. We built MYCam, an accessible and open-source teachable object recognizer iOS app with descriptors.  We shared our findings, observations, and lessons learned from a remote study with 12 blind participants who trained MYCam in their homes to recognize three distinct but visually similar objects. 

Our results showed that participants who choose to iterate their training for an object, were able to provide fewer photos where the object was cropped, included no hand in their photos, and had slightly less blurry photos that overall had more variation in terms of object perspective and size but less in terms of background. Overall, participants increased the variation among their training examples and reduced the number of photos with cropped objects as they moved in training from one object to the next. Some of these changes are reflected in their model performance that somewhat relate to their satisfaction scores. However, errors in descriptor estimates seem to affect overall participants' perception and trust of model performance. Participants' responses indicate that even though it was difficult to gauge the meaning of absolute values for some of the descriptors (\eg, variation), they could infer it based on relative changes. However, many found the training being tedious, opening discussions around the need for balance between information, time, and cognitive load.  These results, taken together, indicate that our novel data descriptors, realized in MYCam, hold potential for facilitating quick inspection of training photos among blind individuals. Going forward, we are excited to continue our endeavors towards building more inclusive participatory machine learning experiences both for blind youth and adults.

% Our approach of generating accessible descriptors contributes to our community effort to improve accessibility of interactions with personalizable AI applications using different modalities.

\begin{acks}
We thank the anonymous reviewers for their thoughtful comments on an earlier draft of this paper. This work is supported by NSF (\#1816380). Kyungjun Lee is supported by NIDILRR (\#90REGE0008).
\end{acks}

\bibliographystyle{ACM-Reference-Format}
\bibliography{references}


\begin{thebibliography}{84}

%%% the Web address.

\ifx \showCODEN    \undefined \def \showCODEN     #1{\unskip}     \fi
\ifx \showDOI      \undefined \def \showDOI       #1{#1}\fi
\ifx \showISBNx    \undefined \def \showISBNx     #1{\unskip}     \fi
\ifx \showISBNxiii \undefined \def \showISBNxiii  #1{\unskip}     \fi
\ifx \showISSN     \undefined \def \showISSN      #1{\unskip}     \fi
\ifx \showLCCN     \undefined \def \showLCCN      #1{\unskip}     \fi
\ifx \shownote     \undefined \def \shownote      #1{#1}          \fi
\ifx \showarticletitle \undefined \def \showarticletitle #1{#1}   \fi
\ifx \showURL      \undefined \def \showURL       {\relax}        \fi
% invisible to TeX
\providecommand\bibfield[2]{#2}
\providecommand\bibinfo[2]{#2}
\providecommand\natexlab[1]{#1}
\providecommand\showeprint[2][]{arXiv:#2}

\bibitem[\protect\citeauthoryear{Abdolrahmani, Easley, Williams, Branham, and
  Hurst}{Abdolrahmani et~al\mbox{.}}{2017}]%
        {abdolrahmani2017embracing}
\bibfield{author}{\bibinfo{person}{Ali Abdolrahmani}, \bibinfo{person}{William
  Easley}, \bibinfo{person}{Michele Williams}, \bibinfo{person}{Stacy Branham},
  {and} \bibinfo{person}{Amy Hurst}.} \bibinfo{year}{2017}\natexlab{}.
\newblock \showarticletitle{Embracing Errors: Examining How Context of Use
  Impacts Blind Individuals' Acceptance of Navigation Aid Errors}. In
  \bibinfo{booktitle}{\emph{Proceedings of the 2017 CHI Conference on Human
  Factors in Computing Systems}} (Denver, Colorado, USA)
  \emph{(\bibinfo{series}{CHI '17})}. \bibinfo{publisher}{Association for
  Computing Machinery}, \bibinfo{address}{New York, NY, USA},
  \bibinfo{pages}{4158–4169}.
\newblock
\showISBNx{9781450346559}
\urldef\tempurl%
\url{https://doi.org/10.1145/3025453.3025528}
\showDOI{\tempurl}


\bibitem[\protect\citeauthoryear{Ahmetovic, Bernareggi, Gerino, and
  Mascetti}{Ahmetovic et~al\mbox{.}}{2014}]%
        {ahmetovic2014zebrarecognizer}
\bibfield{author}{\bibinfo{person}{Dragan Ahmetovic}, \bibinfo{person}{Cristian
  Bernareggi}, \bibinfo{person}{Andrea Gerino}, {and} \bibinfo{person}{Sergio
  Mascetti}.} \bibinfo{year}{2014}\natexlab{}.
\newblock \showarticletitle{ZebraRecognizer: Efficient and Precise Localization
  of Pedestrian Crossings}. In \bibinfo{booktitle}{\emph{2014 22nd
  International Conference on Pattern Recognition}}.
  \bibinfo{pages}{2566--2571}.
\newblock
\urldef\tempurl%
\url{https://doi.org/10.1109/ICPR.2014.443}
\showDOI{\tempurl}


\bibitem[\protect\citeauthoryear{Ahmetovic, Sato, Oh, Ishihara, Kitani, and
  Asakawa}{Ahmetovic et~al\mbox{.}}{2020}]%
        {ahmetovic2020recog}
\bibfield{author}{\bibinfo{person}{Dragan Ahmetovic}, \bibinfo{person}{Daisuke
  Sato}, \bibinfo{person}{Uran Oh}, \bibinfo{person}{Tatsuya Ishihara},
  \bibinfo{person}{Kris Kitani}, {and} \bibinfo{person}{Chieko Asakawa}.}
  \bibinfo{year}{2020}\natexlab{}.
\newblock \bibinfo{booktitle}{\emph{ReCog: Supporting Blind People in
  Recognizing Personal Objects}}.
\newblock \bibinfo{publisher}{Association for Computing Machinery},
  \bibinfo{address}{New York, NY, USA}, \bibinfo{pages}{1–12}.
\newblock
\showISBNx{9781450367080}
\urldef\tempurl%
\url{https://doi.org/10.1145/3313831.3376143}
\showURL{%
\tempurl}


\bibitem[\protect\citeauthoryear{Aira}{Aira}{2017}]%
        {Aira}
\bibfield{author}{\bibinfo{person}{Aira}.} \bibinfo{year}{2017}\natexlab{}.
\newblock \bibinfo{booktitle}{\emph{Your Life, Your Schedule, Right Now.}}
\newblock
\urldef\tempurl%
\url{https://aira.io}
\showURL{%
\tempurl}


\bibitem[\protect\citeauthoryear{Akter, Dosono, Ahmed, Kapadia, and
  Semaan}{Akter et~al\mbox{.}}{2020}]%
        {akter2020privacy}
\bibfield{author}{\bibinfo{person}{Taslima Akter}, \bibinfo{person}{Bryan
  Dosono}, \bibinfo{person}{Tousif Ahmed}, \bibinfo{person}{Apu Kapadia}, {and}
  \bibinfo{person}{Bryan Semaan}.} \bibinfo{year}{2020}\natexlab{}.
\newblock \showarticletitle{"I am uncomfortable sharing what I
  can{\textquoteright}t see": Privacy Concerns of the Visually Impaired with
  Camera Based Assistive Applications}. In \bibinfo{booktitle}{\emph{29th
  USENIX Security Symposium (USENIX Security 20)}}. \bibinfo{publisher}{USENIX
  Association}, \bibinfo{pages}{1929--1948}.
\newblock
\showISBNx{978-1-939133-17-5}
\urldef\tempurl%
\url{https://www.usenix.org/conference/usenixsecurity20/presentation/akter}
\showURL{%
\tempurl}


\bibitem[\protect\citeauthoryear{Amershi, Weld, Vorvoreanu, Fourney, Nushi,
  Collisson, Suh, Iqbal, Bennett, Inkpen, and et~al.}{Amershi
  et~al\mbox{.}}{2019}]%
        {amershi2019guidelines}
\bibfield{author}{\bibinfo{person}{Saleema Amershi}, \bibinfo{person}{Dan
  Weld}, \bibinfo{person}{Mihaela Vorvoreanu}, \bibinfo{person}{Adam Fourney},
  \bibinfo{person}{Besmira Nushi}, \bibinfo{person}{Penny Collisson},
  \bibinfo{person}{Jina Suh}, \bibinfo{person}{Shamsi Iqbal},
  \bibinfo{person}{Paul~N. Bennett}, \bibinfo{person}{Kori Inkpen}, {and}
  \bibinfo{person}{et al.}} \bibinfo{year}{2019}\natexlab{}.
\newblock \showarticletitle{Guidelines for Human-AI Interaction}. In
  \bibinfo{booktitle}{\emph{Proceedings of the 2019 CHI Conference on Human
  Factors in Computing Systems}} (Glasgow, Scotland Uk)
  \emph{(\bibinfo{series}{CHI '19})}. \bibinfo{publisher}{Association for
  Computing Machinery}, \bibinfo{address}{New York, NY, USA}, Article
  \bibinfo{articleno}{3}, \bibinfo{numpages}{13}~pages.
\newblock
\showISBNx{9781450359702}
\urldef\tempurl%
\url{https://doi.org/10.1145/3290605.3300233}
\showDOI{\tempurl}


\bibitem[\protect\citeauthoryear{Bauer, Rojas-Carulla, {\'S}wi{\k{a}}tkowski,
  Sch{\"o}lkopf, and Turner}{Bauer et~al\mbox{.}}{2017}]%
        {bauer2017discriminative}
\bibfield{author}{\bibinfo{person}{Matthias Bauer}, \bibinfo{person}{Mateo
  Rojas-Carulla}, \bibinfo{person}{Jakub~Bart{\l}omiej {\'S}wi{\k{a}}tkowski},
  \bibinfo{person}{Bernhard Sch{\"o}lkopf}, {and} \bibinfo{person}{Richard~E
  Turner}.} \bibinfo{year}{2017}\natexlab{}.
\newblock \showarticletitle{Discriminative k-shot learning using probabilistic
  models}.
\newblock \bibinfo{journal}{\emph{arXiv preprint arXiv:1706.00326}}
  (\bibinfo{year}{2017}).
\newblock
\urldef\tempurl%
\url{https://doi.org/10.48550/ARXIV.1706.00326}
\showDOI{\tempurl}


\bibitem[\protect\citeauthoryear{BeMyEyes}{BeMyEyes}{2016}]%
        {BeMyEyes}
\bibfield{author}{\bibinfo{person}{BeMyEyes}.} \bibinfo{year}{2016}\natexlab{}.
\newblock \bibinfo{booktitle}{\emph{Lend you eyes to the blind}}.
\newblock
\urldef\tempurl%
\url{http://www.bemyeyes.org/}
\showURL{%
\tempurl}


\bibitem[\protect\citeauthoryear{Bragg, Huynh, and Ladner}{Bragg
  et~al\mbox{.}}{2016}]%
        {bragg2016personalizable}
\bibfield{author}{\bibinfo{person}{Danielle Bragg}, \bibinfo{person}{Nicholas
  Huynh}, {and} \bibinfo{person}{Richard~E. Ladner}.}
  \bibinfo{year}{2016}\natexlab{}.
\newblock \showarticletitle{A Personalizable Mobile Sound Detector App Design
  for Deaf and Hard-of-Hearing Users}. In \bibinfo{booktitle}{\emph{Proceedings
  of the 18th International ACM SIGACCESS Conference on Computers and
  Accessibility}} (Reno, Nevada, USA) \emph{(\bibinfo{series}{ASSETS '16})}.
  \bibinfo{publisher}{Association for Computing Machinery},
  \bibinfo{address}{New York, NY, USA}, \bibinfo{pages}{3--13}.
\newblock
\showISBNx{9781450341240}
\urldef\tempurl%
\url{https://doi.org/10.1145/2982142.2982171}
\showDOI{\tempurl}


\bibitem[\protect\citeauthoryear{Bragg, Koller, Bellard, Berke, Boudreault,
  Braffort, Caselli, Huenerfauth, Kacorri, Verhoef, Vogler, and
  Ringel~Morris}{Bragg et~al\mbox{.}}{2019}]%
        {bragg2019sign}
\bibfield{author}{\bibinfo{person}{Danielle Bragg}, \bibinfo{person}{Oscar
  Koller}, \bibinfo{person}{Mary Bellard}, \bibinfo{person}{Larwan Berke},
  \bibinfo{person}{Patrick Boudreault}, \bibinfo{person}{Annelies Braffort},
  \bibinfo{person}{Naomi Caselli}, \bibinfo{person}{Matt Huenerfauth},
  \bibinfo{person}{Hernisa Kacorri}, \bibinfo{person}{Tessa Verhoef},
  \bibinfo{person}{Christian Vogler}, {and} \bibinfo{person}{Meredith
  Ringel~Morris}.} \bibinfo{year}{2019}\natexlab{}.
\newblock \showarticletitle{Sign Language Recognition, Generation, and
  Translation: An Interdisciplinary Perspective}. In
  \bibinfo{booktitle}{\emph{The 21st International ACM SIGACCESS Conference on
  Computers and Accessibility}} (Pittsburgh, PA, USA)
  \emph{(\bibinfo{series}{ASSETS '19})}. \bibinfo{publisher}{Association for
  Computing Machinery}, \bibinfo{address}{New York, NY, USA},
  \bibinfo{pages}{16–31}.
\newblock
\showISBNx{9781450366762}
\urldef\tempurl%
\url{https://doi.org/10.1145/3308561.3353774}
\showDOI{\tempurl}


\bibitem[\protect\citeauthoryear{Bronskill, Gordon, Requeima, Nowozin, and
  Turner}{Bronskill et~al\mbox{.}}{2020}]%
        {bronskill2020tasknorm}
\bibfield{author}{\bibinfo{person}{John Bronskill}, \bibinfo{person}{Jonathan
  Gordon}, \bibinfo{person}{James Requeima}, \bibinfo{person}{Sebastian
  Nowozin}, {and} \bibinfo{person}{Richard Turner}.}
  \bibinfo{year}{2020}\natexlab{}.
\newblock \showarticletitle{{T}ask{N}orm: Rethinking Batch Normalization for
  Meta-Learning}. In \bibinfo{booktitle}{\emph{Proceedings of the 37th
  International Conference on Machine Learning}}
  \emph{(\bibinfo{series}{Proceedings of Machine Learning Research},
  Vol.~\bibinfo{volume}{119})}, \bibfield{editor}{\bibinfo{person}{Hal~Daumé
  III} {and} \bibinfo{person}{Aarti Singh}} (Eds.). \bibinfo{publisher}{PMLR},
  \bibinfo{pages}{1153--1164}.
\newblock
\urldef\tempurl%
\url{https://proceedings.mlr.press/v119/bronskill20a.html}
\showURL{%
\tempurl}


\bibitem[\protect\citeauthoryear{Caine}{Caine}{2016}]%
        {caine2016local}
\bibfield{author}{\bibinfo{person}{Kelly Caine}.}
  \bibinfo{year}{2016}\natexlab{}.
\newblock \showarticletitle{Local Standards for Sample Size at CHI}. In
  \bibinfo{booktitle}{\emph{Proceedings of the 2016 CHI Conference on Human
  Factors in Computing Systems}} (San Jose, California, USA)
  \emph{(\bibinfo{series}{CHI '16})}. \bibinfo{publisher}{Association for
  Computing Machinery}, \bibinfo{address}{New York, NY, USA},
  \bibinfo{pages}{981–992}.
\newblock
\showISBNx{9781450333627}
\urldef\tempurl%
\url{https://doi.org/10.1145/2858036.2858498}
\showDOI{\tempurl}


\bibitem[\protect\citeauthoryear{Campbell, Carpenter, Hashemi, Espinosa,
  Marsan, Borg, Chang, Qiu, Vermeer, Adler, Tepper, Egger, Baker, Sapiro, and
  Dawson}{Campbell et~al\mbox{.}}{2019}]%
        {campbell2019computer}
\bibfield{author}{\bibinfo{person}{Kathleen Campbell},
  \bibinfo{person}{Kimberly~LH Carpenter}, \bibinfo{person}{Jordan Hashemi},
  \bibinfo{person}{Steven Espinosa}, \bibinfo{person}{Samuel Marsan},
  \bibinfo{person}{Jana~Schaich Borg}, \bibinfo{person}{Zhuoqing Chang},
  \bibinfo{person}{Qiang Qiu}, \bibinfo{person}{Saritha Vermeer},
  \bibinfo{person}{Elizabeth Adler}, \bibinfo{person}{Mariano Tepper},
  \bibinfo{person}{Helen~L Egger}, \bibinfo{person}{Jeffery~P Baker},
  \bibinfo{person}{Guillermo Sapiro}, {and} \bibinfo{person}{Geraldine
  Dawson}.} \bibinfo{year}{2019}\natexlab{}.
\newblock \showarticletitle{Computer vision analysis captures atypical
  attention in toddlers with autism}.
\newblock \bibinfo{journal}{\emph{Autism}} \bibinfo{volume}{23},
  \bibinfo{number}{3} (\bibinfo{year}{2019}), \bibinfo{pages}{619--628}.
\newblock
\urldef\tempurl%
\url{https://doi.org/10.1177/1362361318766247}
\showDOI{\tempurl}
\showeprint{https://doi.org/10.1177/1362361318766247}
\newblock
\shownote{PMID: 29595333.}


\bibitem[\protect\citeauthoryear{Carney, Webster, Alvarado, Phillips, Howell,
  Griffith, Jongejan, Pitaru, and Chen}{Carney et~al\mbox{.}}{2020}]%
        {carney2020teachable}
\bibfield{author}{\bibinfo{person}{Michelle Carney}, \bibinfo{person}{Barron
  Webster}, \bibinfo{person}{Irene Alvarado}, \bibinfo{person}{Kyle Phillips},
  \bibinfo{person}{Noura Howell}, \bibinfo{person}{Jordan Griffith},
  \bibinfo{person}{Jonas Jongejan}, \bibinfo{person}{Amit Pitaru}, {and}
  \bibinfo{person}{Alexander Chen}.} \bibinfo{year}{2020}\natexlab{}.
\newblock \showarticletitle{Teachable Machine: Approachable Web-Based Tool for
  Exploring Machine Learning Classification}. In
  \bibinfo{booktitle}{\emph{Extended Abstracts of the 2020 CHI Conference on
  Human Factors in Computing Systems}} (Honolulu, HI, USA)
  \emph{(\bibinfo{series}{CHI EA '20})}. \bibinfo{publisher}{Association for
  Computing Machinery}, \bibinfo{address}{New York, NY, USA},
  \bibinfo{pages}{1–8}.
\newblock
\showISBNx{9781450368193}
\urldef\tempurl%
\url{https://doi.org/10.1145/3334480.3382839}
\showDOI{\tempurl}


\bibitem[\protect\citeauthoryear{Deng, Dong, Socher, Li, Li, and Fei-Fei}{Deng
  et~al\mbox{.}}{2009}]%
        {deng2009imagenet}
\bibfield{author}{\bibinfo{person}{Jia Deng}, \bibinfo{person}{Wei Dong},
  \bibinfo{person}{Richard Socher}, \bibinfo{person}{Li-Jia Li},
  \bibinfo{person}{Kai Li}, {and} \bibinfo{person}{Li Fei-Fei}.}
  \bibinfo{year}{2009}\natexlab{}.
\newblock \showarticletitle{ImageNet: A large-scale hierarchical image
  database}. In \bibinfo{booktitle}{\emph{2009 IEEE Conference on Computer
  Vision and Pattern Recognition}}. \bibinfo{pages}{248--255}.
\newblock
\urldef\tempurl%
\url{https://doi.org/10.1109/CVPR.2009.5206848}
\showDOI{\tempurl}


\bibitem[\protect\citeauthoryear{Dream}{Dream}{2022}]%
        {VoiceDreamScanner}
\bibfield{author}{\bibinfo{person}{Voice Dream}.}
  \bibinfo{year}{2022}\natexlab{}.
\newblock \bibinfo{booktitle}{\emph{Scanner – Voice Dream.}}
\newblock
\urldef\tempurl%
\url{https://www.voicedream.com/scanner/}
\showURL{%
\tempurl}


\bibitem[\protect\citeauthoryear{Dwivedi, Gandhi, Parikh, Coenraad, Bonsignore,
  and Kacorri}{Dwivedi et~al\mbox{.}}{2021}]%
        {dwivedi2021exploring}
\bibfield{author}{\bibinfo{person}{Utkarsh Dwivedi}, \bibinfo{person}{Jaina
  Gandhi}, \bibinfo{person}{Raj Parikh}, \bibinfo{person}{Merijke Coenraad},
  \bibinfo{person}{Elizabeth Bonsignore}, {and} \bibinfo{person}{Hernisa
  Kacorri}.} \bibinfo{year}{2021}\natexlab{}.
\newblock \showarticletitle{Exploring Machine Teaching with Children}. In
  \bibinfo{booktitle}{\emph{2021 IEEE Symposium on Visual Languages and
  Human-Centric Computing (VL/HCC)}}. \bibinfo{pages}{1--11}.
\newblock
\urldef\tempurl%
\url{https://doi.org/10.1109/VL/HCC51201.2021.9576171}
\showDOI{\tempurl}


\bibitem[\protect\citeauthoryear{Fiannaca, Apostolopoulous, and
  Folmer}{Fiannaca et~al\mbox{.}}{2014}]%
        {fiannaca2014headlock}
\bibfield{author}{\bibinfo{person}{Alexander Fiannaca}, \bibinfo{person}{Ilias
  Apostolopoulous}, {and} \bibinfo{person}{Eelke Folmer}.}
  \bibinfo{year}{2014}\natexlab{}.
\newblock \showarticletitle{Headlock: A Wearable Navigation Aid That Helps
  Blind Cane Users Traverse Large Open Spaces}. In
  \bibinfo{booktitle}{\emph{Proceedings of the 16th International ACM SIGACCESS
  Conference on Computers and Accessibility}} (Rochester, New York, USA)
  \emph{(\bibinfo{series}{ASSETS '14})}. \bibinfo{publisher}{Association for
  Computing Machinery}, \bibinfo{address}{New York, NY, USA},
  \bibinfo{pages}{19–26}.
\newblock
\showISBNx{9781450327206}
\urldef\tempurl%
\url{https://doi.org/10.1145/2661334.2661453}
\showDOI{\tempurl}


\bibitem[\protect\citeauthoryear{Giudice, Guenther, Kaplan, Anderson, Knuesel,
  and Cioffi}{Giudice et~al\mbox{.}}{2020}]%
        {giudice2020use}
\bibfield{author}{\bibinfo{person}{Nicholas~A. Giudice},
  \bibinfo{person}{Benjamin~A. Guenther}, \bibinfo{person}{Toni~M. Kaplan},
  \bibinfo{person}{Shane~M. Anderson}, \bibinfo{person}{Robert~J. Knuesel},
  {and} \bibinfo{person}{Joseph~F. Cioffi}.} \bibinfo{year}{2020}\natexlab{}.
\newblock \showarticletitle{Use of an Indoor Navigation System by Sighted and
  Blind Travelers: Performance Similarities across Visual Status and Age}.
\newblock \bibinfo{journal}{\emph{ACM Trans. Access. Comput.}}
  \bibinfo{volume}{13}, \bibinfo{number}{3}, Article \bibinfo{articleno}{11}
  (\bibinfo{date}{aug} \bibinfo{year}{2020}), \bibinfo{numpages}{27}~pages.
\newblock
\showISSN{1936-7228}
\urldef\tempurl%
\url{https://doi.org/10.1145/3407191}
\showDOI{\tempurl}


\bibitem[\protect\citeauthoryear{Goodman, Liu, Jain, McDonnell, Froehlich, and
  Findlater}{Goodman et~al\mbox{.}}{2021}]%
        {goodman2021toward}
\bibfield{author}{\bibinfo{person}{Steven~M. Goodman}, \bibinfo{person}{Ping
  Liu}, \bibinfo{person}{Dhruv Jain}, \bibinfo{person}{Emma~J. McDonnell},
  \bibinfo{person}{Jon~E. Froehlich}, {and} \bibinfo{person}{Leah Findlater}.}
  \bibinfo{year}{2021}\natexlab{}.
\newblock \showarticletitle{Toward User-Driven Sound Recognizer Personalization
  with People Who Are d/Deaf or Hard of Hearing}.
\newblock \bibinfo{journal}{\emph{Proc. ACM Interact. Mob. Wearable Ubiquitous
  Technol.}} \bibinfo{volume}{5}, \bibinfo{number}{2}, Article
  \bibinfo{articleno}{63} (\bibinfo{date}{jun} \bibinfo{year}{2021}),
  \bibinfo{numpages}{23}~pages.
\newblock
\urldef\tempurl%
\url{https://doi.org/10.1145/3463501}
\showDOI{\tempurl}


\bibitem[\protect\citeauthoryear{Google}{Google}{2022a}]%
        {GoogleLookout}
\bibfield{author}{\bibinfo{person}{Google}.} \bibinfo{year}{2022}\natexlab{a}.
\newblock \bibinfo{booktitle}{\emph{Lookout - Assisted vision - Apps on Google
  Play.}}
\newblock
\urldef\tempurl%
\url{https://play.google.com/store/apps/details?id=com.google.android.apps.accessibility.reveal&hl=en_US&gl=US}
\showURL{%
\tempurl}


\bibitem[\protect\citeauthoryear{Google}{Google}{2022b}]%
        {google2022teachablev2}
\bibfield{author}{\bibinfo{person}{Google}.} \bibinfo{year}{2022}\natexlab{b}.
\newblock \bibinfo{title}{Teachable Machine}.
\newblock
  \bibinfo{howpublished}{\url{https://teachablemachine.withgoogle.com/}}.
\newblock


\bibitem[\protect\citeauthoryear{Guerreiro, Sato, Asakawa, Dong, Kitani, and
  Asakawa}{Guerreiro et~al\mbox{.}}{2019}]%
        {guerreiro2019cabot}
\bibfield{author}{\bibinfo{person}{Jo\~{a}o Guerreiro},
  \bibinfo{person}{Daisuke Sato}, \bibinfo{person}{Saki Asakawa},
  \bibinfo{person}{Huixu Dong}, \bibinfo{person}{Kris~M. Kitani}, {and}
  \bibinfo{person}{Chieko Asakawa}.} \bibinfo{year}{2019}\natexlab{}.
\newblock \showarticletitle{CaBot: Designing and Evaluating an Autonomous
  Navigation Robot for Blind People}. In \bibinfo{booktitle}{\emph{The 21st
  International ACM SIGACCESS Conference on Computers and Accessibility}}
  (Pittsburgh, PA, USA) \emph{(\bibinfo{series}{ASSETS '19})}.
  \bibinfo{publisher}{Association for Computing Machinery},
  \bibinfo{address}{New York, NY, USA}, \bibinfo{pages}{68–82}.
\newblock
\showISBNx{9781450366762}
\urldef\tempurl%
\url{https://doi.org/10.1145/3308561.3353771}
\showDOI{\tempurl}


\bibitem[\protect\citeauthoryear{Guo, Chen, Qi, White, Ghosh, Asakawa, and
  Bigham}{Guo et~al\mbox{.}}{2016}]%
        {guo2016vizlens}
\bibfield{author}{\bibinfo{person}{Anhong Guo},
  \bibinfo{person}{Xiang~'Anthony' Chen}, \bibinfo{person}{Haoran Qi},
  \bibinfo{person}{Samuel White}, \bibinfo{person}{Suman Ghosh},
  \bibinfo{person}{Chieko Asakawa}, {and} \bibinfo{person}{Jeffrey~P. Bigham}.}
  \bibinfo{year}{2016}\natexlab{}.
\newblock \showarticletitle{VizLens: A Robust and Interactive Screen Reader for
  Interfaces in the Real World}. In \bibinfo{booktitle}{\emph{Proceedings of
  the 29th Annual Symposium on User Interface Software and Technology}} (Tokyo,
  Japan) \emph{(\bibinfo{series}{UIST '16})}. \bibinfo{publisher}{Association
  for Computing Machinery}, \bibinfo{address}{New York, NY, USA},
  \bibinfo{pages}{651–664}.
\newblock
\showISBNx{9781450341899}
\urldef\tempurl%
\url{https://doi.org/10.1145/2984511.2984518}
\showDOI{\tempurl}


\bibitem[\protect\citeauthoryear{Hitron, Orlev, Wald, Shamir, Erel, and
  Zuckerman}{Hitron et~al\mbox{.}}{2019}]%
        {hitron2019can}
\bibfield{author}{\bibinfo{person}{Tom Hitron}, \bibinfo{person}{Yoav Orlev},
  \bibinfo{person}{Iddo Wald}, \bibinfo{person}{Ariel Shamir},
  \bibinfo{person}{Hadas Erel}, {and} \bibinfo{person}{Oren Zuckerman}.}
  \bibinfo{year}{2019}\natexlab{}.
\newblock \showarticletitle{Can Children Understand Machine Learning Concepts?
  The Effect of Uncovering Black Boxes}. In
  \bibinfo{booktitle}{\emph{Proceedings of the 2019 CHI Conference on Human
  Factors in Computing Systems}} (Glasgow, Scotland Uk)
  \emph{(\bibinfo{series}{CHI '19})}. \bibinfo{publisher}{Association for
  Computing Machinery}, \bibinfo{address}{New York, NY, USA},
  \bibinfo{pages}{1–11}.
\newblock
\showISBNx{9781450359702}
\urldef\tempurl%
\url{https://doi.org/10.1145/3290605.3300645}
\showDOI{\tempurl}


\bibitem[\protect\citeauthoryear{Hong, Lee, Xu, and Kacorri}{Hong
  et~al\mbox{.}}{2019}]%
        {hong2019exploring}
\bibfield{author}{\bibinfo{person}{Jonggi Hong}, \bibinfo{person}{Kyungjun
  Lee}, \bibinfo{person}{June Xu}, {and} \bibinfo{person}{Hernisa Kacorri}.}
  \bibinfo{year}{2019}\natexlab{}.
\newblock \showarticletitle{Exploring Machine Teaching for Object Recognition
  with the Crowd}. In \bibinfo{booktitle}{\emph{Extended Abstracts of the 2019
  CHI Conference on Human Factors in Computing Systems}} (Glasgow, Scotland Uk)
  \emph{(\bibinfo{series}{CHI EA '19})}. \bibinfo{publisher}{Association for
  Computing Machinery}, \bibinfo{address}{New York, NY, USA},
  \bibinfo{pages}{1–6}.
\newblock
\showISBNx{9781450359719}
\urldef\tempurl%
\url{https://doi.org/10.1145/3290607.3312873}
\showDOI{\tempurl}


\bibitem[\protect\citeauthoryear{Hong, Lee, Xu, and Kacorri}{Hong
  et~al\mbox{.}}{2020}]%
        {hong2020crowdsourcing}
\bibfield{author}{\bibinfo{person}{Jonggi Hong}, \bibinfo{person}{Kyungjun
  Lee}, \bibinfo{person}{June Xu}, {and} \bibinfo{person}{Hernisa Kacorri}.}
  \bibinfo{year}{2020}\natexlab{}.
\newblock \bibinfo{booktitle}{\emph{Crowdsourcing the Perception of Machine
  Teaching}}.
\newblock \bibinfo{publisher}{Association for Computing Machinery},
  \bibinfo{address}{New York, NY, USA}, \bibinfo{pages}{1–14}.
\newblock
\showISBNx{9781450367080}
\urldef\tempurl%
\url{https://doi.org/10.1145/3313831.3376428}
\showURL{%
\tempurl}


\bibitem[\protect\citeauthoryear{H{\"o}{\"o}k}{H{\"o}{\"o}k}{2000}]%
        {hook2000steps}
\bibfield{author}{\bibinfo{person}{Kristina H{\"o}{\"o}k}.}
  \bibinfo{year}{2000}\natexlab{}.
\newblock \showarticletitle{Steps to take before intelligent user interfaces
  become real}.
\newblock \bibinfo{journal}{\emph{Interacting with computers}}
  \bibinfo{volume}{12}, \bibinfo{number}{4} (\bibinfo{year}{2000}),
  \bibinfo{pages}{409--426}.
\newblock
\urldef\tempurl%
\url{https://doi.org/10.1016/S0953-5438(99)00006-5}
\showDOI{\tempurl}


\bibitem[\protect\citeauthoryear{Inc.}{Inc.}{2022}]%
        {SuperLidar}
\bibfield{author}{\bibinfo{person}{Virtual Collaboration~Research Inc.}}
  \bibinfo{year}{2022}\natexlab{}.
\newblock \bibinfo{booktitle}{\emph{Super Lidar - Lidar for Blind.}}
\newblock
\urldef\tempurl%
\url{https://apps.apple.com/us/app/super-lidar-lidar-for-blind/id1543706309}
\showURL{%
\tempurl}


\bibitem[\protect\citeauthoryear{Jiang, Zhang, Wachs, and Duerstock}{Jiang
  et~al\mbox{.}}{2016}]%
        {jiang2016enhanced}
\bibfield{author}{\bibinfo{person}{Hairong Jiang}, \bibinfo{person}{Ting
  Zhang}, \bibinfo{person}{Juan~P Wachs}, {and} \bibinfo{person}{Bradley~S
  Duerstock}.} \bibinfo{year}{2016}\natexlab{}.
\newblock \showarticletitle{Enhanced control of a wheelchair-mounted robotic
  manipulator using 3-D vision and multimodal interaction}.
\newblock \bibinfo{journal}{\emph{Computer Vision and Image Understanding}}
  \bibinfo{volume}{149} (\bibinfo{year}{2016}), \bibinfo{pages}{21--31}.
\newblock
\urldef\tempurl%
\url{https://doi.org/10.1016/j.cviu.2016.03.015}
\showDOI{\tempurl}


\bibitem[\protect\citeauthoryear{Kacorri}{Kacorri}{2017}]%
        {kacorri2017teachable}
\bibfield{author}{\bibinfo{person}{Hernisa Kacorri}.}
  \bibinfo{year}{2017}\natexlab{}.
\newblock \showarticletitle{Teachable Machines for Accessibility}.
\newblock \bibinfo{journal}{\emph{SIGACCESS Access. Comput.}}
  \bibinfo{number}{119} (\bibinfo{date}{nov} \bibinfo{year}{2017}),
  \bibinfo{pages}{10–18}.
\newblock
\showISSN{1558-2337}
\urldef\tempurl%
\url{https://doi.org/10.1145/3167902.3167904}
\showDOI{\tempurl}


\bibitem[\protect\citeauthoryear{Kacorri, Dwivedi, Amancherla, Jha, and
  Chanduka}{Kacorri et~al\mbox{.}}{2020a}]%
        {kacorri2020incluset}
\bibfield{author}{\bibinfo{person}{Hernisa Kacorri}, \bibinfo{person}{Utkarsh
  Dwivedi}, \bibinfo{person}{Sravya Amancherla}, \bibinfo{person}{Mayanka Jha},
  {and} \bibinfo{person}{Riya Chanduka}.} \bibinfo{year}{2020}\natexlab{a}.
\newblock \showarticletitle{IncluSet: A Data Surfacing Repository for
  Accessibility Datasets}. In \bibinfo{booktitle}{\emph{The 22nd International
  ACM SIGACCESS Conference on Computers and Accessibility}} (Virtual Event,
  Greece) \emph{(\bibinfo{series}{ASSETS '20})}.
  \bibinfo{publisher}{Association for Computing Machinery},
  \bibinfo{address}{New York, NY, USA}, Article \bibinfo{articleno}{72},
  \bibinfo{numpages}{4}~pages.
\newblock
\showISBNx{9781450371032}
\urldef\tempurl%
\url{https://doi.org/10.1145/3373625.3418026}
\showDOI{\tempurl}


\bibitem[\protect\citeauthoryear{Kacorri, Dwivedi, and Kamikubo}{Kacorri
  et~al\mbox{.}}{2020b}]%
        {kacorri2020data}
\bibfield{author}{\bibinfo{person}{Hernisa Kacorri}, \bibinfo{person}{Utkarsh
  Dwivedi}, {and} \bibinfo{person}{Rie Kamikubo}.}
  \bibinfo{year}{2020}\natexlab{b}.
\newblock \showarticletitle{Data Sharing in Wellness, Accessibility, and
  Aging}.
\newblock \bibinfo{journal}{\emph{NeurIPS 2020 Workshop on Dataset Curation and
  Security}} (\bibinfo{year}{2020}).
\newblock


\bibitem[\protect\citeauthoryear{Kacorri, Kitani, Bigham, and Asakawa}{Kacorri
  et~al\mbox{.}}{2017}]%
        {kacorri2017people}
\bibfield{author}{\bibinfo{person}{Hernisa Kacorri}, \bibinfo{person}{Kris~M.
  Kitani}, \bibinfo{person}{Jeffrey~P. Bigham}, {and} \bibinfo{person}{Chieko
  Asakawa}.} \bibinfo{year}{2017}\natexlab{}.
\newblock \showarticletitle{People with Visual Impairment Training Personal
  Object Recognizers: Feasibility and Challenges}. In
  \bibinfo{booktitle}{\emph{Proceedings of the 2017 CHI Conference on Human
  Factors in Computing Systems}} (Denver, Colorado, USA)
  \emph{(\bibinfo{series}{CHI '17})}. \bibinfo{publisher}{Association for
  Computing Machinery}, \bibinfo{address}{New York, NY, USA},
  \bibinfo{pages}{5839–5849}.
\newblock
\showISBNx{9781450346559}
\urldef\tempurl%
\url{https://doi.org/10.1145/3025453.3025899}
\showDOI{\tempurl}


\bibitem[\protect\citeauthoryear{Kayukawa, Higuchi, Guerreiro, Morishima, Sato,
  Kitani, and Asakawa}{Kayukawa et~al\mbox{.}}{2019}]%
        {kayukawa2019bbeep}
\bibfield{author}{\bibinfo{person}{Seita Kayukawa}, \bibinfo{person}{Keita
  Higuchi}, \bibinfo{person}{Jo\~{a}o Guerreiro}, \bibinfo{person}{Shigeo
  Morishima}, \bibinfo{person}{Yoichi Sato}, \bibinfo{person}{Kris Kitani},
  {and} \bibinfo{person}{Chieko Asakawa}.} \bibinfo{year}{2019}\natexlab{}.
\newblock \showarticletitle{BBeep: A Sonic Collision Avoidance System for Blind
  Travellers and Nearby Pedestrians}. In \bibinfo{booktitle}{\emph{Proceedings
  of the 2019 CHI Conference on Human Factors in Computing Systems}} (Glasgow,
  Scotland Uk) \emph{(\bibinfo{series}{CHI '19})}.
  \bibinfo{publisher}{Association for Computing Machinery},
  \bibinfo{address}{New York, NY, USA}, \bibinfo{pages}{1–12}.
\newblock
\showISBNx{9781450359702}
\urldef\tempurl%
\url{https://doi.org/10.1145/3290605.3300282}
\showDOI{\tempurl}


\bibitem[\protect\citeauthoryear{Khan, Nyholm, Westin, and Dougherty}{Khan
  et~al\mbox{.}}{2014}]%
        {khan2014computer}
\bibfield{author}{\bibinfo{person}{Taha Khan}, \bibinfo{person}{Dag Nyholm},
  \bibinfo{person}{Jerker Westin}, {and} \bibinfo{person}{Mark Dougherty}.}
  \bibinfo{year}{2014}\natexlab{}.
\newblock \showarticletitle{A computer vision framework for finger-tapping
  evaluation in Parkinson's disease}.
\newblock \bibinfo{journal}{\emph{Artificial intelligence in medicine}}
  \bibinfo{volume}{60}, \bibinfo{number}{1} (\bibinfo{year}{2014}),
  \bibinfo{pages}{27--40}.
\newblock
\urldef\tempurl%
\url{https://doi.org/10.1016/j.artmed.2013.11.004}
\showDOI{\tempurl}


\bibitem[\protect\citeauthoryear{Kimmel and Bruckstein}{Kimmel and
  Bruckstein}{2003}]%
        {kimmel2003regularized}
\bibfield{author}{\bibinfo{person}{Ron Kimmel} {and} \bibinfo{person}{Alfred~M
  Bruckstein}.} \bibinfo{year}{2003}\natexlab{}.
\newblock \showarticletitle{Regularized Laplacian zero crossings as optimal
  edge integrators}.
\newblock \bibinfo{journal}{\emph{International Journal of Computer Vision}}
  \bibinfo{volume}{53}, \bibinfo{number}{3} (\bibinfo{year}{2003}),
  \bibinfo{pages}{225--243}.
\newblock
\urldef\tempurl%
\url{https://doi.org/10.1023/A:1023030907417}
\showDOI{\tempurl}


\bibitem[\protect\citeauthoryear{Kuribayashi, Kayukawa, Takagi, Asakawa, and
  Morishima}{Kuribayashi et~al\mbox{.}}{2021}]%
        {kuribayashi2021linechaser}
\bibfield{author}{\bibinfo{person}{Masaki Kuribayashi}, \bibinfo{person}{Seita
  Kayukawa}, \bibinfo{person}{Hironobu Takagi}, \bibinfo{person}{Chieko
  Asakawa}, {and} \bibinfo{person}{Shigeo Morishima}.}
  \bibinfo{year}{2021}\natexlab{}.
\newblock \showarticletitle{LineChaser: A Smartphone-Based Navigation System
  for Blind People to Stand in Lines}. In \bibinfo{booktitle}{\emph{Proceedings
  of the 2021 CHI Conference on Human Factors in Computing Systems}} (Yokohama,
  Japan) \emph{(\bibinfo{series}{CHI '21})}. \bibinfo{publisher}{Association
  for Computing Machinery}, \bibinfo{address}{New York, NY, USA}, Article
  \bibinfo{articleno}{33}, \bibinfo{numpages}{13}~pages.
\newblock
\showISBNx{9781450380966}
\urldef\tempurl%
\url{https://doi.org/10.1145/3411764.3445451}
\showDOI{\tempurl}


\bibitem[\protect\citeauthoryear{Lab}{Lab}{2017}]%
        {google2017teachable}
\bibfield{author}{\bibinfo{person}{Google~Creative Lab}.}
  \bibinfo{year}{2017}\natexlab{}.
\newblock \bibinfo{title}{Teachable Machine}.
\newblock
  \bibinfo{howpublished}{\url{https://teachablemachine.withgoogle.com/v1/}}.
\newblock


\bibitem[\protect\citeauthoryear{Lee, Hong, Jarjue, Mensah, and Kacorri}{Lee
  et~al\mbox{.}}{2022}]%
        {lee2022lab}
\bibfield{author}{\bibinfo{person}{Kyungjun Lee}, \bibinfo{person}{Jonggi
  Hong}, \bibinfo{person}{Ebrima Jarjue}, \bibinfo{person}{Ernest~Essuah
  Mensah}, {and} \bibinfo{person}{Hernisa Kacorri}.}
  \bibinfo{year}{2022}\natexlab{}.
\newblock \showarticletitle{From the Lab to People's Home: Lessons from
  Accessing Blind Participants' Interactions via Smart Glasses in Remote
  Studies}. In \bibinfo{booktitle}{\emph{Proceedings of the 19th International
  Web for All Conference}} (Lyon, France) \emph{(\bibinfo{series}{W4A '22})}.
  \bibinfo{publisher}{Association for Computing Machinery},
  \bibinfo{address}{New York, NY, USA}, Article \bibinfo{articleno}{24},
  \bibinfo{numpages}{11}~pages.
\newblock
\showISBNx{9781450391702}
\urldef\tempurl%
\url{https://doi.org/10.1145/3493612.3520448}
\showDOI{\tempurl}


\bibitem[\protect\citeauthoryear{Lee, Hong, Pimento, Jarjue, and Kacorri}{Lee
  et~al\mbox{.}}{2019}]%
        {lee2019revisiting}
\bibfield{author}{\bibinfo{person}{Kyungjun Lee}, \bibinfo{person}{Jonggi
  Hong}, \bibinfo{person}{Simone Pimento}, \bibinfo{person}{Ebrima Jarjue},
  {and} \bibinfo{person}{Hernisa Kacorri}.} \bibinfo{year}{2019}\natexlab{}.
\newblock \showarticletitle{Revisiting Blind Photography in the Context of
  Teachable Object Recognizers}. In \bibinfo{booktitle}{\emph{The 21st
  International ACM SIGACCESS Conference on Computers and Accessibility}}
  (Pittsburgh, PA, USA) \emph{(\bibinfo{series}{ASSETS '19})}.
  \bibinfo{publisher}{Association for Computing Machinery},
  \bibinfo{address}{New York, NY, USA}, \bibinfo{pages}{83–95}.
\newblock
\showISBNx{9781450366762}
\urldef\tempurl%
\url{https://doi.org/10.1145/3308561.3353799}
\showDOI{\tempurl}


\bibitem[\protect\citeauthoryear{Lee and Kacorri}{Lee and Kacorri}{2019}]%
        {lee2019hands}
\bibfield{author}{\bibinfo{person}{Kyungjun Lee} {and} \bibinfo{person}{Hernisa
  Kacorri}.} \bibinfo{year}{2019}\natexlab{}.
\newblock \showarticletitle{Hands Holding Clues for Object Recognition in
  Teachable Machines}. In \bibinfo{booktitle}{\emph{Proceedings of the 2019 CHI
  Conference on Human Factors in Computing Systems}} (Glasgow, Scotland Uk)
  \emph{(\bibinfo{series}{CHI '19})}. \bibinfo{publisher}{Association for
  Computing Machinery}, \bibinfo{address}{New York, NY, USA},
  \bibinfo{pages}{1–12}.
\newblock
\showISBNx{9781450359702}
\urldef\tempurl%
\url{https://doi.org/10.1145/3290605.3300566}
\showDOI{\tempurl}


\bibitem[\protect\citeauthoryear{Lee, Sato, Asakawa, Kacorri, and Asakawa}{Lee
  et~al\mbox{.}}{2020}]%
        {lee2020pedestrian}
\bibfield{author}{\bibinfo{person}{Kyungjun Lee}, \bibinfo{person}{Daisuke
  Sato}, \bibinfo{person}{Saki Asakawa}, \bibinfo{person}{Hernisa Kacorri},
  {and} \bibinfo{person}{Chieko Asakawa}.} \bibinfo{year}{2020}\natexlab{}.
\newblock \bibinfo{booktitle}{\emph{Pedestrian Detection with Wearable Cameras
  for the Blind: A Two-Way Perspective}}.
\newblock \bibinfo{publisher}{Association for Computing Machinery},
  \bibinfo{address}{New York, NY, USA}, \bibinfo{pages}{1–12}.
\newblock
\showISBNx{9781450367080}
\urldef\tempurl%
\url{https://doi.org/10.1145/3313831.3376398}
\showURL{%
\tempurl}


\bibitem[\protect\citeauthoryear{Lewis}{Lewis}{1995}]%
        {lewis1995ibm}
\bibfield{author}{\bibinfo{person}{James~R Lewis}.}
  \bibinfo{year}{1995}\natexlab{}.
\newblock \showarticletitle{IBM computer usability satisfaction questionnaires:
  psychometric evaluation and instructions for use}.
\newblock \bibinfo{journal}{\emph{International Journal of Human-Computer
  Interaction}} \bibinfo{volume}{7}, \bibinfo{number}{1}
  (\bibinfo{year}{1995}), \bibinfo{pages}{57--78}.
\newblock
\urldef\tempurl%
\url{https://doi.org/10.1080/10447319509526110}
\showDOI{\tempurl}


\bibitem[\protect\citeauthoryear{MacLeod, Bennett, Morris, and Cutrell}{MacLeod
  et~al\mbox{.}}{2017}]%
        {macleod2017understanding}
\bibfield{author}{\bibinfo{person}{Haley MacLeod}, \bibinfo{person}{Cynthia~L.
  Bennett}, \bibinfo{person}{Meredith~Ringel Morris}, {and}
  \bibinfo{person}{Edward Cutrell}.} \bibinfo{year}{2017}\natexlab{}.
\newblock \showarticletitle{Understanding Blind People's Experiences with
  Computer-Generated Captions of Social Media Images}. In
  \bibinfo{booktitle}{\emph{Proceedings of the 2017 CHI Conference on Human
  Factors in Computing Systems}} (Denver, Colorado, USA)
  \emph{(\bibinfo{series}{CHI '17})}. \bibinfo{publisher}{Association for
  Computing Machinery}, \bibinfo{address}{New York, NY, USA},
  \bibinfo{pages}{5988–5999}.
\newblock
\showISBNx{9781450346559}
\urldef\tempurl%
\url{https://doi.org/10.1145/3025453.3025814}
\showDOI{\tempurl}


\bibitem[\protect\citeauthoryear{Manresa-Yee, Varona, Perales, and
  Salinas}{Manresa-Yee et~al\mbox{.}}{2014}]%
        {manresa2014design}
\bibfield{author}{\bibinfo{person}{Cristina Manresa-Yee},
  \bibinfo{person}{Javier Varona}, \bibinfo{person}{Francisco~J Perales}, {and}
  \bibinfo{person}{Iosune Salinas}.} \bibinfo{year}{2014}\natexlab{}.
\newblock \showarticletitle{Design recommendations for camera-based
  head-controlled interfaces that replace the mouse for motion-impaired users}.
\newblock \bibinfo{journal}{\emph{Universal access in the information society}}
  \bibinfo{volume}{13}, \bibinfo{number}{4} (\bibinfo{year}{2014}),
  \bibinfo{pages}{471--482}.
\newblock
\urldef\tempurl%
\url{https://doi.org/10.1007/s10209-013-0326-z}
\showDOI{\tempurl}


\bibitem[\protect\citeauthoryear{Massiceti, Theodorou, Zintgraf, Harris,
  Stumpf, Morrison, Cutrell, and Hofmann}{Massiceti et~al\mbox{.}}{2021}]%
        {massiceti2021orbit}
\bibfield{author}{\bibinfo{person}{Daniela Massiceti}, \bibinfo{person}{Lida
  Theodorou}, \bibinfo{person}{Luisa Zintgraf}, \bibinfo{person}{Matthew~Tobias
  Harris}, \bibinfo{person}{Simone Stumpf}, \bibinfo{person}{Cecily Morrison},
  \bibinfo{person}{Edward Cutrell}, {and} \bibinfo{person}{Katja Hofmann}.}
  \bibinfo{year}{2021}\natexlab{}.
\newblock \bibinfo{title}{ORBIT: A real-world few-shot dataset for teachable
  object recognition collected from people who are blind or low vision}.
\newblock
\newblock
\urldef\tempurl%
\url{https://doi.org/10.25383/CITY.14294597}
\showDOI{\tempurl}


\bibitem[\protect\citeauthoryear{Mediate}{Mediate}{2022}]%
        {Supersense}
\bibfield{author}{\bibinfo{person}{Mediate}.} \bibinfo{year}{2022}\natexlab{}.
\newblock \bibinfo{booktitle}{\emph{Supersense - AI for Blind / Scan text,
  money and objects.}}
\newblock
\urldef\tempurl%
\url{https://www.supersense.app/}
\showURL{%
\tempurl}


\bibitem[\protect\citeauthoryear{Morales}{Morales}{2019}]%
        {morales2019what}
\bibfield{author}{\bibinfo{person}{Carrie Morales}.}
  \bibinfo{year}{2019}\natexlab{}.
\newblock \bibinfo{booktitle}{\emph{What’s Better for the Blind and Low
  Vision? Android or iPhone?}}
\newblock
\urldef\tempurl%
\url{https://liveaccessible.com/2019/03/03/whats-better-for-the-blind-and-low-vision-android-or-iphone/}
\showURL{%
\tempurl}


\bibitem[\protect\citeauthoryear{Morris and Mueller}{Morris and
  Mueller}{2014}]%
        {morris2014blind}
\bibfield{author}{\bibinfo{person}{John Morris} {and} \bibinfo{person}{James
  Mueller}.} \bibinfo{year}{2014}\natexlab{}.
\newblock \showarticletitle{Blind and deaf consumer preferences for android and
  iOS smartphones}.
\newblock In \bibinfo{booktitle}{\emph{Inclusive designing}}.
  \bibinfo{publisher}{Springer}, \bibinfo{pages}{69--79}.
\newblock
\urldef\tempurl%
\url{https://doi.org/10.1007/978-3-319-05095-9_7}
\showDOI{\tempurl}


\bibitem[\protect\citeauthoryear{Morris}{Morris}{2020}]%
        {morris2020ai}
\bibfield{author}{\bibinfo{person}{Meredith~Ringel Morris}.}
  \bibinfo{year}{2020}\natexlab{}.
\newblock \showarticletitle{AI and Accessibility}.
\newblock \bibinfo{journal}{\emph{Commun. ACM}} \bibinfo{volume}{63},
  \bibinfo{number}{6} (\bibinfo{year}{2020}), \bibinfo{pages}{35--37}.
\newblock
\urldef\tempurl%
\url{https://doi.org/10.1145/3356727}
\showDOI{\tempurl}


\bibitem[\protect\citeauthoryear{Norman}{Norman}{1994}]%
        {norman1994might}
\bibfield{author}{\bibinfo{person}{Donald~A Norman}.}
  \bibinfo{year}{1994}\natexlab{}.
\newblock \showarticletitle{How might people interact with agents}.
\newblock \bibinfo{journal}{\emph{Commun. ACM}} \bibinfo{volume}{37},
  \bibinfo{number}{7} (\bibinfo{year}{1994}), \bibinfo{pages}{68--71}.
\newblock
\urldef\tempurl%
\url{https://doi.org/10.1145/176789.176796}
\showDOI{\tempurl}


\bibitem[\protect\citeauthoryear{Olson, Kambhamettu, and McCoy}{Olson
  et~al\mbox{.}}{2021}]%
        {olson2021livephoto}
\bibfield{author}{\bibinfo{person}{Lauren Olson}, \bibinfo{person}{Chandra
  Kambhamettu}, {and} \bibinfo{person}{Kathleen McCoy}.}
  \bibinfo{year}{2021}\natexlab{}.
\newblock \bibinfo{booktitle}{\emph{Towards Using Live Photos to Mitigate Image
  Quality Issues In VQA Photography}}.
\newblock \bibinfo{publisher}{Association for Computing Machinery},
  \bibinfo{address}{New York, NY, USA}.
\newblock
\showISBNx{9781450383066}
\urldef\tempurl%
\url{https://doi.org/10.1145/3441852.3476541}
\showURL{%
\tempurl}


\bibitem[\protect\citeauthoryear{Palmeri and Gauthier}{Palmeri and
  Gauthier}{2004}]%
        {palmeri2004visual}
\bibfield{author}{\bibinfo{person}{Thomas~J. Palmeri} {and}
  \bibinfo{person}{Isabel Gauthier}.} \bibinfo{year}{2004}\natexlab{}.
\newblock \showarticletitle{Visual object understanding}.
\newblock \bibinfo{journal}{\emph{Nature Reviews Neuroscience}}
  \bibinfo{volume}{5}, \bibinfo{number}{4} (\bibinfo{year}{2004}),
  \bibinfo{pages}{291--303}.
\newblock
\showISBNx{1471-0048}
\urldef\tempurl%
\url{https://doi.org/10.1038/nrn1364}
\showDOI{\tempurl}


\bibitem[\protect\citeauthoryear{Patel and Roy}{Patel and Roy}{1998}]%
        {patel1998teachable}
\bibfield{author}{\bibinfo{person}{Rupal Patel} {and} \bibinfo{person}{Deb
  Roy}.} \bibinfo{year}{1998}\natexlab{}.
\newblock \showarticletitle{Teachable interfaces for individuals with
  dysarthric speech and severe physical disabilities}. In
  \bibinfo{booktitle}{\emph{Proceedings of the AAAI Workshop on Integrating
  Artificial Intelligence and Assistive Technology}}. Citeseer,
  \bibinfo{pages}{40--47}.
\newblock


\bibitem[\protect\citeauthoryear{Queiroz, Sampaio, Lima, and Lima}{Queiroz
  et~al\mbox{.}}{2020}]%
        {queiroz2020ai}
\bibfield{author}{\bibinfo{person}{Rubens~Lacerda Queiroz},
  \bibinfo{person}{Fábio~Ferrentini Sampaio}, \bibinfo{person}{Cabral Lima},
  {and} \bibinfo{person}{Priscila Machado~Vieira Lima}.}
  \bibinfo{year}{2020}\natexlab{}.
\newblock \bibinfo{title}{AI from concrete to abstract: demystifying artificial
  intelligence to the general public}.
\newblock
\newblock
\urldef\tempurl%
\url{https://doi.org/10.1007/s00146-021-01151-x}
\showDOI{\tempurl}
\showeprint[arxiv]{2006.04013}~[cs.CY]


\bibitem[\protect\citeauthoryear{Redmon and Farhadi}{Redmon and
  Farhadi}{2018}]%
        {redmon2018yolov3}
\bibfield{author}{\bibinfo{person}{Joseph Redmon} {and} \bibinfo{person}{Ali
  Farhadi}.} \bibinfo{year}{2018}\natexlab{}.
\newblock \bibinfo{title}{YOLOv3: An Incremental Improvement}.
\newblock
\newblock
\urldef\tempurl%
\url{https://doi.org/10.48550/ARXIV.1804.02767}
\showDOI{\tempurl}


\bibitem[\protect\citeauthoryear{Reyes-Amaro, Fadraga-Gonz{\'a}lez,
  Vera-P{\'e}rez, Dom{\'\i}nguez-Campillo, Nodarse-Ravelo, Mesejo-Chiong,
  Moy{\`a}-Alcover, and Jaume-i Cap{\'o}}{Reyes-Amaro et~al\mbox{.}}{2012}]%
        {reyes2012rehabilitation}
\bibfield{author}{\bibinfo{person}{Alejandro Reyes-Amaro},
  \bibinfo{person}{Yanet Fadraga-Gonz{\'a}lez}, \bibinfo{person}{Oscar~Luis
  Vera-P{\'e}rez}, \bibinfo{person}{Elizabeth Dom{\'\i}nguez-Campillo},
  \bibinfo{person}{Jenny Nodarse-Ravelo}, \bibinfo{person}{Alejandro
  Mesejo-Chiong}, \bibinfo{person}{Biel Moy{\`a}-Alcover}, {and}
  \bibinfo{person}{Antoni Jaume-i Cap{\'o}}.} \bibinfo{year}{2012}\natexlab{}.
\newblock \showarticletitle{Rehabilitation of patients with motor disabilities
  using computer vision based techniques}.
\newblock \bibinfo{journal}{\emph{Journal of accessibility and design for all}}
  \bibinfo{volume}{2}, \bibinfo{number}{1} (\bibinfo{year}{2012}),
  \bibinfo{pages}{62--70}.
\newblock
\urldef\tempurl%
\url{https://doi.org/10.17411/jacces.v2i1.87}
\showDOI{\tempurl}


\bibitem[\protect\citeauthoryear{Rodrigues, Santos, Montague, Nicolau, and
  Guerreiro}{Rodrigues et~al\mbox{.}}{2019}]%
        {rodrigues2019understanding}
\bibfield{author}{\bibinfo{person}{Andr{\'e} Rodrigues},
  \bibinfo{person}{Andr{\'e} Santos}, \bibinfo{person}{Kyle Montague},
  \bibinfo{person}{Hugo Nicolau}, {and} \bibinfo{person}{Tiago Guerreiro}.}
  \bibinfo{year}{2019}\natexlab{}.
\newblock \showarticletitle{Understanding the Authoring and Playthrough of
  Nonvisual Smartphone Tutorials}. In \bibinfo{booktitle}{\emph{Human-Computer
  Interaction -- INTERACT 2019}}, \bibfield{editor}{\bibinfo{person}{David
  Lamas}, \bibinfo{person}{Fernando Loizides}, \bibinfo{person}{Lennart Nacke},
  \bibinfo{person}{Helen Petrie}, \bibinfo{person}{Marco Winckler}, {and}
  \bibinfo{person}{Panayiotis Zaphiris}} (Eds.). \bibinfo{publisher}{Springer
  International Publishing}, \bibinfo{address}{Cham}, \bibinfo{pages}{42--62}.
\newblock
\showISBNx{978-3-030-29381-9}


\bibitem[\protect\citeauthoryear{Saha, Fiannaca, Kneisel, Cutrell, and
  Morris}{Saha et~al\mbox{.}}{2019}]%
        {saha2019closing}
\bibfield{author}{\bibinfo{person}{Manaswi Saha}, \bibinfo{person}{Alexander~J.
  Fiannaca}, \bibinfo{person}{Melanie Kneisel}, \bibinfo{person}{Edward
  Cutrell}, {and} \bibinfo{person}{Meredith~Ringel Morris}.}
  \bibinfo{year}{2019}\natexlab{}.
\newblock \showarticletitle{Closing the Gap: Designing for the Last-Few-Meters
  Wayfinding Problem for People with Visual Impairments}. In
  \bibinfo{booktitle}{\emph{The 21st International ACM SIGACCESS Conference on
  Computers and Accessibility}} (Pittsburgh, PA, USA)
  \emph{(\bibinfo{series}{ASSETS '19})}. \bibinfo{publisher}{Association for
  Computing Machinery}, \bibinfo{address}{New York, NY, USA},
  \bibinfo{pages}{222–235}.
\newblock
\showISBNx{9781450366762}
\urldef\tempurl%
\url{https://doi.org/10.1145/3308561.3353776}
\showDOI{\tempurl}


\bibitem[\protect\citeauthoryear{Salisbury, Kamar, and Morris}{Salisbury
  et~al\mbox{.}}{2017}]%
        {Salisbury2017TowardSS}
\bibfield{author}{\bibinfo{person}{Elliot Salisbury}, \bibinfo{person}{Ece
  Kamar}, {and} \bibinfo{person}{Meredith~Ringel Morris}.}
  \bibinfo{year}{2017}\natexlab{}.
\newblock \showarticletitle{Toward Scalable Social Alt Text: Conversational
  Crowdsourcing as a Tool for Refining Vision-to-Language Technology for the
  Blind}. In \bibinfo{booktitle}{\emph{HCOMP}}.
\newblock


\bibitem[\protect\citeauthoryear{Samek, Wiegand, and M{\"{u}}ller}{Samek
  et~al\mbox{.}}{2017}]%
        {samek2017explainable}
\bibfield{author}{\bibinfo{person}{Wojciech Samek}, \bibinfo{person}{Thomas
  Wiegand}, {and} \bibinfo{person}{Klaus{-}Robert M{\"{u}}ller}.}
  \bibinfo{year}{2017}\natexlab{}.
\newblock \showarticletitle{Explainable Artificial Intelligence: Understanding,
  Visualizing and Interpreting Deep Learning Models}.
\newblock \bibinfo{journal}{\emph{CoRR}}  \bibinfo{volume}{abs/1708.08296}
  (\bibinfo{year}{2017}).
\newblock
\showeprint[arxiv]{1708.08296}
\urldef\tempurl%
\url{http://arxiv.org/abs/1708.08296}
\showURL{%
\tempurl}


\bibitem[\protect\citeauthoryear{SeeingAI}{SeeingAI}{2017}]%
        {SeeingAI}
\bibfield{author}{\bibinfo{person}{SeeingAI}.} \bibinfo{year}{2017}\natexlab{}.
\newblock \bibinfo{booktitle}{\emph{An app for visually impaired people that
  narrates the world around you}}.
\newblock
\urldef\tempurl%
\url{https://www.microsoft.com/en-us/seeing-ai}
\showURL{%
\tempurl}


\bibitem[\protect\citeauthoryear{Shannon}{Shannon}{2001}]%
        {shannon2001mathematical}
\bibfield{author}{\bibinfo{person}{Claude~Elwood Shannon}.}
  \bibinfo{year}{2001}\natexlab{}.
\newblock \showarticletitle{A mathematical theory of communication}.
\newblock \bibinfo{journal}{\emph{ACM SIGMOBILE mobile computing and
  communications review}} \bibinfo{volume}{5}, \bibinfo{number}{1}
  (\bibinfo{year}{2001}), \bibinfo{pages}{3--55}.
\newblock


\bibitem[\protect\citeauthoryear{Simard, Amershi, Chickering, Pelton, Ghorashi,
  Meek, Ramos, Suh, Verwey, Wang, and Wernsing}{Simard et~al\mbox{.}}{2017}]%
        {simard2017machine}
\bibfield{author}{\bibinfo{person}{Patrice~Y. Simard}, \bibinfo{person}{Saleema
  Amershi}, \bibinfo{person}{David~M. Chickering},
  \bibinfo{person}{Alicia~Edelman Pelton}, \bibinfo{person}{Soroush Ghorashi},
  \bibinfo{person}{Christopher Meek}, \bibinfo{person}{Gonzalo Ramos},
  \bibinfo{person}{Jina Suh}, \bibinfo{person}{Johan Verwey},
  \bibinfo{person}{Mo Wang}, {and} \bibinfo{person}{John Wernsing}.}
  \bibinfo{year}{2017}\natexlab{}.
\newblock \bibinfo{title}{Machine Teaching: A New Paradigm for Building Machine
  Learning Systems}.
\newblock
\newblock
\urldef\tempurl%
\url{https://doi.org/10.48550/ARXIV.1707.06742}
\showDOI{\tempurl}


\bibitem[\protect\citeauthoryear{Simonyan, Vedaldi, and Zisserman}{Simonyan
  et~al\mbox{.}}{2013}]%
        {simonyan2014deep}
\bibfield{author}{\bibinfo{person}{Karen Simonyan}, \bibinfo{person}{Andrea
  Vedaldi}, {and} \bibinfo{person}{Andrew Zisserman}.}
  \bibinfo{year}{2013}\natexlab{}.
\newblock \bibinfo{title}{Deep Inside Convolutional Networks: Visualising Image
  Classification Models and Saliency Maps}.
\newblock
\newblock
\urldef\tempurl%
\url{https://doi.org/10.48550/ARXIV.1312.6034}
\showDOI{\tempurl}


\bibitem[\protect\citeauthoryear{Son, Krishnagiri, Jeganathan, and Weiland}{Son
  et~al\mbox{.}}{2020}]%
        {son2020crosswalk}
\bibfield{author}{\bibinfo{person}{Hojun Son}, \bibinfo{person}{Divya
  Krishnagiri}, \bibinfo{person}{V.~Swetha Jeganathan}, {and}
  \bibinfo{person}{James Weiland}.} \bibinfo{year}{2020}\natexlab{}.
\newblock \showarticletitle{Crosswalk Guidance System for the Blind}. In
  \bibinfo{booktitle}{\emph{2020 42nd Annual International Conference of the
  IEEE Engineering in Medicine \& Biology Society (EMBC)}}.
  \bibinfo{pages}{3327--3330}.
\newblock
\urldef\tempurl%
\url{https://doi.org/10.1109/EMBC44109.2020.9176623}
\showDOI{\tempurl}


\bibitem[\protect\citeauthoryear{Sosa-Garc\'{\i}a and Odone}{Sosa-Garc\'{\i}a
  and Odone}{2017}]%
        {sosagarcia2017hands}
\bibfield{author}{\bibinfo{person}{Joan Sosa-Garc\'{\i}a} {and}
  \bibinfo{person}{Francesca Odone}.} \bibinfo{year}{2017}\natexlab{}.
\newblock \showarticletitle{“Hands On” Visual Recognition for Visually
  Impaired Users}.
\newblock \bibinfo{journal}{\emph{ACM Trans. Access. Comput.}}
  \bibinfo{volume}{10}, \bibinfo{number}{3}, Article \bibinfo{articleno}{8}
  (\bibinfo{date}{Aug.} \bibinfo{year}{2017}), \bibinfo{numpages}{30}~pages.
\newblock
\showISSN{1936-7228}
\urldef\tempurl%
\url{https://doi.org/10.1145/3060056}
\showDOI{\tempurl}


\bibitem[\protect\citeauthoryear{Stock and Cisse}{Stock and Cisse}{2018}]%
        {stock2018convnets}
\bibfield{author}{\bibinfo{person}{Pierre Stock} {and}
  \bibinfo{person}{Moustapha Cisse}.} \bibinfo{year}{2018}\natexlab{}.
\newblock \showarticletitle{ConvNets and ImageNet Beyond Accuracy:
  Understanding Mistakes and Uncovering Biases}. In
  \bibinfo{booktitle}{\emph{The European Conference on Computer Vision
  (ECCV)}}.
\newblock


\bibitem[\protect\citeauthoryear{Sun, Liu, Chua, and Schiele}{Sun
  et~al\mbox{.}}{2019}]%
        {sun2019meta}
\bibfield{author}{\bibinfo{person}{Qianru Sun}, \bibinfo{person}{Yaoyao Liu},
  \bibinfo{person}{Tat-Seng Chua}, {and} \bibinfo{person}{Bernt Schiele}.}
  \bibinfo{year}{2019}\natexlab{}.
\newblock \showarticletitle{Meta-transfer learning for few-shot learning}. In
  \bibinfo{booktitle}{\emph{Proceedings of the IEEE Conference on Computer
  Vision and Pattern Recognition}}. \bibinfo{pages}{403--412}.
\newblock


\bibitem[\protect\citeauthoryear{Tapu, Mocanu, and Zaharia}{Tapu
  et~al\mbox{.}}{2019}]%
        {tapu2019deep}
\bibfield{author}{\bibinfo{person}{Ruxandra Tapu}, \bibinfo{person}{Bogdan
  Mocanu}, {and} \bibinfo{person}{Titus Zaharia}.}
  \bibinfo{year}{2019}\natexlab{}.
\newblock \showarticletitle{DEEP-HEAR: A Multimodal Subtitle Positioning System
  Dedicated to Deaf and Hearing-Impaired People}.
\newblock \bibinfo{journal}{\emph{IEEE Access}}  \bibinfo{volume}{7}
  (\bibinfo{year}{2019}), \bibinfo{pages}{88150--88162}.
\newblock
\urldef\tempurl%
\url{https://doi.org/10.1109/ACCESS.2019.2925806}
\showDOI{\tempurl}


\bibitem[\protect\citeauthoryear{Theodorou, Massiceti, Zintgraf, Stumpf,
  Morrison, Cutrell, Harris, and Hofmann}{Theodorou et~al\mbox{.}}{2021}]%
        {theodorou2021disability}
\bibfield{author}{\bibinfo{person}{Lida Theodorou}, \bibinfo{person}{Daniela
  Massiceti}, \bibinfo{person}{Luisa Zintgraf}, \bibinfo{person}{Simone
  Stumpf}, \bibinfo{person}{Cecily Morrison}, \bibinfo{person}{Edward Cutrell},
  \bibinfo{person}{Matthew~Tobias Harris}, {and} \bibinfo{person}{Katja
  Hofmann}.} \bibinfo{year}{2021}\natexlab{}.
\newblock \showarticletitle{Disability-First Dataset Creation: Lessons from
  Constructing a Dataset for Teachable Object Recognition with Blind and Low
  Vision Data Collectors}. In \bibinfo{booktitle}{\emph{The 23rd International
  ACM SIGACCESS Conference on Computers and Accessibility}} (Virtual Event,
  USA) \emph{(\bibinfo{series}{ASSETS '21})}. \bibinfo{publisher}{Association
  for Computing Machinery}, \bibinfo{address}{New York, NY, USA}, Article
  \bibinfo{articleno}{27}, \bibinfo{numpages}{12}~pages.
\newblock
\showISBNx{9781450383066}
\urldef\tempurl%
\url{https://doi.org/10.1145/3441852.3471225}
\showDOI{\tempurl}


\bibitem[\protect\citeauthoryear{Touretzky}{Touretzky}{2019}]%
        {touretzky2019ai}
\bibfield{author}{\bibinfo{person}{David Touretzky}.}
  \bibinfo{year}{2019}\natexlab{}.
\newblock \bibinfo{title}{The AI4K12 Initiative: Developing National Guidelines
  for Teaching AI In K-12}.
\newblock
  \bibinfo{howpublished}{\url{https://github.com/touretzkyds/ai4k12/blob/master/documents/CSTA_2019_How_To_Teach_AI_Across_K-12.pdf
  }}.
\newblock


\bibitem[\protect\citeauthoryear{Touretzky}{Touretzky}{2020}]%
        {touretzky2020ai4k12talk}
\bibfield{author}{\bibinfo{person}{David Touretzky}.}
  \bibinfo{year}{2020}\natexlab{}.
\newblock \bibinfo{title}{The AI4K12 Initiative: Developing National Guidelines
  for Teaching AI In K-12}.
\newblock
  \bibinfo{howpublished}{\url{https://raw.githubusercontent.com/touretzkyds/ai4k12/master/documents/GlobalSWEdu2020_Touretzky.pdf}}.
\newblock


\bibitem[\protect\citeauthoryear{Vartiainen, Tedre, and Valtonen}{Vartiainen
  et~al\mbox{.}}{2020}]%
        {vartiainen2020learning}
\bibfield{author}{\bibinfo{person}{Henriikka Vartiainen},
  \bibinfo{person}{Matti Tedre}, {and} \bibinfo{person}{Teemu Valtonen}.}
  \bibinfo{year}{2020}\natexlab{}.
\newblock \showarticletitle{Learning machine learning with very young children:
  {Who} is teaching whom?}
\newblock \bibinfo{journal}{\emph{International Journal of Child-Computer
  Interaction}}  \bibinfo{volume}{25} (\bibinfo{date}{Sept.}
  \bibinfo{year}{2020}), \bibinfo{pages}{1--11}.
\newblock
\showISSN{22128689}
\urldef\tempurl%
\url{https://linkinghub.elsevier.com/retrieve/pii/S2212868920300155}
\showURL{%
\tempurl}


\bibitem[\protect\citeauthoryear{Vinyals, Blundell, Lillicrap, kavukcuoglu, and
  Wierstra}{Vinyals et~al\mbox{.}}{2016}]%
        {vinyals2016matching}
\bibfield{author}{\bibinfo{person}{Oriol Vinyals}, \bibinfo{person}{Charles
  Blundell}, \bibinfo{person}{Timothy Lillicrap}, \bibinfo{person}{koray
  kavukcuoglu}, {and} \bibinfo{person}{Daan Wierstra}.}
  \bibinfo{year}{2016}\natexlab{}.
\newblock \showarticletitle{Matching Networks for One Shot Learning}.
\newblock In \bibinfo{booktitle}{\emph{Advances in Neural Information
  Processing Systems 29}}, \bibfield{editor}{\bibinfo{person}{D.~D. Lee},
  \bibinfo{person}{M.~Sugiyama}, \bibinfo{person}{U.~V. Luxburg},
  \bibinfo{person}{I.~Guyon}, {and} \bibinfo{person}{R.~Garnett}} (Eds.).
  \bibinfo{publisher}{Curran Associates, Inc.}, \bibinfo{pages}{3630--3638}.
\newblock
\urldef\tempurl%
\url{http://papers.nips.cc/paper/6385-matching-networks-for-one-shot-learning.pdf}
\showURL{%
\tempurl}


\bibitem[\protect\citeauthoryear{Vuzix}{Vuzix}{2021}]%
        {VuzixBlade}
\bibfield{author}{\bibinfo{person}{Vuzix}.} \bibinfo{year}{2021}\natexlab{}.
\newblock \bibinfo{booktitle}{\emph{Vuzix Blade Smart Glasses}}.
\newblock
\urldef\tempurl%
\url{https://www.vuzix.com/products/blade-smart-glasses-upgraded}
\showURL{%
\tempurl}


\bibitem[\protect\citeauthoryear{Yamanaka, Kayukawa, Takagi, Nagaoka,
  Hiratsuka, and Kurihara}{Yamanaka et~al\mbox{.}}{2022}]%
        {yamanaka2022one-shot}
\bibfield{author}{\bibinfo{person}{Yutaro Yamanaka}, \bibinfo{person}{Seita
  Kayukawa}, \bibinfo{person}{Hironobu Takagi}, \bibinfo{person}{Yuichi
  Nagaoka}, \bibinfo{person}{Yoshimune Hiratsuka}, {and}
  \bibinfo{person}{Satoshi Kurihara}.} \bibinfo{year}{2022}\natexlab{}.
\newblock \showarticletitle{One-Shot Wayfinding Method for Blind People
  via OCR and Arrow Analysis with a 360-Degree Smartphone Camera}. In
  \bibinfo{booktitle}{\emph{Mobile and Ubiquitous Systems: Computing,
  Networking and Services}}, \bibfield{editor}{\bibinfo{person}{Takahiro Hara}
  {and} \bibinfo{person}{Hirozumi Yamaguchi}} (Eds.).
  \bibinfo{publisher}{Springer International Publishing},
  \bibinfo{address}{Cham}, \bibinfo{pages}{150--168}.
\newblock
\showISBNx{978-3-030-94822-1}
\urldef\tempurl%
\url{https://doi.org/10.1007/978-3-030-94822-1_9}
\showDOI{\tempurl}


\bibitem[\protect\citeauthoryear{Zhang, Jiang, and Davis}{Zhang
  et~al\mbox{.}}{2012}]%
        {zhang2012online}
\bibfield{author}{\bibinfo{person}{Guangxiao Zhang}, \bibinfo{person}{Zhuolin
  Jiang}, {and} \bibinfo{person}{Larry~S Davis}.}
  \bibinfo{year}{2012}\natexlab{}.
\newblock \showarticletitle{Online semi-supervised discriminative dictionary
  learning for sparse representation}. In \bibinfo{booktitle}{\emph{Asian
  conference on computer vision}}. Springer, \bibinfo{pages}{259--273}.
\newblock
\urldef\tempurl%
\url{https://doi.org/10.1007/978-3-642-37331-2_20}
\showDOI{\tempurl}


\bibitem[\protect\citeauthoryear{Zhang, Luo, Cui, and Lu}{Zhang
  et~al\mbox{.}}{2021}]%
        {zhang2021meta}
\bibfield{author}{\bibinfo{person}{Gongjie Zhang}, \bibinfo{person}{Zhipeng
  Luo}, \bibinfo{person}{Kaiwen Cui}, {and} \bibinfo{person}{Shijian Lu}.}
  \bibinfo{year}{2021}\natexlab{}.
\newblock \showarticletitle{Meta-detr: Few-shot object detection via unified
  image-level meta-learning}.
\newblock \bibinfo{journal}{\emph{arXiv preprint arXiv:2103.11731}}
  (\bibinfo{year}{2021}).
\newblock


\bibitem[\protect\citeauthoryear{Zhao, Kupferstein, Castro, Feiner, and
  Azenkot}{Zhao et~al\mbox{.}}{2019}]%
        {zhao2019designing}
\bibfield{author}{\bibinfo{person}{Yuhang Zhao}, \bibinfo{person}{Elizabeth
  Kupferstein}, \bibinfo{person}{Brenda~Veronica Castro},
  \bibinfo{person}{Steven Feiner}, {and} \bibinfo{person}{Shiri Azenkot}.}
  \bibinfo{year}{2019}\natexlab{}.
\newblock \showarticletitle{Designing AR Visualizations to Facilitate Stair
  Navigation for People with Low Vision}. In
  \bibinfo{booktitle}{\emph{Proceedings of the 32nd Annual ACM Symposium on
  User Interface Software and Technology}} (New Orleans, LA, USA)
  \emph{(\bibinfo{series}{UIST '19})}. \bibinfo{publisher}{Association for
  Computing Machinery}, \bibinfo{address}{New York, NY, USA},
  \bibinfo{pages}{387–402}.
\newblock
\showISBNx{9781450368162}
\urldef\tempurl%
\url{https://doi.org/10.1145/3332165.3347906}
\showDOI{\tempurl}


\bibitem[\protect\citeauthoryear{Zhao, Wu, Reynolds, and Azenkot}{Zhao
  et~al\mbox{.}}{2018}]%
        {zhao2018face}
\bibfield{author}{\bibinfo{person}{Yuhang Zhao}, \bibinfo{person}{Shaomei Wu},
  \bibinfo{person}{Lindsay Reynolds}, {and} \bibinfo{person}{Shiri Azenkot}.}
  \bibinfo{year}{2018}\natexlab{}.
\newblock \showarticletitle{A Face Recognition Application for People with
  Visual Impairments: Understanding Use Beyond the Lab}. In
  \bibinfo{booktitle}{\emph{Proceedings of the 2018 CHI Conference on Human
  Factors in Computing Systems}} (Montreal QC, Canada)
  \emph{(\bibinfo{series}{CHI '18})}. \bibinfo{publisher}{Association for
  Computing Machinery}, \bibinfo{address}{New York, NY, USA},
  \bibinfo{pages}{1–14}.
\newblock
\showISBNx{9781450356206}
\urldef\tempurl%
\url{https://doi.org/10.1145/3173574.3173789}
\showDOI{\tempurl}


\bibitem[\protect\citeauthoryear{Zhu, Singla, Zilles, and Rafferty}{Zhu
  et~al\mbox{.}}{2018}]%
        {zhu2018overview}
\bibfield{author}{\bibinfo{person}{Xiaojin Zhu}, \bibinfo{person}{Adish
  Singla}, \bibinfo{person}{Sandra Zilles}, {and} \bibinfo{person}{Anna~N.
  Rafferty}.} \bibinfo{year}{2018}\natexlab{}.
\newblock \showarticletitle{An Overview of Machine Teaching}.
\newblock \bibinfo{journal}{\emph{CoRR}}  \bibinfo{volume}{abs/1801.05927}
  (\bibinfo{year}{2018}).
\newblock
\showeprint[arxiv]{1801.05927}
\urldef\tempurl%
\url{http://arxiv.org/abs/1801.05927}
\showURL{%
\tempurl}


\bibitem[\protect\citeauthoryear{Zoom}{Zoom}{2022}]%
        {zoom}
\bibfield{author}{\bibinfo{person}{Zoom}.} \bibinfo{year}{2022}\natexlab{}.
\newblock \bibinfo{booktitle}{\emph{Video Conferencing, Cloud Phones, Webinars,
  Chat, Virtual Events}}.
\newblock
\urldef\tempurl%
\url{https://zoom.us/}
\showURL{%
\tempurl}


\end{thebibliography}

% \appendix


% Lorem ipsum 

\end{document}